\documentclass[aps,prd,preprint,groupedaddress,nofootinbib]{revtex4-2}

\usepackage{amsmath}
\usepackage{mathtools}
\usepackage[mathscr]{euscript}
\usepackage{relsize}
\usepackage[]{hyperref}
\usepackage{cleveref}
\usepackage{relsize}
\allowdisplaybreaks[2]

\renewcommand*{\Re}{\mathop{\mathfrak{Re}}}
\renewcommand*{\Im}{\mathop{\mathfrak{Im}}}

\DeclareMathOperator{\arccot}{arccot}

\usepackage{hhline}
\usepackage{bm}
\usepackage{bbm}
\usepackage[utf8]{inputenc}
\newcommand\id{\ensuremath{\mathbbm{1}}} 
\usepackage[T1]{fontenc} 
\usepackage{bbm}
\usepackage{graphicx}
\usepackage{subfig}
\usepackage{booktabs} 
\usepackage{xcolor}
\definecolor{olivegreen}{RGB}{0,110,0}

\begin{document}

\title{\boldmath Low-scale Leptogenesis with Minimal Lepton Flavour Violation}

\author{Matthew J. Dolan}
\email[]{matthew.dolan@unimelb.edu.au}
\affiliation{University of Melbourne}

\author{Tomasz P. Dutka}
\email[]{tdutka@student.unimelb.edu.au}
\altaffiliation{Corresponding Author}
\affiliation{University of Melbourne}

\author{Raymond R. Volkas}
\email[]{raymondv@unimelb.edu.au}
\affiliation{University of Melbourne}

\begin{abstract}
We analyse the feasibility of low-scale leptogenesis where the inverse seesaw (ISS) and linear seesaw (LSS) terms are not simultaneously present. In order to generate the necessary mass splittings, we adopt a Minimal Lepton Flavour Violation (MLFV) hypothesis where a sterile neutrino mass degeneracy is broken by flavour effects. We find that resonant leptogenesis is feasible in both scenarios. However, because of a flavour alignment issue, MLFV-ISS leptogenesis succeeds only with a highly tuned choice of Majorana masses. For MLFV-LSS, on the other hand, a large portion of parameter space is able to generate sufficient asymmetry. In both scenarios we find that the lightest neutrino mass must be of order $10^{-2}\text{ eV}$ or below for successful leptogenesis.  We briefly explore implications for low-energy flavour violation experiments, in particular $\mu \rightarrow e\,\gamma$. We find that the future MEG-II experiment, while sensitive to MLFV in our setup, will not be sensitive to the specific regions required for resonant leptogenesis.
\end{abstract}

\maketitle
\newpage

\tableofcontents

\section{Introduction}
\label{sec:intro}

The type-I seesaw model~\cite{Minkowski:1977sc,Yanagida:1979as,GellMann:1980vs,Mohapatra:1979ia,Glashow:1979nm} is the simplest known extension of the Standard Model (SM) that simultaneously addresses the origin of neutrino mass and the generation of the cosmological matter-antimatter asymmetry, the latter through thermal leptogenesis~\cite{Fukugita:1986hr} (for reviews see~\cite{Buchmuller:2004nz,Buchmuller:2005eh,DiBari:2012fz}). One extra heavy Majorana sterile fermion is added per generation, where their masses are required to be higher than the critical temperature of the electroweak phase transition.

This scenario has been extensively studied and it has long been known that in the standard hierarchical\footnote{Hierarchical is typically taken to mean $200 m_{N_1} \lesssim m_{N_2} \lesssim m_{N_3}$.} type-I scenario of thermal leptogenesis, generation of the necessary asymmetry places a \textit{lower-bound}~\cite{Davidson:2002qv,Giudice:2003jh} on the heavy sterile neutrinos (SNs) of order $10^{9}$ GeV. This prevents low-scale realisations of this model in its simplest form. However, an \textit{upper-bound}~\cite{Vissani:1997ys,Clarke:2015gwa} can be placed on the SN neutrino mass of order $10^{7}$ GeV by requiring that radiative corrections to the Higgs mass parameter $\mu^2$ driven by the high seesaw scale does not exceed $1\text{ TeV}^2$. A clear tension exists as there is no overlap between these two bounds. This can be resolved in a number of ways; by invoking supersymmetry to stabilise the Higgs mass, permitting a departure from strictly hierarchical heavy sterile neutrino masses~\cite{Racker:2012vw,Moffat:2018smo}, adding a second Higgs doublet~\cite{Clarke:2015hta,Clarke:2015bea} or adopting a modified lepton sector which allows for low-scale leptogenesis\footnote{We define low-scale leptogenesis to occur at $\mathcal{O}(10-100\text{ TeV})$ or below.} from much lighter SN masses. Note that the first solution, while preventing large corrections, may generate other problems, e.g.~cosmological gravitino overproduction~\cite{Weinberg:1982zq,Baer:2010gr,Baer:2010kw}.

Lowering the thermal leptogenesis scale is an attractive and obvious way to resolve the above tension, and forms the basis of the present analysis. The scale of leptogenesis can be significantly lowered if a quasi-degenerate (meaning not exactly degenerate) spectrum of masses for the SNs is assumed~\cite{Pilaftsis:2003gt,Pilaftsis:2005rv}. In the minimal type-I scenario this has the added consequence of suppressing the Yukawa couplings and therefore reducing discovery signals such as charged lepton flavour violation (cLFV). Additionally, a quasi-degeneracy should be motivated by some symmetry if such models are to be taken seriously. It is natural to expect that the origin of such a theory would tie in with the flavour problem of the SM.

A possible method of lowering the scale of thermal leptogenesis with potential cLFV signals is by introducing two SN states (with opposite lepton number assignments) per light neutrino and simultaneously promoting lepton number to being a ``good symmetry'' of the theory. For certain regimes of the couplings a double suppression of the active neutrino masses occurs, allowing significantly larger Yukawa couplings for SNs at TeV-scale masses compared to the minimal scenario. Additionally, due to the weakly broken lepton number symmetry, the heavy SNs form pseudo-Dirac states with mass splittings proportional to the small lepton number breaking terms. For specific choices of parameters, two popular limiting cases constitute the ``inverse seesaw (ISS)''~\cite{Wyler:1982dd,Leung:1983ti,Mohapatra:1986aw,Mohapatra:1986bd,Ma:1987zm,GonzalezGarcia:1988rw,GonzalezGarcia:1990fb,Deppisch:2004fa} and the ``linear seesaw (LSS)''~\cite{Akhmedov:1995vm,Barr:2003nn,Malinsky:2005bi,Dib:2014fua} models. The two cases can be linked by a rotation~\cite{Ma:2009du} only in the case where no additional symmetry precludes such a rotation (e.g.\ left-right or Pati-Salam symmetry~\cite{Pati:1974yy,Mohapatra:1974hk,Senjanovic:1975rk,Mohapatra:1979ia,Mohapatra:1980yp}). We will therefore be treating the two as independent scenarios in what follows.

In the specific case of the LSS above the electroweak phase transition, the heavy SNs are degenerate in mass and therefore self-energy diagrams vanish, leaving only the highly-suppressed vertex contributions to produce the CP asymmetry~\cite{Pilaftsis:1998pd}. In the case of the ISS, a mass splitting does exist allowing for a resonant enhancement in the self-energy contribution for a specific choice of the mass splitting. However, decreasing the mass splitting also increases the efficiency of washout making asymmetry generation difficult in its simplest realisations~\cite{Gu:2010xc,Blanchet:2010kw,Dolan:2018qpy,Agashe:2018oyk} for ISS at the TeV scale.

In addition to the intra-family SN degeneracies, there is also the possibility of a quasi-degeneracy between SNs from different families, a flavour degeneracy in other words. To that end we explore a scenario in which an $SU(3)^2\times SO(3)^2$ flavour symmetry, to be defined below, is utilised in the lepton sector. Due to the $SO(3)$ symmetry, mass terms for the heavy SNs are proportional to the identity matrix and lead to a flavour degeneracy amongst the SN states in both the ISS and LSS scenarios. However, radiative effects induce $SO(3)$-breaking terms which in turn break these degeneracies, enabling a resonant enhancement in the asymmetry generation for both the LSS and ISS scenarios.

We work in the framework of an extended version of the Minimal Lepton Flavour Violation (MLFV)~\cite{Cirigliano:2005ck,Dinh:2017smk} scenario, which is itself an extension of the well-established idea of Minimal Flavour Violation (MFV) within the quark sector~\cite{Chivukula:1987py,Buras:2000dm,DAmbrosio:2002vsn}. The MFV scheme is an ansatz designed to address the very stringent bounds placed on new physics from various rare hadronic decays and neutral meson oscillations. It is motivated by the idea that taking the Yukawa sector couplings to zero recovers five separate\footnote{In the absence of neutrino mass. If neutrino mass is included more rotations are present.} flavour space rotations which leave the Lagrangian invariant. Therefore an MFV or MLFV Ansatz is based on the core principle that these are ``good'' flavour symmetries which are broken \textit{solely} by the SM Yukawa fields.

Under this ansatz, Yukawa couplings are said to arise from dynamical fields typically known as `spurions' which transform under the flavour symmetries such that the Lagrangian is flavour invariant. These spurion fields are taken to have non-zero vacuum expectation values (VEVs) which spontaneously break the flavour symmetries and thereby induce the masses and flavour mixing we observe. As a result, in its most minimal realisation, any flavour changing process can be predicted from the well-measured Cabibbo-Kobayashi-Maskawa (CKM) mixing matrix and the fermion masses.

Due to the multiple distinct ways in which neutrino masses can be incorporated into the SM there is no unique way of defining an MLFV Ansatz, in contrast with MFV in the quark sector. The potential connection between an MLFV Ansatz and resonant leptogenesis\footnote{See~\cite{Merlo:2018rin} for an example of an MLFV Ansatz which incorporates leptogensis without resonance effects.} has been identified previously~\cite{Cirigliano:2006nu,Branco:2006hz,Cirigliano:2007hb} in the type-I seesaw context. Since then it has been discovered that, due to developments within flavoured leptogenesis, there is a significant reduction in the allowed parameter space\footnote{Recently it was found that flavour effects can still be relevant at much higher temperatures than thought previously, potentially reducing the allowed parameters space even further~\cite{Moffat:2018smo}.}~\cite{Dev:2014laa,Dev:2015wpa,Pilaftsis:2015bja} specific to MLFV leptogenesis. Additionally, in order to produce large leptonic flavour violation whilst suppressing neutrino masses in such models, a separation between the scale of lepton-number violation (LNV) and lepton-flavour violation (LFV) should exist which is not present in the type-I seesaw alone~\cite{Branco:2006hz,Gavela:2009cd}. Neither of these issues are present for MLFV in the ISS and LSS, allowing the possibility of MLFV-induced resonant leptogenesis in these scenarios. We assume an ansatz such that the heavy sterile Majorana neutrinos introduced have exactly degenerate masses in the Lagrangian, motivated by the flavour symmetry of the theory. Radiative effects will break this down to a quasi-degeneracy between the SNs from which a resonant enhancement in the asymmetry generated per decay of SN will occur. For some choices of parameters which we will identify, this enhancement is able generate the necessary baryon asymmetry observed in the universe. We will explore the viability of leptogenesis both from PMNS phases (particularly the Dirac phase) alone~\cite{Pascoli:2006ie,Pascoli:2006ci,Anisimov:2007mw,Dolan:2018qpy} as well as from CPV parameters related to the heavy SNs.

This paper is organised as follows. Section~\ref{sec:Model} reviews the basic idea behind the MFV and MLFV Ans\"{a}tze including justification from hadronic observables for the MFV hypothesis. Section~\ref{sec:leptogenesis} describes our technique for calculation of the baryon asymmetry in the ISS and LSS models with MLFV, while Sec.~\ref{sec:numericalresults} provides a description of our choice of scans and a discussion of the results. Here we also briefly assess the discovery potential of these models at current and future cLFV experiments. We conclude in Sec.~\ref{sec:conclusion}.

\section{Model}
\label{sec:Model}

\subsection{MFV in the quark sector}

The Minimal Flavour Violation ansatz~\cite{Chivukula:1987py,Buras:2000dm,DAmbrosio:2002vsn} recognises that, in the massless limit of absent Yukawa couplings, the quark sector of the three-family SM is invariant under a product of $U(3)$ flavour groups, where an individual $U(3)$ acts in flavour space on a given type of SM multiplet: $q_L$, $u_R$ and $d_R$. Utilising $U(3) = SU(3) \times U(1)$, the quark-sector flavour group may be expressed as
\begin{eqnarray}
\label{eqn:quarkMFV}
\mathcal{G}^Q \;\;\;\;=\;\;\;\; \underbrace{U(1)_B^{\vphantom{T}} \times U(1)_{A^u}^{\vphantom{T}} \times U(1)_{A^d_{\vphantom{a}}}^{\vphantom{T}}} \;\;\;\; &\times & \;\;\;\;\underbrace{SU(3)_{q_L} \times SU(3)_{u_R} \times SU(3)_{d_R}}.\\
\mathcal{G}^Q_{\text{A}} \;\qquad\qquad\quad\quad &\mbox{} & \;\quad\qquad\qquad\quad\quad \mathcal{G}^Q_{\text{NA}}\nonumber
\end{eqnarray}
The Abelian transformations are flavour-blind and we have the freedom to identify them with baryon number and two axial rotations~\cite{Alonso:2013dba}. The non-Abelian transformations do act in flavour space; they govern interactions between the different flavours and hence are responsible for the flavour violation in the theory. Table~\ref{table0} defines the representations of the SM quark fields under the flavour symmetries and specifies a basis for the Abelian generators.

\begin{table}[t]
\begin{center}
\scalebox{1.0}{
\begin{tabular}{cccccccccc}
\toprule
 & & $U(1)_B^{\vphantom{T}}$ & $U(1)_{A^u}^{\vphantom{T}}$ & $U(1)_{A^d}^{\vphantom{T}}$ & & $SU(3)_{q_L}$ & $SU(3)_{u_R}^{\vphantom{T}}$ & $SU(3)_{d_R}^{\vphantom{T}}$ &	 \\ 
\midrule
\midrule
$q_L^{\vphantom{T}}$ & & $1/3$ & $1$ & $1$ & & $\textbf{3}$ & $\textbf{1}$ & $\textbf{1}$ & \\[3pt]
$u_R^{\vphantom{T}}$ & & $1/3$ & $-1$ & $0$ & & $\textbf{1}$ & $\textbf{3}$ & $\textbf{1}$ &\\[3pt]
$d_R^{\vphantom{T}}$ & & $1/3$ & $0$ & $-1$ & & $\textbf{1}$ & $\textbf{1}$ & $\textbf{3}$ &\\
\bottomrule
\end{tabular}
}
\end{center}
\caption{Representations of the SM quark fields under the Abelian and non-Abelian parts of $\mathcal{G}^Q$ which leave the kinetic terms of the SM Lagrangian invariant.} 
\label{table0}
\end{table}

The Yukawa terms in the SM are not invariant under these flavour transformations, but can be made invariant if the Yukawa matrices are `promoted' to be \textit{spurionic} fields.\footnote{In this work, promoting a Yukawa coupling to be a dynamical field will be represented by $Y_i \rightarrow \mathcal{Y}_i$.} Unique transformations under the flavour symmetries are assigned to the spurions in order to make the would-be flavour-breaking terms invariant. This is a hypothesis motivated by the lack of experimental evidence for new flavour-violating physics; it should be treated as the low-energy limit of an unspecified higher-scale, renormalisable theory. The necessary transformations under the non-Abelian symmetries are summarised in~\cref{table-yukspur}.

Quark masses are generated in this framework when the spurionic fields acquire nonzero VEVs alongside the SM Higgs doublet. These background values relate to the \textit{physically measurable} quark masses and mixings where we have the freedom to choose a basis such that
\begin{equation}
\langle \mathcal{Y}_u \rangle = \frac{\sqrt{2}}{v} V_{\text{CKM}}^\dagger \hat{m}_u \qquad\qquad \text{and} \qquad\qquad \langle \mathcal{Y}_d \rangle = \frac{\sqrt{2}}{v} \hat{m}_d
\end{equation}
where
\begin{equation}
\quad\quad\hat{m}_u = \text{diag}(m_u,\,m_c,\,m_t)\;\; \;\;\qquad \text{and}\;\; \qquad \hat{m}_d = \text{diag}(m_d,\,m_s,\,m_b),
\end{equation}
$V_{\text{CKM}}$ is the quark mixing matrix measured by experiment and here, by the usual convention, we have chosen the VEVs of the spurion fields such that the down-type matrix is diagonal and the up-type is not.

The MFV ansatz is based on two assumptions about the high-scale, renormalisable theory:
\begin{itemize}
\item The flavour symmetry present in the SM quark kinetic terms is a `good' symmetry which the high-scale sector respects. The \textit{only} sources of flavour symmetry breaking within the theory at the high scale are from the VEVs of the Yukawa coupling spurions.\footnote{Therefore under this Ansatz flavour violation is completely dictated by the flavour structure of the Yukawa terms in the Lagrangian; any new physics introduced into the theory should not contribute to flavour-changing processes.} The dynamics by which these VEVs are generated lie in the unspecified high-scale theory.
\item The SM is an effective theory for which all renormalisable and non-renormalisable operators must respect both the gauge and \textit{flavour} symmetries of the theory. All operators which are not formally invariant under the flavour symmetry are made so by insertions of the appropriate combinations of Yukawa fields.
\end{itemize}
A number of consequences follow from this ansatz. 

\begin{table}[t]
\begin{center}
\scalebox{1.0}{
\begin{tabular}{ccccc}
\toprule
 & $SU(3)_{q_L}$ & $SU(3)_{u_R}^{\vphantom{T}}$ & $SU(3)_{d_R}^{\vphantom{T}}$ &	 \\ 
\midrule
\midrule
$\mathcal{Y}_{u}^{\vphantom{\dagger}}$ & $\textbf{3}$ & $\bm{\overline{3}}$ & $\textbf{1}$ & \\[3pt]
$\mathcal{Y}_{d}^{\vphantom{\dagger}}$ & $\textbf{3}$ & $\textbf{1}$ & $\bm{\overline{3}}$ &\\[3pt]
\bottomrule
\end{tabular}
}
\end{center}
\caption{Representation assignments for Yukawa spurions such that the Yukawa terms in the MFV effective theory respect the flavour symmetry which exists in their absence.} 
\label{table-yukspur}
\end{table}

SM effective field theory operators which describe flavour changing processes require insertions of spurion combinations in order to become flavour invariant. For example, the four fermion operator $(\overline{Q_L}  \gamma_\mu Q_L )(\overline{Q_L} \gamma^\mu Q_L)$ is forced to be flavour preserving by the flavour symmetry.  In order to get a related $\Delta F =1$ operator, a spurion insertion must be made: $\mathcal{O}_{q1} = (\overline{Q_L} \mathcal{Y}_{u}^{\vphantom{\dagger}} \mathcal{Y}_u^{\dagger} \gamma_\mu Q_L )(\overline{Q_L} \gamma^\mu Q_L)$ is flavour invariant due to the insertion of $\mathcal{Y}_{u}^{\vphantom{\dagger}} \mathcal{Y}_u^{\dagger}$, but gives rise to flavour-changing processes when the spurions acquire VEVs. Due to the hierarchy of the couplings within $\mathcal{Y}_{u}^{\vphantom{\dagger}}$ there is an in-built suppression of flavour-changing processes for the first two generations of quarks. A list of relevant spurion insertions for operators involving fermions at dimension six can be found in~\cite{DAmbrosio:2002vsn}.

Of particular importance when adapted to resonant leptogenesis, which will be discussed in more detail later, all terms in the lowest-order MFV Lagrangian will receive corrections to their couplings from spurion terms which transform in the same way as the basic spurion under the flavour symmetry. For example, in
\begin{equation}
\overline{Q_L}\left( \mathcal{Y}_d^{\vphantom{\dagger}} + \epsilon_1 \mathcal{Y}_d^{\vphantom{\dagger}} \mathcal{Y}_d^{\dagger} \mathcal{Y}_d^{\vphantom{\dagger}} + \dots \right) d_R H,
\end{equation}
the higher-order term $\mathcal{Y}_d^{\vphantom{\dagger}} \mathcal{Y}_d^{\dagger} \mathcal{Y}_d^{\vphantom{\dagger}}$ transforms in the same way as $\mathcal{Y}_d^{\vphantom{\dagger}}$ and therefore under the ansatz has to be included. As Yukawa spurion VEVs are small in magnitude these corrections are generally not significant.

\subsection{MFV vs Experiment}

The motivation behind adopting the MFV ansatz is due to the strong limits placed on the mass scale of new physics from flavour-violating hadronic processes involving the first generation, as illustrated in~\cref{table-mesonmixingbounds}. If new physics appears at scales lower than these bounds it would indicate that either flavour violation arises solely, or dominantly, due to the SM CKM matrix indicating an MFV-like theory, or that it could couple dominantly to the heavier generations which have less stringent bounds.

\begin{table}[t]
\begin{center}
\scalebox{1.0}{
\begin{tabular}{ccc}
\toprule
Hadronic Process & Bound on $\Lambda_{\text{NP}}$ 	 \\ 
\midrule 
\midrule
$K^0 - \overline{K^0}$ & $9 \times 10^3 \text{ TeV}$  \\
\midrule
$B_\text{d} - \overline{B_\text{d}}$ & $4 \times 10^2 \text{ TeV}$  \\
\midrule
$B_\text{s}- \overline{B_\text{s}}$ & $7 \times 10^1 \text{ TeV}$  \\
\bottomrule
\end{tabular}
}
\end{center}
\caption{Approximate lower bounds on the scale $\Lambda_{\text{NP}}$ of new physics from precision measurements of neutral meson oscillations assuming dimension six and $\mathcal{O}(1)$ Wilson coefficients~\cite{Isidori:2013ez}. The bounds are stronger when first-generation quarks are involved.} 
\label{table-mesonmixingbounds}
\end{table}

A key advantage of the former assumption is that strict relations between different flavour changing processes arise due to precise measurements of the CKM parameters. Therefore MFV-like theories have a degree of predictivity which can aid experimental searches. In the case of an exact MFV ansatz, relations amongst different flavour transitions, e.g.\ $b\rightarrow s, b\rightarrow d \text{ and } s\rightarrow d$, can be used to make predictions of unmeasured (or poorly measured) observables~\cite{Hurth:2008jc,Isidori:2010kg,Hurth:2012jn}, some of which are summarised in~\cref{table-MFVpredictions}. They can be compared to their corresponding experimental measurement or limit~\cite{Crivellin:2011ba,CMS:2014xfa,Tanabashi:2018oca,Watanuki:2018xxg,ATLAS-CONF-2018-046} showing strong correlation.

Most channels have measurements (or upper limits) in agreement with MFV predictions, however a discrepancy exists for $B_d \rightarrow \mu^+ \mu^-$ between the CMS/LHCb~\cite{CMS:2014xfa} and ATLAS~\cite{ATLAS-CONF-2018-046} collaborations. CMS/LHCb measure the rate for this process at the limit of the MFV prediction, whereas ATLAS place an upper-bound consistent with MFV. More precise measurements are required and could potentially suggest a deviation from an exact MFV theory. Certainly, however, we have strong evidence that flavour violation is dominantly generated by the Yukawa couplings.

\begin{table}[t]
\begin{center}
{
\scalebox{1.0}{
\begin{tabular}{ccc}
\toprule
Hadronic Process & Measured value or upper limit & MFV prediction	 \\ 
\midrule 
\midrule
$\mathcal{A}_{CP} (\,\,B \rightarrow X_s \gamma\,\,)$ & $0.0144^{+0.0139}_{-0.0139}$ & $<0.02$ \\
\midrule
$\mathcal{B}(\,\,K_L \rightarrow \pi^0\nu\overline{\nu}\,\,)$ & $<2.6\times 10^{-10}$ & $<2.9\times 10^{-10}$\\
\midrule
$\mathcal{B}(B\rightarrow X_s \tau^+ \tau^-)$ & - & $[0.2,3.7]\times 10^{-7}$ \\
\midrule
$\mathcal{B}(\,B\,\,\,\rightarrow\,\,\, X_d \gamma\,\,\,\,)$ & $(1.41^{+0.57}_{-0.57}) \times 10^{-5}$ & $[1.0,4.0]\times 10^{-5}$\\
\midrule
$\mathcal{B}(B_d\,\,\rightarrow\,\, \mu^+ \mu^-)$ & $(0.39^{+0.16}_{-0.14}\times 10^{-9})_{\text{\tiny{CMS/LHCb}}}$ & $< 0.38 \times 10^{-9}$\\
 & < $(0.21\times 10^{-9})_{\text{\tiny{ATLAS}}}$ & \\
\bottomrule
\end{tabular}
}
}
\end{center}
\caption{Predictions for some rare hadronic observables under the MFV ansatz~\cite{Hurth:2008jc,Isidori:2010kg,Hurth:2012jn}, based on other precisely measured processes, compared with their current experimental limits~\cite{Crivellin:2011ba,CMS:2014xfa,Tanabashi:2018oca,Watanuki:2018xxg,ATLAS-CONF-2018-046}.} 
\label{table-MFVpredictions}
\end{table}

\subsection{MLFV}

While the MFV ansatz for the quark sector can be uniquely implemented (with strong agreement with measurement) there is an issue when extending this concept to leptons. Currently we do not know the origin of LFV and most importantly whether the light neutrinos are predominantly Dirac or Majorana. One may expand the MFV ansatz to the lepton sector, which is termed MLFV, but a dependence exists on the specific SM extension adopted for neutrino mass generation and a freedom exists in choosing which couplings are flavour violating~\cite{Davidson:2006bd}. The simplest approach is to extend the SM by three right-handed neutrinos $N_R$ in the usual way where lepton number violation occurs through gauge-invariant Majorana neutrino mass terms,
\begin{equation}
\mathcal{L}_{\textsc{seesaw}}^{\vphantom{m}} = \mathcal{L}_{\textsc{ss}}^{\textsc{k}} - \mathcal{L}_{\textsc{ss}}^{\textsc{m}} ,
\end{equation}
where $\mathcal{L}_{\textsc{ss}}^{\textsc{k}}$ is the usual kinetic term and
\begin{equation}
\label{eqn:type1}
\mathcal{L}_{\textsc{ss}}^{\textsc{m}} = \epsilon_e \overline{\ell}_L Y_e e_R H + \epsilon_D \overline{\ell}_L Y_{D} N_R \tilde{H} + \frac{1}{2} \mu_N \overline{N_R^c} Y_N N_R + \text{ h.c.}
\end{equation}
The \textit{numbers} $\epsilon_{e,D}$ and $\mu_N$ are to be distinguished from the \textit{matrices} (in flavour space) $Y_{e,D,N}$ which will ultimately act as sources for the breaking of the Abelian and non-Abelian flavour symmetries respectively once promoted to spurions having non-zero VEVs.

Similar to the quarks and MFV, the gauge sector of the leptons obey a flavour-transformation invariance which can be defined analogously to eq.~(\ref{eqn:quarkMFV}),
\begin{eqnarray}
\label{eqn:MLFVsymm}
\mathcal{G}^L\quad = \;\;\underbrace{U(1)_{L^{\vphantom{\dagger}}}^{\vphantom{\dagger}}\times U(1)_{A^e_{\vphantom{\dagger}}}^{\vphantom{\dagger}}\times U(1)_{A^N_{\vphantom{\dagger}}}^{\vphantom{\dagger}}} \,\, &\times & \,\,\underbrace{SU(3)_{\ell_L^{\vphantom{\dagger}}}^{\vphantom{\dagger}} \times SU(3)_{e_R^{\vphantom{\dagger}}}^{\vphantom{\dagger}}\times SU(3)_{N_R^{\vphantom{\dagger}}}^{\vphantom{\dagger}}}.\\
\mathcal{G}^L_{\text{A}} \qquad\qquad\quad\;\; &\mbox{} & \;\;\qquad\qquad\qquad\mathcal{G}^L_{\text{NA}}\nonumber
\end{eqnarray}
In the absence of  $\mathcal{L}_{\textsc{ss}}^{\textsc{m}}$, the Lagrangian is invariant under the following $SU(3)$ rotations
\begin{equation}
\ell_L^{\vphantom{\dagger}} \rightarrow \mathcal{U}_{\ell_L}^{\vphantom{\dagger}}\ell_L^{\vphantom{\dagger}} \;\;\;\qquad\qquad\qquad e_R^{\vphantom{\dagger}} \rightarrow \mathcal{U}_{e_R}^{\vphantom{\dagger}}e_R^{\vphantom{\dagger}} \;\;\qquad\qquad\qquad N_R^{\vphantom{\dagger}} \rightarrow \mathcal{U}_{N_R}^{\vphantom{\dagger}}N_R^{\vphantom{\dagger}}
\end{equation}
implying the following representations for the fields:
\begin{equation} 
\ell_L^{\vphantom{\dagger}} \sim \bm{(3,1,1)} \;\qquad\qquad\qquad e_R^{\vphantom{\dagger}} \sim \bm{(1,3,1)} \qquad\qquad\qquad N_R^{\vphantom{\dagger}} \sim \bm{(1,1,3)}.
\end{equation}
Once again, flavour invariance can be recovered (once $\mathcal{L}_{\textsc{ss}}^{\textsc{m}}$ is reintroduced) by promoting the relevant couplings to spurions with fixed transformation properties under the flavour symmetries,
\begin{equation}
\mathcal{Y}_e^{\vphantom{\dagger}} \rightarrow \mathcal{U}_{\ell_L}^{\vphantom{\dagger}}\mathcal{Y}_e^{\vphantom{\dagger}}\, \mathcal{U}_{e_R}^\dagger \qquad\qquad \mathcal{Y}_D^{\vphantom{\dagger}} \rightarrow \mathcal{U}_{\ell_L}^{\vphantom{\dagger}} \mathcal{Y}_D^{\vphantom{\dagger}}\, \mathcal{U}_{N_R}^\dagger \qquad\qquad \mathcal{Y}_N^{\vphantom{\dagger}} \rightarrow \mathcal{U}_{N_R}^* \mathcal{Y}_N^{\vphantom{\dagger}}\, \mathcal{U}_{N_R}^\dagger
\end{equation}
implying the following representations for the fields
\begin{equation}
\;\;\quad\mathcal{Y}_{e}^{\vphantom{\dagger}} \sim \bm{(3,\overline{3},1)} \,\,\quad\qquad\qquad \mathcal{Y}_D^{\vphantom{\dagger}} \sim \bm{(3,1,\overline{3})} \,\,\,\quad\qquad\qquad \mathcal{Y}_N^{\vphantom{\dagger}} \sim \bm{(1,1,\overline{6})}.
\end{equation}
Lepton masses and mixings appear once these spurionic fields acquire non-zero VEVs along with the SM Higgs. We are free to choose a basis such that
\begin{equation}
\label{eqn:T1masses}
\qquad\langle \mathcal{Y}_e \rangle = \frac{\sqrt{2}}{v}\hat{m}_\ell, \qquad \quad \qquad\qquad \langle \mathcal{Y}_D \rangle \langle \mathcal{Y}_N^{-1} \rangle \langle \mathcal{Y}_D^T \rangle = \frac{2 \mu_N}{v^2} U_{\text{PMNS}} \,\hat{m}_{\nu}\, U_{\text{PMNS}}^T
\end{equation} 
where
\begin{equation}
\hat{m}_\ell = \text{diag}(m_e,\,m_\mu,\,m_\tau), \quad\qquad\qquad \hat{m}_\nu = \text{diag}(m_{\nu_1},\,m_{\nu_2},\,m_{\nu_3}).\qquad\quad
\end{equation}
It is clear from~\cref{eqn:T1masses} that it is not possible to assign unique background values to the spurions $\mathcal{Y}_D \text{ and } \mathcal{Y}_N$ in terms of physically measurable parameters. Rather, it is the combination in the seesaw formula that is fixed by the physical observables. As a consequence each individual spurion cannot be uniquely written in terms of physical masses and mixings. In the quark sector, under the MFV ansatz, higher-dimensional operators which control rare processes are made flavour invariant from the necessary combination of spurion fields. As their background values are only dependent on quark sector masses and mixings, this prevents any sources of new physics from contributing in a way not aligned with the SM flavour violation. For the lepton sector, however, rare flavour-violating processes will not in general be made invariant by the exact combination $\mathcal{Y}_D^{\vphantom{\dagger}} \mathcal{Y}_N^{-1} \mathcal{Y}_D{\vphantom{\dagger}}$, but through different combinations of these two spurions (see~\cite{Dinh:2017smk} for a list of examples). Therefore specific flavour-changing processes cannot be uniquely written in terms of physically measurable parameters in the same way they can be for MFV.

Predictivity can be recovered if only one of the spurion fields is taken to have non-trivial flavour transformations, with the other not acting as a source of flavour-symmetry breaking. Usually, it is assumed that the Majorana mass term does not act as a source of lepton-flavour breaking and, and as in the quark sector, only the Yukawa couplings are responsible. If it is assumed that the Majorana mass term is lepton-flavour blind, then $Y_N$ must necessarily be equal to the identity matrix (in flavour space). Under this assumption the non-Abelian flavour symmetry of the theory\footnote{A alternative approach is to assume $\mathcal{G}^L_{\text{NA}} = SU(3)_{V = \left(\ell_L + N_R\right)} \times SU(3)_{e_R}$ and then via Schur's Lemma one of the flavour space matrices must be unitary which can be rotated to the identity matrix with a field redefinition. Under this assumption no degeneracy is implied amongst the right-handed neutrinos and CP invariance is not necessary.~\cite{Alonso:2011jd}} reduces to
\begin{equation}
\label{eqn:GLNA-I}
\mathcal{G}^L_{\text{NA}} = SU(3)_{\ell_L} \times SU(3)_{e_R} \times SO(3)_{N_R} \times CP.
\end{equation}
The lepton-flavour symmetry is broken by $\mathcal{Y}_D^{\vphantom{\dagger}} \text{ and } \mathcal{Y}_e^{\vphantom{\dagger}}$ which can now be written uniquely in terms of lepton masses and mixings
\begin{equation}
\label{eqn:type-I-back}
\qquad\langle \mathcal{Y}_e \rangle = \frac{\sqrt{2}}{v}\hat{m}_\ell,  \quad \qquad\langle \mathcal{Y}_D \rangle \langle \mathcal{Y}_D^T \rangle = \frac{2 \mu_N}{v^2} U_{\text{PMNS}}\, \hat{m}_{\nu} \,U_{\text{PMNS}}^T,\qquad \langle \mathcal{Y}_N \rangle =  \id_3.
\end{equation} 
The measured masses and mixings in the lepton sector fix the combination $\mathcal{Y}_D^{\vphantom{T}} \mathcal{Y}_D^T$ in this setup and not the spurion $\mathcal{Y}_D^{\vphantom{T}}$ itself. However, often the combination $\mathcal{Y}_D^{\vphantom{\dagger}} \mathcal{Y}_D^\dagger$ (which transforms as a \textbf{(8,1,1)} under $\mathcal{G}^L$) is required to be inserted for many operators. Without the imposition of CP invariance, additional and yet unmeasured phases will appear in the predictions for flavour-changing processes. However, in general we expect CP violation (CPV) to be present in the lepton sector.

Allowing for leptonic CPV, as necessary for leptogenesis, a slightly less minimal flavour ansatz is required, namely
\begin{equation}
\label{eqn:GLNA-I-CPV}
\mathcal{G}^L_{\text{NA}} = SU(3)_{\ell_L} \times SU(3)_{e_R} \times SO(3)_{N_R}.
\end{equation}
Lifting the assumption of CP invariance introduces new non-SM phases, as well as allowing non-trivial phases in the Pontecorvo-Maki-Nakagawa-Sakata (PMNS) mixing matrix. While they spoil the absolute predictivity of the theory, they generally lead to only small deviations in cLFV observables~\cite{Cirigliano:2006nu,Branco:2006hz,Cirigliano:2007hb}. This is not in conflict with experiment (as it might be in the quark sector) as currently no cLFV processes have been measured. Additionally, although not definitive, there is interesting evidence for CPV in neutrino oscillations~\cite{Esteban:2018azc} motivating the less-minimal definition of the MLFV principle where phases are allowed and their effect on measurable processes are taken into account.

As already emphasised, MFV and MLFV should be understood as hypotheses motivated by experiment and do not provide a complete UV description of the flavour sector. Attempts have been made in moving from this ansatz to a renormalisable model. In particular in the case of the simplest type-I MLFV Ansatz, the spurion scalar potentials have been studied~\cite{Alonso:2012fy,Alonso:2013mca,Alonso:2013nca} with suggestive conclusions. For hierarchical masses amongst the charged leptons, \textit{large} mixing angles appear when Majorana neutrinos are considered, whereas in the quark sector \textit{small} mixing angles are predicted. It has been suggested that the disparity in the mixing between the lepton and quark sector is therefore due to the Majorana properties of the right-handed neutrinos. These results suggest that UV completions of the MFV and MLFV Ans\"{a}tze could ultimately explain the origin of the flavour structure of the SM.

\subsection{MLFV and the inverse and linear seesaw models}

It was noted in~\cite{Cirigliano:2005ck,Branco:2006hz,Gavela:2009cd} that in order to generate measurable LFV effects whilst explaining the smallness of the active neutrino masses, a decoupling between the LFV scale and LNV scale is required such that $\Lambda_\text{LFV}\ll \Lambda_\text{LN}$. This observation, coupled with any future measurement of cLFV, motivates the incorporation of such a scale separation into the MLFV hypothesis. Note that LNV arises from the breaking of Abelian symmetries in eq.~(\ref{eqn:MLFVsymm}), whereas LFV arises from the breaking of the non-Abelian symmetries.

Scale separation, however, does not occur in the minimal version of the type-I and type-III seesaw mechanisms where only one new scale is introduced such that $\Lambda_{\text{LFV}} = \Lambda_{\text{LNV}} = \mu_N^{\vphantom{T}}$. Under an MLFV Ansatz the flavour spurions decouple when the LNV scale is taken to infinity~\cite{Gavela:2009cd}, preventing significant LFV. In contrast it was argued that the type-II seesaw does exhibit such a behaviour as $\Lambda_{\text{LNV}} \sim  M_\Delta^2/\mu_\Delta$ while $\Lambda_{\text{LFV}} \sim M_\Delta$, where $M_\Delta$ is the mass of the triplet and $\mu_\Delta$ is the dimensionful cubic coupling constant between the triplet and the SM Higgs doublet. This therefore defines the only minimal seesaw model for which an MLFV Ansatz may lead to significant LFV.

The simplest\footnote{In~\cite{Branco:2006hz} it was suggested the two scales could be made distinct by implementing the minimal type-I seesaw mechanism through an extended theory such as the MSSM, where $\Lambda_{\text{LFV}} = m_{\tilde{\ell}} \sim \mathcal{O}(\text{TeV})$ is proportional to the soft-SUSY breaking terms.} \textit{fermionic} completion which achieves a natural scale separation features in the ISS and LSS mechanisms (and a combination of the two) where small LNV parameters are introduced in order to ensure an \textit{approximate} $U(1)_L$ symmetry. Here additional sterile neutrino degrees of freedom $S_L$ are introduced alongside the right-handed neutrinos $N_R$,
\begin{equation}
\label{eqn:ISSlag}
\mathcal{L}_{\textsc{ess}}^{\vphantom{m}} = \mathcal{L}_{\textsc{ess}}^{\textsc{k}} - \mathcal{L}_{\textsc{ess}}^{\textsc{m}},
\end{equation}
\begin{equation}
\label{eqn:ISSlagm}
\begin{split}
\mathcal{L}_{\textsc{ess}}^{\textsc{m}} = \epsilon_e \overline{\ell_{L}} \mathcal{Y}_e^{\vphantom{T}} H e_{R} &+ \epsilon_D \overline{\ell_{L}} \mathcal{Y}_D^{\vphantom{T}} \tilde{H} N_{R} + \epsilon_L \overline{\ell_{L}} \mathcal{Y}_L^{\vphantom{T}} \tilde{H} \left( S_L \right)^{c}
 + m_{R}\overline{\left( N_{R}  \right)^c }\, \mathcal{Y}_R^{\vphantom{T}} \left( S_{L} \right)^{c} \\
 &+ \frac{1}{2}\mu_S^{\vphantom{T}} \,\overline{S_{L}^{\vphantom{T}} } \, \mathcal{Y}_{\mu_S}^{\vphantom{T}} \left( S_{L}^{\vphantom{T}}\right)^{c}+ \frac{1}{2} \mu_N^{\vphantom{T}}\, \overline{(N_{R}^{\vphantom{T}})^{c} }\, \mathcal{Y}_{\mu_N}^{\vphantom{T}} N_{R}^{\vphantom{T}} +\text{h.c},
 \end{split}
\end{equation}
where it is conventional to choose $L(\ell_L) = L(N_R) = 1 \text{ and } L(S_L)=-1$. 

As before, $\mathcal{Y}_i$ correspond to dimensionless $3 \times 3$ matrices in flavour space associated with the breaking of $\mathcal{G}^L_{\text{NA}}$. We impose an $SO(3)$ rather than $SU(3)$ flavour invariance for the heavy singlets to ensure that only the Yukawa interactions with the Higgs doublet act as sources of flavour-symmetry breaking. Therefore all bare mass terms related to the heavy singlets are proportional to the identity matrix. The flavour symmetry of the theory is defined analogously to eq.~(\ref{eqn:MLFVsymm}),
\begin{eqnarray}
\label{eqn:ISSMLFVsymm}
\mathscr{G}^\textsc{l} = \,\underbrace{U(1)_L\times U(1)_{A^e}\times U(1)_{A^N}\times U(1)_{A^S}}  &\times & \underbrace{SU(3)_{\ell_L} \times SU(3)_{e_R}\times SO(3)_{N_R}\times SO(3)_{S_L}}.\nonumber\\
\mathscr{G}^\textsc{l}_{\textsc{a}} \qquad\qquad\qquad\quad\; &\mbox{} & \qquad\;\;\qquad\qquad\qquad\mathscr{G}^\textsc{l}_{\textsc{na}}
\end{eqnarray}
The mass matrix has the form
\begin{equation}
\label{fullmatrix}
M_{\nu}=\begin{pmatrix}
0 & \frac{1}{\sqrt{2}} \epsilon_D^{\vphantom{T}} \mathcal{Y}_D^{\vphantom{T}} v & \frac{1}{\sqrt{2}}\epsilon_L^{\vphantom{T}} \mathcal{Y}_L^{\vphantom{T}} v \\
\frac{1}{\sqrt{2}} \epsilon_D^{\vphantom{T}} \mathcal{Y}_D^T v & \mu_N^{\vphantom{T}} \mathcal{Y}_{\mu_N}^{\vphantom{T}} & m_{R} \mathcal{Y}_R^{\vphantom{T}} \\
\frac{1}{\sqrt{2}}\epsilon_L \mathcal{Y}_L^T v & m_{R} \mathcal{Y}_R^{\vphantom{T}} & \mu_S^{\vphantom{T}} \mathcal{Y}_{\mu_S}^{\vphantom{T}} \end{pmatrix}
\end{equation}
where for the ISS ($\epsilon_L^{\vphantom{T}} = 0$)
\begin{equation}
\Lambda_{\text{LNV}} \sim \frac{m_R^2}{\mu} \;\;\qquad\qquad\text{ and }\qquad\qquad\;\; \Lambda_{\text{LFV}} \sim m_R^{\vphantom{2}}
\end{equation}
and for the LSS ($\mu_N^{\vphantom{T}},\,\mu_S^{\vphantom{T}} = 0$) 
\begin{equation}
\Lambda_{\text{LNV}} \sim \frac{m_R^{\vphantom{2}}}{\sqrt{\epsilon_L}} \;\;\qquad\qquad\text{ and } \;\;\qquad\qquad\Lambda_{\text{LFV}} \sim m_R^{\vphantom{2}}.
\end{equation}
The imposition of small lepton number violation means that $\mu_S^{\vphantom{T}},\,\mu_N^{\vphantom{T}} \text{ and } \epsilon_L^{\vphantom{T}}$ are small. This provides the desired separation of scales.

As with the quark sector and the example of the minimal type-I seesaw model, the kinetic terms in~\cref{eqn:ISSlag} are invariant under the following flavour rotations
\begin{eqnarray}
\quad\ell_L^{\vphantom{T}} \rightarrow \mathcal{U}_{\ell_L} \ell_L^{\vphantom{T}} \quad\quad\quad\quad
e_R^{\vphantom{T}} & \rightarrow & \mathcal{U}_{e_R} e_R^{\vphantom{T}} \quad\quad\quad\quad
N_R^{\vphantom{T}} \rightarrow \mathcal{O}_{N_R} N_R^{\vphantom{T}}\quad\quad\quad
S_L^{\vphantom{T}} \rightarrow \mathcal{O}_{S_L} S_L^{\vphantom{T}}\quad\quad\quad\quad\quad
\end{eqnarray}
where $\mathcal{U}_i$ are $3 \times 3$ special unitary matrices, $\mathcal{O}_i$ are $3\times 3$ real-orthogonal matrices and for example $\ell_L^{\vphantom{T}} = (e_L^{\vphantom{T}},\mu_L^{\vphantom{T}},\tau_L^{\vphantom{T}})$. The leptons transform as fundamentals under their respective non-Abelian rotations
\begin{eqnarray}
\label{eqn:ISSleptontrans}
\ell_L \sim \bm{(3,1,1,1)}\quad\quad e_R \sim \bm{(1,3,1,1)}\quad\quad N_R \sim \bm{(1,1,3,1)}\quad\quad S_L \sim \bm{(1,1,1,3)}.\qquad
\end{eqnarray}
It follows that~\cref{eqn:ISSlagm} can be made invariant by promoting the relevant couplings to dynamical spurions with the flavour transformation properties,
\begin{eqnarray}
\label{eqn:ISSspuriontrans}
\quad\;\;\;\mathcal{Y}_e \rightarrow \mathcal{U}_{\ell_L} \mathcal{Y}_e \,\mathcal{U}^{\dagger}_{e_R}\quad\quad
\mathcal{Y}_D & \rightarrow & \mathcal{U}_{\ell_L} \mathcal{Y}_D \,\mathcal{O}^{T}_{N_R}\quad\quad
\mathcal{Y}_L \rightarrow \mathcal{U}_{\ell_L} \mathcal{Y}_L \,\mathcal{O}^{T}_{S_L}\quad\quad
\end{eqnarray}
implying that
\begin{eqnarray}
\quad\;\;\mathcal{Y}_e \sim  \bm{(3,\overline{3},1,1)} \quad\quad \mathcal{Y}_D &\sim & \bm{(3,1,3,1)} \quad\quad\;\; \mathcal{Y}_L \sim \bm{(3,1,1,3)}. \quad\quad 
\end{eqnarray}
Lepton masses and mixings result as usual from non-zero VEVs for the spurion fields where, similarly to~\cref{eqn:T1masses}, a basis can be chosen such that
\begin{equation}
\langle \mathcal{Y}_e \rangle = \frac{\sqrt{2}}{v} \hat{m}_{\ell}\quad\quad\quad\quad\quad\quad\quad\quad\quad \langle \mathcal{Y}_D \rangle  \langle \mathcal{Y}_D^T \rangle = \frac{2 \,m_R^2}{ v^2 \mu_S^{\vphantom{T}}} U_{\text{PMNS}}\, \hat{m}_{\nu}\, U^T_{\text{PMNS}}
\end{equation}
in the case of the inverse seesaw and
\begin{equation}
\langle \mathcal{Y}_e \rangle = \frac{\sqrt{2}}{v} \hat{m}_{\ell} \;\;\; \quad\quad\quad \langle \mathcal{Y}_D \rangle  \langle \mathcal{Y}_L^T \rangle + \langle Y_L \rangle \langle \mathcal{Y}_D^T \rangle = \frac{2 \,m_R}{ v^2\epsilon_L^{\vphantom{T}}} U_{\text{PMNS}} \,\hat{m}_{\nu}\, U^T_{\text{PMNS}}
\end{equation}
in the case of the linear seesaw.

\section{Leptogenesis}
\label{sec:leptogenesis}

The leading-order flavour degeneracy amongst the heavy SNs is built into the ansatz, with higher order spurion effects breaking the degeneracy.  This allows for a resonant enhancement in asymmetry generation. 

The case of the simplest MLFV scenario has been explored in detail~\cite{Cirigliano:2006nu,Branco:2006hz,Uhlig:2006xf,Cirigliano:2007hb}. Recent developments related to flavoured leptogenesis~\cite{Dev:2014laa,Dev:2015wpa,Pilaftsis:2015bja} have, however, highlighted the importance of a term not previously included, such that the flavoured CP-asymmetry is proportional to
\begin{equation}
\label{eqn:offdiageps}
\epsilon_i^\alpha \propto 2\sum_{j\neq i} \Im{\left\{ \big(\mathcal{Y}_D^{\vphantom{T}}\big)_{\alpha i} \big( \mathcal{Y}_D^{\vphantom{T}}\big)_{\alpha j} \right\}} \Re{\left\{ \big(\mathcal{Y}_D^\dagger \mathcal{Y}_D^{\vphantom{T}} \big)_{i j} \right\}} + \mathcal{O}(\mathcal{Y}_D^6).
\end{equation}
Equation~\ref{eqn:offdiageps} requires the existence of real, off-diagonal entries in $\mathcal{Y}_D^\dagger \mathcal{Y}_D^{\vphantom{T}}$ in order for non-zero asymmetry generation to appear at lowest order in the couplings. It turns out that this presents a problem, as we now review, and provides an independent motivation for the ISS and LSS extended seesaw schemes in the MLFV and leptogenesis context.

In the case of the MLFV Ansatz implemented in the type-I seesaw scenario, the radiative corrections to the SN Majorana masses are incorporated through
\begin{eqnarray}
\tilde{\mu}_{N} = \mu_N^{\vphantom{T}}\id_3 + \delta\mu_N^{\vphantom{T}} ,
\end{eqnarray}
where $\mu_N^{\vphantom{T}}$ is defined in~\cref{eqn:type1} and the $SO(3)$ breaking source $\delta \mu_N^{\vphantom{T}}$ arises from all spurion insertions that transform such that the combination $\overline{N_R^{\text{c}}}\, \delta \mu_N^{\vphantom{T}} \,N_R^{\vphantom{\text{c}}}$ starts off flavour invariant. From here on, Majorana masses with tildes will denote that corrections have been included whereas without will denote the $SO(3)$ conserving bare mass. At lowest order they are 
\begin{eqnarray}
\label{eqn:type-I-radiative}
\delta \mu_N^{\vphantom{T}} &=& \mu_N^{\vphantom{T}} \left[ n_1 \left( \mathcal{Y}_D^\dagger \mathcal{Y}_D^{\vphantom{\dagger}}  + \mathcal{Y}_D^T \mathcal{Y}_D^* \right)\right]\nonumber\\
& = & 2\, \mu_N^{\vphantom{T}} \, n_1 \, \Re\left\{\mathcal{Y}_D^{\dagger} \mathcal{Y}_D^{\vphantom{T}} \right\}
\end{eqnarray}
where $n_1$ is a Wilson coefficient which at the effective level is a free parameter. The two terms appearing in eq.~(\ref{eqn:type-I-radiative}) can be checked to transform in the appropriate way by applying the rotations defined in eqs.~(\ref{eqn:ISSleptontrans}) and~(\ref{eqn:ISSspuriontrans}) to the terms of eq.~(\ref{eqn:type-I-radiative}). For the MLFV Ansatz these terms must be included, but we remain agnostic as to how they are generated.

In order to calculate the CP-asymmetry one must first rotate the SN into their mass basis. The unitary matrix responsible for diagonalising the heavy SNs in this case is real (and therefore orthogonal):
\begin{equation}
\begin{pmatrix}
0 & m_D \\[5pt]
m_D^T \,\,&\,\, \mu_N^{\vphantom{T}}\id_3 + \delta \mu_N^{\vphantom{T}} \end{pmatrix} \rightarrow \begin{pmatrix}
0 & m_D \mathcal{O} \\[5pt]
\mathcal{O}^T m_D^T \,\,& \,\,\mathcal{O}^T(\mu_N^{\vphantom{T}}\id_3 + \delta \mu_N^{\vphantom{T}}) \mathcal{O} \end{pmatrix} = \begin{pmatrix}
0 & m_D \mathcal{O} \\[5pt]
\mathcal{O}^T m_D^T \,\,& \,\hat{\tilde{\mu}}_N \end{pmatrix}
\end{equation}
where $\hat{\tilde{\mu}}_N = \text{diag}(\mu_N^{\vphantom{T}}+\delta_1,\,\mu_N^{\vphantom{T}}+\delta_2,\,\mu_N^{\vphantom{T}}+\delta_3)$. Due to the degeneracy required in $\mathcal{Y}_N^{\vphantom{T}}$ and the allowed flavour invariant terms of eq.~(\ref{eqn:type-I-radiative}) the rotation matrix $\mathcal{O}$ is exactly the matrix which diagonalizes $\Re{\left\{ \mathcal{Y}_D^\dagger \mathcal{Y}_D^{\vphantom{T}} \right\}}$.
Therefore the CP asymmetry for flavoured leptogenesis shifts to
\begin{equation}
\Re{\left\{\mathcal{Y}_D^\dagger \mathcal{Y}_D^{\vphantom{\dagger}} \right\}} \rightarrow \Re{\left\{\mathcal{O}^T \mathcal{Y}_D^\dagger \mathcal{Y}_D^{\vphantom{\dagger}} \mathcal{O}\right\}} = \mathcal{O}^T\Re{\left\{ \mathcal{Y}_D^\dagger \mathcal{Y}_D^{\vphantom{\dagger}} \right\}}\mathcal{O} = \text{diag}(\dots)
\end{equation}
which has no off-diagonal terms and thus the CP asymmetry vanishes exactly. A non-zero asymmetry could potentially be recovered by including higher-order terms of $\mathcal{O}(\mathcal{Y}_D^6)$ of appropriate form. However, due to the suppression from the additional powers of couplings in the assumed perturbative regime, a substantial asymmetry will be difficult to generate even if resonance effects are present. 

Alternatively, the consideration of higher-order corrections to the Majorana mass which are not all aligned in flavour space will cause non-zero real entries in the relevant term of~\cref{eqn:offdiageps}. Higher-order terms which transform in the same way as in eq.~(\ref{eqn:type-I-radiative}) can be constructed from the charged-lepton spurion, as per
\begin{equation}
\label{eqn:type-I-radiative-higher}
\delta \mu_N^{\vphantom{\dagger}} = \mu_N^{\vphantom{\dagger}} \left[ c_1 \left( \mathcal{Y}_D^\dagger \mathcal{Y}_D^{\vphantom{\dagger}}  + \mathcal{Y}_D^T \mathcal{Y}_D^* \right) + \dots + c_i \left( \mathcal{Y}_D^\dagger \mathcal{Y}_e^{\vphantom{\dagger}} \mathcal{Y}_e^\dagger \mathcal{Y}_D + \mathcal{Y}_D^T \mathcal{Y}_e^* \mathcal{Y}_e^T \mathcal{Y}_D^*\right) + \dots\right]
\end{equation}
where, for clarity, only the terms not flavour aligned have been written explicitly. (The additional terms in~\cref{eqn:type-I-radiative-higher} will be fully written out later on.) Due to the inclusion of $\mathcal{Y}_e^{\vphantom{\dagger}}$, $\delta \mu_N^{\vphantom{T}}$ cannot be diagonalised by the same orthogonal matrix $\mathcal{O}$, since $\mathcal{Y}_e^{\vphantom{\dagger}} \mathcal{Y}_e^\dagger \not\propto \Re\left\{\mathcal{Y}_D^{\dagger} \mathcal{Y}_D^{\vphantom{T}} \right\}$. Changing to the mass basis now allows off-diagonal real terms and therefore the generation of a non-zero asymmetry. The necessary off-diagonal entries, however, are generated by the misalignment between the two spurions, and thus they will be suppressed compared to if they had been generated at leading order.

Due to the extra fields introduced, the above flavour alignment problem does not occur in general for the extended seesaw models. Consider the full ISS + LSS scenario where the equivalent radiative corrections to the relevant Majorana masses are included,
\begin{equation}
\label{eqn:ISSLSSrot}
\begin{pmatrix}
0 & \,m_D^{\vphantom{T}}\, & m_L^{\vphantom{T}}\\[5pt]
m_D^T \,&\,\, \mu_N^{\vphantom{\dagger}}\id_3 + \delta \mu_N^{\vphantom{\dagger}}\, \,&\, m_R \id_3\\[5pt]
m_L^T &\, m_R \id_3 \,& \mu_S^{\vphantom{\dagger}}\id_3 + \delta \mu_S^{\vphantom{T}} \end{pmatrix} \rightarrow \begin{pmatrix}
0 & \,m_D^{\vphantom{T}}  \mathcal{O}\, & m_L^{\vphantom{T}} \mathcal{V}\\[5pt]
\mathcal{O}^T m_D^T \,&\,\, \hat{\tilde{\mu}}_N \,\,&\, m_R \mathcal{V} \mathcal{O}^T\\[5pt]
\mathcal{V}^T m_L^T &\, m_R\mathcal{V}^{T}  \mathcal{O} \,& \hat{\tilde{\mu}}_S \end{pmatrix}.
\end{equation}
Here $\hat{\tilde{\mu}}_\text{i} = \text{diag}(\dots)$ and the rotation matrices $\mathcal{O}$ and $\mathcal{V}$ diagonalise $\delta \mu_N^{\vphantom{T}}$ and $\delta \mu_S^{\vphantom{T}}$ respectively. The spoiling of the flavour alignment should be clear: in the process of diagonalising the Majorana mass terms $\mu_i^{\vphantom{\dagger}}$, the Dirac mass between the two heavy SNs $m_R \id_3$ has been rotated as well. Diagonalising the lower right $2 \times 2$ block will now no longer rotate the Dirac mass matrices $m_D^{\vphantom{T}}$ and $m_L^{\vphantom{T}}$ by the same orthogonal rotation that diagonalises $\mathcal{Y}_D^{\dagger} \mathcal{Y}_D^{\vphantom{\dagger}}$. The same alignment effect can occur only for very specific choices between the entries of~\cref{eqn:ISSLSSrot} and is not a general feature. We will provide an illustrative example below.

\subsection{Baryon Asymmetry}
\label{subsec:baryo}

In this section we briefly describe the procedure and formalism we employ for estimating the efficiency of asymmetry generation. A more detailed explanation of our conventions, definitions and numerical calculations can be found in~\cite{Dolan:2018qpy}. These, in turn, are based on the Boltzmann equations (BEs) derived in~\cite{Pilaftsis:2003gt,Pilaftsis:2005rv} specifically for flavoured leptogenesis in the resonant regime.

As is conventional, we work in the heavy sterile neutrino mass basis, i.e.~the bottom-right sub-matrix of Eq.~\eqref{fullmatrix} is real, positive and diagonal. The block diagonalisation of the lower right $2 \times 2$ block defines the rotation,
\begin{equation}
\label{blockrotation}
M_{\nu} \rightarrow U_{\text{rot}}^T\, M_{\nu}\, U_{\text{rot}}^{\vphantom{T}} \equiv M_{\nu}^{\prime} = 
\renewcommand\arraystretch{1.2}
\begin{pmatrix}
m_\nu^{\text{loop}} & \hphantom{.} m_D^{\prime} \hphantom{.} & m_L^{\prime} \\
\left(m_D^{\prime}\right)^{T} & \hphantom{.}  \hat{M}_{+} \hphantom{.}  & 0 \\
\left(m_L^{\prime}\right)^{T} &  \hphantom{,}  0 \hphantom{,} & \hat{M}_{-}\end{pmatrix},
\end{equation}
where primed parameters indicate rotated matrices and $\hat{M}_i = \text{diag}(\dots)$. We have included the one-loop contribution to the active neutrino masses which may significantly contribute in some regions of parameter space. In the most general case, a simple analytic expression for the rotated matrices $m_i^\prime$ is not possible. However, in our setup, where an $SO(3)^2$ flavour symmetry is employed (and before radiative effects are included) the entries of the bottom-right $2 \times 2$ sub-block are all commuting. Under this simplification the rotation can be generalised from the simple one-generation case, giving
\begin{eqnarray}
U_{\text{rot}}^{\vphantom{T}} &=& \begin{pmatrix}
\id_{3} & 0 & 0 \\
0 & \hphantom{.}\cos\left[\frac{\pi}{4} + \frac{1}{2} \arctan(\frac{\mu_S^{\vphantom{T}} - \mu_N^{\vphantom{T}}}{2 m_R})\right] \id_3  \hphantom{.}& -i \sin\left[\frac{\pi}{4} + \frac{1}{2} \arctan(\frac{\mu_S^{\vphantom{T}} - \mu_N^{\vphantom{T}}}{2 m_R})\right] \id_3 \\[8pt]
0 &\hphantom{.} \sin\left[\frac{\pi}{4} + \frac{1}{2} \arctan(\frac{\mu_S^{\vphantom{T}} - \mu_N^{\vphantom{T}}}{2 m_R})\right] \id_3\hphantom{.} & \hphantom{-}i \cos\left[\frac{\pi}{4} + \frac{1}{2} \arctan(\frac{\mu_S^{\vphantom{T}} - \mu_N^{\vphantom{T}}}{2 m_R})\right] \id_3 
\end{pmatrix}
\end{eqnarray}
leading to
\begin{eqnarray}
\label{eqn:rotmatrices}
m_D^{\prime} &=&  \cos\left[\frac{1}{4}\left(\pi + 2\arccot\left[\frac{2 m_R}{\mu_S^{\vphantom{T}} - \mu_N^{\vphantom{T}}}\right]\right)\right] m_D + \sin \left[\frac{1}{4}\left(\pi + 2\arccot\left[\frac{2 m_R}{\mu_S^{\vphantom{T}} - \mu_N^{\vphantom{T}}}\right]\right)\right]  m_L \nonumber\\[5pt]
&\simeq &  \left( \frac{1}{\sqrt{2}} - \frac{1}{4\sqrt{2}}\left(\frac{\mu_S^{\vphantom{T}}-\mu_N^{\vphantom{T}}}{2 m_R} \right)\right) m_D +  \left( \frac{1}{\sqrt{2}}  + \frac{1}{4\sqrt{2}}\left(\frac{\mu_S^{\vphantom{T}}-\mu_N^{\vphantom{T}}}{2 m_R} \right) \right) m_L \nonumber\\[10pt]
m_L^{\prime} &=& -i  \sin\left[\frac{1}{4}\left(\pi + 2\arccot\left(\frac{2 m_R}{\mu_S^{\vphantom{T}} - \mu_N^{\vphantom{T}}}\right)\right)\right] m_D + i  \cos \left[\frac{1}{4}\left(\pi + 2\arccot\left(\frac{2 m_R}{\mu_S^{\vphantom{T}} - \mu_N^{\vphantom{T}}}\right)\right)\right] m_L \nonumber\\[5pt]
&\simeq & - i  \left( \frac{1}{\sqrt{2}}  + \frac{1}{4\sqrt{2}}\left(\frac{\mu_S^{\vphantom{T}}-\mu_N^{\vphantom{T}}}{2 m_R} \right) \right) m_D + i \left( \frac{1}{\sqrt{2}}  - \frac{1}{4\sqrt{2}}\left(\frac{\mu_S^{\vphantom{T}}-\mu_N^{\vphantom{T}}}{2 m_R} \right)\right)  m_L \nonumber\\[10pt]
\hat{M}_{+} &=& \left( m_R \sqrt{1 + \frac{(\mu_S^{\vphantom{T}} - \mu_N^{\vphantom{T}})^2}{4 \,m_R^2}} + \frac{\mu_S^{\vphantom{T}}}{2} + \frac{\mu_N^{\vphantom{T}}}{2}\right)\id_3\nonumber\\[5pt]
& \simeq & \left( m_R + \frac{\mu_S^{\vphantom{T}}}{2} + \frac{\mu_N^{\vphantom{T}}}{2}\right) \id_3\nonumber\\[10pt]
\hat{M}_{-} &=& \left( m_R \sqrt{1 + \frac{(\mu_S^{\vphantom{T}} - \mu_N^{\vphantom{T}})^2}{4 \,m_R^2}} - \frac{\mu_S^{\vphantom{T}}}{2} - \frac{\mu_N^{\vphantom{T}}}{2}\right)\id_3\nonumber\\[5pt]
& \simeq & \left( m_R - \frac{\mu_S^{\vphantom{T}}}{2} - \frac{\mu_N^{\vphantom{T}}}{2}\right) \id_3
\end{eqnarray}
where in every second line we have expanded up to $\mathcal{O}(\frac{\mu_S^{\vphantom{T}}-\mu_N^{\vphantom{T}}}{m_R})$. Once the spurionic radiative corrections are included in the Majorana self-energies, the $\mathcal{O}(3)^2$ invariance is lifted and the matrices no longer commute. In the regime where these corrections are small these expressions will remain approximately valid. However, once the corrections are included, we perform the rotations numerically during our scan.

The combination
\begin{equation}
\label{eqn:mddmd}
h^{\dagger} h \simeq \frac{2}{v^2}\begin{pmatrix}
m_D^{\prime}\hphantom{}^{\dagger} \, m_D^{\prime}\ \  &\ \  m_D^{\prime}\hphantom{}^{\dagger}\, m_L^{\prime}  \\
 m_L^{\prime}\hphantom{}^{\dagger} \,m_D^{\prime}\ \  &\ \  m_L^{\prime}\hphantom{}^{\dagger}\,m_L^{\prime} \end{pmatrix}
\end{equation}
where $h = \frac{\sqrt{2}}{v} \left( m^{\prime}_{D}, m^{\prime}_{L} \right)$, is now the relevant combination for flavoured leptogenesis. The entries in the diagonal blocks of $h^{\dagger} h$ control the CP asymmetry generated between  two different generations of heavy SNs with same sign mass splitting
\begin{equation}
\label{eqn:ISS-SSsplitting}
\Delta m_{i,j}^{\text{SS}} = (\hat{M}_{\pm})_{ii} - (\hat{M}_{\pm})_{jj}.
\end{equation}
The off-diagonal blocks control the CP asymmetry generated between heavy SNs of opposite mass splitting (same generation or otherwise)
\begin{equation}
\label{eqn:ISS-OSsplitting}
\Delta m_{i,j}^{\text{OS}} = (\hat{M}_{\pm})_{ii} - (\hat{M}_{\mp})_{jj}.
\end{equation}
Clearly, in the absence of radiative effects where we have $\mu_N^{\vphantom{T}}, \,\mu_S^{\vphantom{T}} \propto \id_3$ only $\Delta m_{i,j}^{\text{OS}} \neq 0$, with $\Delta m_{i,j}^{\text{SS}} = 0$. Once the corrections are included, however, the $SO(3)^2$ invariance is broken allowing for both $\Delta m_{i,j}^{\text{OS}} \neq 0$  and $\Delta m_{i,j}^{\text{SS}} \neq 0$. From eqs.~(\ref{eqn:rotmatrices}) and~(\ref{eqn:mddmd}) it can be seen that for example if $m_D^{\vphantom{T}} = m_L^{\vphantom{T}}$\footnote{This would imply that  $\delta \mu_N^{\vphantom{T}} = \delta \mu_S^{\vphantom{T}}$ and therefore $ \mathcal{O}=\mathcal{V}$.} along with $\mu_N^{\vphantom{T}} = \mu_S^{\vphantom{T}}$ then $m_D^{\prime} \neq 0$ but $m_L^{\prime} = 0$. No real, off-diagonal entries would exist and the same alignment problem of the type-I seesaw would occur. This is not a general feature in the extended MLFV seesaw (as it is in the minimal type-I MLFV seesaw) and would require an additional symmetry to enforce the necessary relations between the couplings and masses such that this flavour alignment would occur.

We consider a low-scale scenario in which the sterile masses are set to $\mathcal{O}(\text{TeV})$. This implies a low-temperature regime of lepton asymmetry generation where individual lepton flavours are in equilibrium with the thermal bath and distinguishable. We employ \textit{flavour-dependent} Boltzmann equations~\cite{Pilaftsis:2005rv,Abada:2006ea} for which we define the CP asymmetry generated from the decays of an SN, $N_i$, to a specific lepton flavour $\ell_\alpha$,
\begin{equation}
\label{cpasymmetry1}
	\varepsilon^{i}_{\alpha} = \frac{\Gamma\left(N_i \rightarrow l_{\alpha} \Phi \right) - \Gamma\left( N_i \rightarrow \left(l_{\alpha}\right)^{c} \Phi^{\dag}\right)}{\sum\limits_{\alpha}^{ } \Big[\, \Gamma\left(N_i \rightarrow l_{\alpha} \Phi \right) + \Gamma\left( N_i \rightarrow \left(l_{\alpha}\right)^{c} \Phi^{\dag}\right) \Big]}.
\end{equation}
These have been calculated previously~\cite{Pilaftsis:2003gt,Pilaftsis:2005rv} for the general case relevant for both hierarchical and degenerate scenarios with potential resonance effects~\cite{Adhikary:2014qba} included. The result, separated into vertex $\varepsilon_{\text{V}}$ and self-energy $\varepsilon_{\text{S}}$ contributions, is
\begin{eqnarray}
\label{cpasymmetry2}
\varepsilon^{\alpha}_{i} &=& \frac{1}{8\pi \left(h^{\dag} h\right)_{ii}}   \,\mathlarger{\mathlarger{\sum\limits_{j\ne i}}} \bigg(\, \underbrace{\mathcal{A}_{ij}^{\alpha}\, f(x_{ij})\vphantom{\frac{\sqrt{x_{ij}} \left(1 - x_{ij} \right)}{\left(1-x_{ij}\right)^{2} + \frac{1}{64\pi^{2}}\left(h^{\dag} h\right)^{2}_{jj}}}}  \,+ \,\underbrace{(\mathcal{A}_{ij}^{\alpha} \sqrt{x_{ij}} + \mathcal{B}_{ij}^{\alpha}) \frac{\sqrt{x_{ij}} \left(1 - x_{ij} \right)}{\left(1-x_{ij}\right)^{2} + \frac{1}{64\pi^{2}}\left(h^{\dag} h\right)^{2}_{jj}}} \bigg)\,+\,\, \mathcal{O}(h^6) \ldots \nonumber\\
&\mbox{}& \,\quad\qquad\qquad\qquad\qquad \varepsilon_{\text{V}}^{\vphantom{T}}\qquad\qquad\qquad\qquad\qquad\quad\,\, \varepsilon_{\text{S}}^{\vphantom{T}}
\end{eqnarray}
where $\mathcal{A}_{ij}^{\alpha} = \Im\left\{ \big( h^{\dagger} h \big)_{ij}^{\vphantom{*}}\, h_{\alpha i}^{*} \,h_{\alpha j}^{\vphantom{*}}\right\}$, $\mathcal{B}_{ij}^{\alpha} = \Im \left\{\big( h^{\dagger} h \big)_{ji}^{\vphantom{*}} \,h_{\alpha i}^{*} \,h_{\alpha j}^{\vphantom{*}}\right\}$, $x_{ij} = \left(\frac{m_{N_j}}{ m_{N_i}} \right)^{2}$ and the loop function for the vertex-diagram contribution $f(x_{ij})$, is given by~\cite{Fukugita:1986hr}
\begin{equation}
\label{loopfunction}
f(x_{ij})=\sqrt{x_{ij}}\left[1-(1+x_{ij})\ln\left(\frac{1+x_{ij}}{x_{ij}}\right)\right].
\end{equation}
A resonant enhancement in the CP asymmetry will occur when
\begin{equation}
\label{resonantcondition}
\quad1-x_{ij} \rightarrow  \frac{1}{8\pi} \left( h^{\dag} h \right)_{jj}
\end{equation}
which leads to the simplified condition
\begin{equation}
\label{simpleres}
m_{N_i} - m_{N_j} \rightarrow \frac{\Gamma_{i,j}}{2}
\end{equation}
where $\Gamma_i \simeq \frac{m_i}{8\pi}\sum_{l} h_{li}^* h_{li}^{\vphantom{*}}$ is the decay width of $N_i$ and we assume $m_{N_i} - m_{N_j} \ll m_{N_{i,j}}$ in order to move from eq.~(\ref{resonantcondition}) to~(\ref{simpleres}).

We choose the regulator of $\varepsilon^i_\alpha$ to be of the form $m_i \Gamma_j$, which as discussed in Appendix A of~\cite{Dev:2014laa} has consistent behaviour for models with approximate lepton-number conservation. For other choices of regulators unphysical behaviour may be encountered in such a scenario. For example choosing a regulator of the form ($m_i \Gamma_i - m_j \Gamma_j$) diverges in the scenario $m_i \rightarrow m_j$ combined with $\Gamma_i \rightarrow \Gamma_j$ which occurs as the LNV parameters are taken to zero.

Specifically for the LSS scenario, it is important to note that the second term in eq.~(\ref{cpasymmetry2}), due to the one-loop \textit{self-energy} correction $\varepsilon_{S}^{\vphantom{T}}$, identically goes to zero when $m_{N_i} \rightarrow m_{N_j}$. By contrast, the first term arises from the one-loop \textit{vertex} correction $\varepsilon_{V}^{\vphantom{T}}$. In the limit $m_{N_i} \rightarrow m_{N_j}$ the asymmetry it generates is non-zero. However, $A_{ij}^{\alpha} = - A_{ji}^{\alpha}$ and therefore
\begin{equation}
\label{eqn:vertexonlyeps}
\varepsilon^{\alpha}_{i} = - \frac{\big( h^{\dagger} h \big)_{ii}}{\big( h^{\dagger} h \big)_{jj}} \varepsilon^{\alpha}_{j}
\end{equation}
for $m_{N_i} = m_{N_j}$. Therefore the asymmetry produced by the decay of $N_i$ to flavour $\alpha$ is almost equal and opposite to that of $N_j$ and in the limit $\big( h^{\dagger} h \big)_{ii} \rightarrow \big( h^{\dagger} h \big)_{jj}$ the asymmetry will vanish identically once the decays of all SNs are included. If the two SNs have different decay widths a non-zero asymmetry is possible, albeit highly suppressed.

We will work under the condition that, due to the strong washout nature of the temperature regime we favour~\cite{Dolan:2018qpy}, asymmetry generation occurs predominately well before the electroweak phase transition crossover and so we need not consider the changing sphaleron rate as the temperature approaches and crosses its critical value. The baryon asymmetry is then expressed as a fraction of the lepton asymmetry generated through
\begin{equation}
\label{leptotobaryo}
\eta_{B} = -\frac{28}{51} \frac{1}{27}  \sum_{\alpha = e,\mu,\tau} \eta_{\alpha}
\end{equation}
where the factor of $28/51$ arises from the fraction of lepton asymmetry reprocessed into a baryon asymmetry by electroweak sphalerons \cite{Harvey:1990qw}, while the dilution factor $1/27$ arises from photon production until the recombination epoch \cite{Deppisch:2010fr}.

\section{Numerical Results}
\label{sec:numericalresults}

In full analogy to the type-I scenario, all appropriate flavour invariant operators must be included. This leads to corrections to the Majorana mass terms from spurion combinations transforming the appropriate way such that when coupled with the heavy SN fields, the term is flavour invariant at the high scale. Now in the case of the extended seesaw both Majorana masses receive corrections, as per
\begin{eqnarray}
\label{eqn:inv-masscorrections}
\tilde{\mu}_{N} &=& \,\mu_N^{\vphantom{\dagger}} \left( \mathcal{Y}_{\mu_N} \,+ n_1^{\vphantom{(1)}} \left(\mathcal{Y}_D^\dagger \mathcal{Y}_D^{\vphantom{\dagger}} + (\mathcal{Y}_D^\dagger \mathcal{Y}_D^{\vphantom{\dagger}})^T\right) \,\,\,+\,\,n_2^{(1)} \left(\mathcal{Y}_D^\dagger \mathcal{Y}_D^{\vphantom{\dagger}} \mathcal{Y}_D^\dagger \mathcal{Y}_D^{\vphantom{\dagger}} \,+ (\mathcal{Y}_D^\dagger \mathcal{Y}_D^{\vphantom{\dagger}} \mathcal{Y}_D^\dagger \mathcal{Y}_D^{\vphantom{\dagger}})^T\right) \,+ \dots    \right)\nonumber\\[5pt]
\tilde{\mu}_{S} &=& \,\,\mu_S^{\vphantom{\dagger}} \left( \mathcal{Y}_{\mu_S} \,\,+ s_1^{\vphantom{(1)}} \left(\mathcal{Y}_L^\dagger \mathcal{Y}_L^{\vphantom{\dagger}} \,+ (\mathcal{Y}_L^\dagger \mathcal{Y}_L^{\vphantom{\dagger}})^T\right) \,\,\,\,\,\,+\,\,\, s_2^{(1)} \left(\mathcal{Y}_L^\dagger \mathcal{Y}_L^{\vphantom{\dagger}} \mathcal{Y}_L^\dagger \mathcal{Y}_L^{\vphantom{\dagger}} \,\,\,+ (\mathcal{Y}_L^\dagger \mathcal{Y}_L \mathcal{Y}_L^\dagger \mathcal{Y}_L^{\vphantom{\dagger}})^T\right) \,\,\,\,\,+ \dots    \right).\nonumber\\[5pt]
\end{eqnarray}
As can be checked explicitly by applying the transformations defined in eq.~(\ref{eqn:ISSspuriontrans}), the combinations of spurions above are flavour invariant when coupled to $\overline{N_R^{\text{c}}} N_R^{\vphantom{T}}$ and $\overline{S_L^{\vphantom{T}}} S_L^{\text{c}}$. All terms at next-to-leading order in the flavout-invariant spurion insertions will be explicitly written below.

The coefficients $n_i\text{ and }s_i$ are dimensionless Wilson coefficients which are treated as free parameters in the absence of an explicit high-scale, renormalisable theory. They are conventionally either taken to be $\mathcal{O}(1)$ numbers~\cite{Cirigliano:2005ck,Cirigliano:2006su,Cirigliano:2006nu,Dinh:2017smk}, or arising from radiative effects~\cite{Branco:2006hz,Cirigliano:2007hb} so that $n_1\simeq s_1 \simeq \frac{1}{16\pi^2}$ and $s_2^{(i)} \sim (s_1)^2$ etc. Note that in the ISS regime where $\mathcal{Y}_L^{\vphantom{\dagger}} = 0$, only the Majorana mass $\mu_N^{\vphantom{\dagger}}$ receives radiative corrections from flavour effects in our setup, whereas in the LSS regime where $\mathcal{Y}_{\mu_N}^{\vphantom{\dagger}} = \mathcal{Y}_{\mu_S}^{\vphantom{\dagger}} =0$ both Majorana masses receive corrections.

We consider the general scenario where one copy of $N_R$ and $S_L$ is added for each generation of active neutrino. The tree-level light neutrino mass matrix can be expressed in powers of the LNV parameters~\cite{Dev:2012sg} which in complete generality (without assuming MLFV) can be expanded to
\begin{eqnarray}
\label{eqn:fullissmass}
m_{\nu}^{\text{tree}} = &\mbox{}& m_D^{\vphantom{T}} \left( M_R^T \right)^{-1} \mu_S^{\vphantom{T}}\, M_R^{-1} \,m_D^T + m_D \left( M_R^T \right)^{-1}  \mu_S^{\vphantom{T}} M_R^{-1} \mu_N^{\vphantom{T}} \left( M_R^T \right)^{-1} \mu_S^{\vphantom{T}} M_R^{-1} m_D^T \nonumber\\
&-& m_L^{\vphantom{T}} M_R^{-1} m_D^T  -  m_L^{\vphantom{T}} M_R^{-1} \mu_N^{\vphantom{T}} \left( M_R^T\right)^{-1}\mu_S^{\vphantom{T}} M_R^{-1} m_D^T\nonumber\\
&-& m_D^{\vphantom{T}} \left( M_R^T\right)^{-1} m_L^T 
- m_D^{\vphantom{T}} \left(M_R^T\right)^{-1} \mu_S M_R^{-1} \mu_N^{\vphantom{T}} \left(M_R^T\right)^{-1}m_L^T \nonumber\\
&+& m_L M_R^{-1} \mu_N^{\vphantom{T}} (M_R^T)^{-1} m_L^T + m_L M_R^{-1} \mu_N^{\vphantom{T}} (M_R^T)^{-1} \mu_S^{\vphantom{T}} M_R^{-1} \mu_N^{\vphantom{T}} (M_R^T)^{-1} m_L^T\nonumber\\
&+&\dots
\end{eqnarray}
where for completeness we have written all leading and next-to-leading terms valid for $||\mu_S^{\vphantom{\dagger}}||,\,||\mu_N^{\vphantom{\dagger}}||,\, ||m_L^{\vphantom{\dagger}}|| \ll ||M_R||$. The above equation is valid for the general case of non-commuting sub-matrices, but for our MLFV setup $M_R^{\vphantom{T}} = m_R \id_3$, $m_L$ is diagonal and $\mu_S^{\vphantom{T}} \propto \mu_N^{\vphantom{T}} \propto \id_3$.

In the limiting case of the ISS, only the first line of eq.~(\ref{eqn:fullissmass}) contributes at leading order to the active neutrino masses. This mass vanishes in the limit $\mu_S{\vphantom{T}} \rightarrow 0$ even if $\mu_N \neq 0$. Even for $\mu_S^{\vphantom{T}} \neq 0$ the contribution from $\mu_N^{\vphantom{\dagger}}$ is highly suppressed, including for a large regime where $\mu_N^{\vphantom{\dagger}} \gg \mu_S^{\vphantom{\dagger}}$.

At one-loop order however, additional terms are generated for the active neutrino masses which can be important~\cite{Dev:2012sg},
\begin{equation}
\label{eqn:oneloopneutrinoinverse}
m_\nu^{\text{loop}} \simeq \frac{f(m_R)}{m_W^2} \left( m_D\, \mu_N^{\vphantom{\dagger}}\, m_D^T + m_L \,\mu_S^{\vphantom{\dagger}} \,m_L^T + m_R \,m_L \,m_D^T + m_R \,m_D\, m_L^T \right)
\end{equation}
where
\begin{equation}
f(x)= \frac{\alpha_W}{16\pi}\left( \frac{m_H^2}{ x^2-m_H^2} \ln \left[ \frac{x^2}{m_H^2} \right] + \frac{3 m_Z^2}{ x^2 - m_Z^2} \ln \left[ \frac{ x^2}{m_Z^2} \right]\right)
\end{equation}
is a one-loop function and~\cref{eqn:oneloopneutrinoinverse} is only valid for $M_R = m_R \id_3$. Different terms in~\cref{eqn:oneloopneutrinoinverse} can contribute significantly for different hierarchies amongst $\mu_N^{\vphantom{dagger}},\,\mu_S^{\vphantom{dagger}}$ and $m_L$ which are free parameters in our scan.

Combining the tree-level and one-loop contributions at leading order leads to a general light-neutrino mass matrix in our MLFV Ansatz of the form
\begin{eqnarray}
&\mbox{}& \quad\qquad\quad\qquad\qquad\qquad\qquad\qquad\quad\quad\,\,\,\,\,\,\,\,\text{one-loop}\nonumber\\
m_\nu &=& \overbrace{\frac{1}{m_R^2} \left( f(m_R)\frac{m_R^2}{m_W^2}\right) \left( m_D\, \mu_N^{\vphantom{\dagger}}\, m_D^T + m_L\, \mu_S^{\vphantom{\dagger}} \,m_L^T \right) + \frac{1}{m_R}\left(f(m_R)\frac{m_R^2}{m_W^2}\right) \left( m_L \,m_D^T + m_D \,m_L^T \right)}\nonumber\\[5pt]
&+& \underbrace{\frac{1}{m_R^2} \left( m_D \,\mu_N^{\vphantom{\dagger}} \,m_D^T + m_L \,\mu_S^{\vphantom{\dagger}}\, m_L^T\right) - \frac{1}{m_R} \left(m_L \,m_D^T + m_D\, m_L^T \right).} \nonumber\\
&\mbox{}& \qquad\qquad\qquad\,\qquad\,\,\quad\qquad\;\;\text{tree}
\end{eqnarray}

For both the ISS and LSS we fix $m_R = 1$ TeV. For the ISS specifically, the LNV parameters are randomly (and independently) scanned over the ranges
\begin{eqnarray}
\mu_S^{\vphantom{\dagger}} \,&\sim &\, \mathcal{O}(10^{-15} - 10^{2} )\id_3\, \text{ GeV}\nonumber\\
\mu_N^{\vphantom{\dagger}} \,&\sim &\, \mathcal{O}(10^{-15} - 10^{4} )\id_3\, \text{ GeV}
\end{eqnarray}
where we allow for larger values of $\mu_N^{\vphantom{\dagger}}$ over $\mu_S^{\vphantom{\dagger}}$ as its contribution to the active neutrino mass is suppressed by a loop factor.
For the LSS we vary
\begin{equation}
(m_L)_{ii} \,\sim\,  \mathcal{O}( 10^{-9} - 10^{1})\, \text{ GeV}
\end{equation}
where, in order to reduce the number of degrees of freedom, we take it to be diagonal but not proportional to the identity matrix in order for its inclusion to break the flavour symmetry.  We fix the diagonal entries of $m_L$ to satisfy $\text{min}(m_L) > \frac{1}{5} \text{max}(m_L)$ such that significant hierarchies between entries of $m_L$ do not occur. In this way the entries of $m_D$ will act as the most significant sources of flavour symmetry breaking in analogy with MFV and minimal MLFV.

We fit active-neutrino data by fixing the Dirac mass matrix $m_D$ with an \textit{approximate}\footnote{The parameterisation is only approximate as we ignore the corrections to the mass terms e.g. $\tilde{\mu}_i = \mu_i^{\vphantom{\dagger}} + \delta \mu_i^{\vphantom{\dagger}}$ generated through spurion insertions when solving for $m_D$. This approximation is valid in the regime where $\delta \mu_i^{\vphantom{\dagger}} < \mu_i^{\vphantom{\dagger}}$. We additionally check that once these corrections are included the active-neutrino mass differences are not spoiled.} Casas-Ibarra parameterisation~\cite{Casas:2001sr} which for the ISS is
\begin{equation}
\label{eqn:ISScasas}
m_D = m_R \,U_{\text{PMNS}} \,\hat{m}_{\nu}^{1/2}\, R\,\mu_{\text{eff}}^{-1/2},
\end{equation}
where we define an effective mass
\begin{equation}
\mu_{\text{eff}} = f(m_R)\,\left(\frac{m_R}{m_W}\right)^2 \, \mu_N^{\vphantom{\dagger}}  \,+ \vphantom{\left(\frac{m_R}{m_W}\right)^2 f(m_R)\, \mu^{\prime}}\mu_S^{\vphantom{\dagger}}.
\end{equation}
For the LSS,
\begin{equation}
\label{eqn:LSScasas}
m_D = - m_R\,U \hat{m}_{\nu}^{1/2}\, C \, \hat{m}_{\nu}^{1/2} \, U_{\text{PMNS}}^T\, (m_{L}^{\text{eff}})^{T-1}
\end{equation}
with an equivalently defined effective mass
\begin{equation}
m_{L}^{\text{eff}} = \left(f(m_R)\,\left(\frac{m_R}{m_W}\right)^2 -1 \right) m_L.
\end{equation}

The matrix $R$ is a complex-orthogonal matrix $R \, R^T = \id_3$ which can be parameterised with three complex mixing angles $\theta_i = \theta_i^{\text{r}} + i \theta_i^{\text{c}}$ where
\begin{eqnarray}
&\hphantom{-} &\qquad\qquad\quad R(\theta_1,\,\theta_2,\,\theta_3) = R_{12}(\theta_1) \, R_{13}(\theta_2)\, R_{23}(\theta_3),\nonumber\\[10pt]
R_{12}(\theta_1) &=& \begin{pmatrix}
\text{c}_{\theta_1} & -\text{s}_{\theta_1} & \,0\\
\text{s}_{\theta_1} & \text{c}_{\theta_1} & \,0\\
0 & 0 & \,1\\
\end{pmatrix},\,\,
R_{13}(\theta_2) = \begin{pmatrix}
\text{c}_{\theta_2} & \,0\, & -\text{s}_{\theta_2}\\
0 & \,1\, & 0\\
\text{s}_{\theta_2} & \,0\, & \text{c}_{\theta_2}\\
\end{pmatrix},\,\,
R_{23}(\theta_3) = \begin{pmatrix}
1\,\, & 0 & 0\\
0\,\, & \text{c}_{\theta_3} & -\text{s}_{\theta_3}\\
0\,\, & \text{s}_{\theta_3} & \text{c}_{\theta_3}\\
\end{pmatrix}\quad
\end{eqnarray}
and we have defined $\cos(x) \equiv \text{c}_x \text{ and }\sin(x) \equiv \text{s}_x$ in the usual way.

In order to reduce the number of free parameters we fix the real components of the mixing angles to
\begin{equation}
\theta_1^{\text{r}} = \frac{\pi}{5},\,\,\theta_2^{\text{r}} =\frac{5\pi}{6},\,\,\theta_3^{\text{r}} = \frac{4\pi}{7}
\end{equation}
where there is no signficance to the values chosen. The asymmetry is not sensitive to the values of the real angles~\cite{Dolan:2018qpy} and therefore should apply for any choice of their values. We fix $\theta_{1}^{\text{c}}=\theta_{3}^{\text{c}}=0$ for simplicity and scan over 
\begin{eqnarray}
\theta^{\text{c}}_{2} &\sim & \mathcal{O}(10^{-3} - 10^{1} ).
\end{eqnarray}

By contrast to the complex-orthogonal $R$, the matrix $C$ instead satisfies $C + C^T = \id_3$ and therefore must be a combination of a skew-symmetric matrix and a diagonal matrix of the specific form,
\begin{equation}
C = \begin{pmatrix}
\frac{1}{2} & a_1 & a_2\\
-a_1 & \frac{1}{2} & a_3\\
-a_2 & -a_3 & \frac{1}{2}
\end{pmatrix}
\end{equation}
with $a_i = a_i^{\text{r}} + i a_i^{\text{c}}$ where we choose to fix
\begin{equation}
a_1^{\text{r}} = \frac{1}{10},\,\,a_2^{\text{r}} =\frac{2}{10},\,\,a_3^{\text{r}} = \frac{1}{2},\,a_1^{\text{c}} = a_3^{\text{c}} = 0
\end{equation}
and scan over
\begin{eqnarray}
a_2^{\text{c}} \sim  \mathcal{O}(10^{-3} - 10^{1} ).
\end{eqnarray}

The active-neutrino masses
\begin{equation}
\hat{m}_{\nu} = \text{diag}(m_{\nu_1},\,m_{\nu_2},\,m_{\nu_3}) = U_{\text{PMNS}}^\dagger (m_\nu^{\text{tree}} + m_\nu^{\text{loop}}) U_{\text{PMNS}}^*
\end{equation}
are inputs for~\cref{eqn:ISScasas,eqn:LSScasas} where for simplicity we assume normal ordering which is currently favoured over inverted ordering~\cite{Esteban:2018azc}, implying
\begin{eqnarray}
m_{\nu_2} &=& \sqrt{m_{\nu_1}^2 + \Delta m_{21}^2}\nonumber\\
m_{\nu_3} &=& \sqrt{m_{\nu_1}^2 + \Delta m_{21}^2 + \Delta m_{31}^2}
\end{eqnarray}
and unless stated otherwise we fix the lightest neutrino mass $m_{\nu_1}$ to $0.01\text{ eV}$. Other active neutrino parameters are fixed to their current best fit values~\cite{Esteban:2018azc} and listed in~\cref{table-oscpars}, while the Dirac phase has been fixed to be maximally CP-violating.

We work in the perturbative regime of the Yukawa couplings generated by the Casas-Ibarra parameterisations such that $|\mathcal{Y}_{ij}|^2 \lesssim 4\pi$, which prevents very large choices for the complex parameters. We fix the unknown Wilson coefficients $s_i\text{ and }n_i$ to $\frac{1}{16\pi^2}$ for definiteness unless stated otherwise.

\begin{table}[t]
\begin{center}
\scalebox{1.0}{
\begin{tabular}{ccc}
\toprule
Parameter & Value \\ 
\midrule
\midrule
$\sin^{2} \theta_{12}$ & 0.310 \\ 
$\sin^{2} \theta_{23}$ & 0.580 \\ 
$\sin^{2} \theta_{13}$ & 0.02241 \\ 
\midrule 
$\Delta m_{21}^{2}/(10^{-5}\; \rm{eV}^{2})$ & $7.39^{+ 0.19}_{-0.17}$\\ 
$\Delta m_{31}^{2}/(10^{-3} \;\rm{eV}^{2})$ & $2.525^{+0.039}_{-0.040}$ \\ 
\midrule 
$\delta_{CP}$ & $3\pi/2$ \\ 	
\bottomrule 
\end{tabular}
}
\end{center}
\caption{List of experimental measurements of the parameters in the PMNS matrix and the light neutrino mass splittings fixed by active neutrino oscillation experiments. The light neutrino mass differences were allowed to vary within 1$\sigma$ of their current best fit values~\cite{Esteban:2018azc}.} 
\label{table-oscpars}
\end{table}

To check our numerical solutions to the Boltzmann equations, we compare the results to a known approximate analytic expression of the baryon asymmetry that is valid in the strong washout regime ($K^\alpha_{\rm{eff}} \geq 5$)~\cite{Deppisch:2010fr,Borah:2017qdu}, where

\begin{eqnarray}
\label{washout}
& \;\;\;\;\;K_{\text{eff}}^{\alpha} = \kappa^{\alpha} \sum_i K_i J_{i}^{\alpha},\nonumber\\[5pt]
&\kappa^{\alpha} = 2 \mathlarger{\sum}\limits_{i,j}\frac{\left( \mathbf{\mathbf{h}}_{\alpha i}^{*} \mathbf{h}_{\alpha j}^{\vphantom{*}} +\, \mathbf{h}_{\alpha i}^{\mathbf{c}^*} \mathbf{h}_{\alpha j}^{\mathbf{c}} \right) \left[ \left(\mathbf{h}^{\dagger} \mathbf{h}\right)_{ij} +\, \left( \mathbf{h^c}^{\dagger} \mathbf{h^c}\right)_{ij} \right] \,\,+\,\, \left( \mathbf{\mathbf{h}}_{\alpha i}^{*} \mathbf{h}_{\alpha j}^{\vphantom{*}} -\, \mathbf{h}_{\alpha i}^{\mathbf{c}^*} \mathbf{h}_{\alpha j}^{\mathbf{c}}  \right)^{2} }{\left[ \left( \mathbf{h} \mathbf{h}^{\dagger} \right)_{\alpha \alpha} +\, \left(\mathbf{h^c} \mathbf{h}^{\mathbf{c} \dagger} \right)_{\alpha \alpha} \right] \left[ \left(\mathbf{h}^{\dagger} \mathbf{h}\right)_{ii}   +\, \left( \mathbf{h^c}^{\dagger} \mathbf{h^c}\right)_{ii} \,+\, \,\left(\mathbf{h}^{\dagger} \mathbf{h}\right)_{jj} +\, \left( \mathbf{h^c}^{\dagger} \mathbf{h^c}\right)_{jj}\right]   } \left( 1 - 2i\frac{m_{N_i}-m_{N_j}}{\Gamma_i+\Gamma_j}\right)^{-1},\nonumber\\[5pt]
&J^{\alpha}_{i}=\frac{\Gamma\left(N_i \rightarrow l_{\alpha} \Phi \right) + \Gamma\left( N_i \rightarrow l^{c}_{\alpha} \Phi^{\dag}\right)}{\sum\limits_{\alpha}^{ } \Big[\, \Gamma\left(N_i \rightarrow \left(l_{\alpha}\right)^{c} \Phi \right) + \Gamma\left( N_i \rightarrow \left(l_{\alpha}\right)^{c} \Phi^{\dag}\right) \Big]},
\end{eqnarray}
with $m_{N_i} = (\hat{M}_{\pm})_{ii}$ and $K_i = \Gamma_{N_i} / H(m_N)$ being the na\"ive washout solely from inverse decays. The inclusion of the scale factor $\kappa^\alpha$ accounts for the numerically significant $2 \leftrightarrow 2$ scattering processes relevant for models with small lepton number violation. The Yukawa couplings $\mathbf{h}$ and $\mathbf{h^c}$ appearing in bold are resummed Yukawa couplings first defined in~\cite{Pilaftsis:2003gt}. They are required to properly account for unstable particle mixing effects amongst the heavy sterile neutrinos. As a consequence of the resummed Yukawa couplings, the scaling parameter $\kappa^{\alpha}$ is real-valued.

The asymmetry is approximated by
\begin{equation}
\label{anal}
\eta_{B}^{\rm{approx.}} \simeq -\frac{28}{51}\frac{1}{27}\frac{3}{2}\sum_{\alpha,i} \frac{\varepsilon_{i}^{ \alpha}}{K^{\alpha}_{\rm{eff}} \rm{min}[z_c,1.25\ln({25 K_{\rm{eff}}^{\alpha}})]}
\end{equation}
where $z_c = m_N/T_c$ is related to the critical temperature of the electroweak phase transition.
The numerical versus analytic approximation comparison illustrated in~\cref{table:approx-etaB} provides strong evidence that our numerical routines are accurate.
\begin{table}[t]
\begin{center}
\scalebox{1.0}{
\begin{tabular}{ccc}
\toprule
Example parameters  & $\left|\eta_{B}\right|$ & $\left|\eta_{B}^{\rm{approx.}}\right|$ \\ 
\midrule 
\midrule
$m_L \simeq \text{diag}(1.5,\,2.6,\,1.2)\times 10^{-3} \,\rm{GeV}$ & & \\
$\quad\,\mu_S^{\vphantom{T}} = \mu_N^{\vphantom{T}} = 0$ & $1.37 \times 10^{-10}$ & $7.23 \times 10^{-10}$ \\ 
$a_1^{\vphantom{r}} = \frac{1}{10},\quad a_2^{\vphantom{r}} \simeq \frac{2}{10} + 0.00028 i,\quad a_3^{\vphantom{r}} = \frac{5}{10}\quad$ & & \\
\midrule
$m_L \simeq \text{diag}(0.70,\,1.33,\,1.26)\times 10^{-5} \,\rm{GeV}$ & & \\
$\quad\,\mu_S^{\vphantom{T}} = \mu_N^{\vphantom{T}} = 0$ & $1.98 \times 10^{-11}$ & $2.31 \times 10^{-11}$ \\ 
$a_1^{\vphantom{r}} = \frac{1}{10},\quad a_2^{\vphantom{r}} \simeq \frac{2}{10} + 2.06 i,\quad a_3^{\vphantom{r}} = \frac{5}{10}\quad$ & & \\
\midrule
$m_L = 0$ & & \\
$\mu_S^{\vphantom{T}} \simeq 9 \times 10^{-11}\, \rm{GeV},\,\,\,\mu_N^{\vphantom{T}} \simeq 225 \, \rm{GeV}$ & $2.57 \times 10^{-11}$ & $1.21 \times 10^{-11}$ \\ 
$\,\,\,\,\,\,\,\theta_1^{\vphantom{T}} = \frac{\pi}{5},\,\quad \theta_2^{\vphantom{T}} \simeq \frac{5\pi}{6} + 0.094 i,\,\quad \theta_3^{\vphantom{T}} = \frac{4\pi}{7}\quad$ & & \\
\midrule
$m_L = 0$ & & \\
$\mu_S^{\vphantom{T}} \simeq 1\, \rm{GeV},\,\,\,\mu_N^{\vphantom{T}} \simeq 414 \, \rm{GeV}$ & $1.71 \times 10^{-10}$ & $5.75 \times 10^{-11}$ \\ 
$\,\,\,\,\,\,\,\theta_1^{\vphantom{T}} = \frac{\pi}{5},\,\quad \theta_2^{\vphantom{T}} \simeq \frac{5\pi}{6} + 0.16 i,\,\quad \theta_3^{\vphantom{T}} = \frac{4\pi}{7}\quad$ & & \\
\bottomrule
\end{tabular}
}
\end{center}
\caption{Comparison between the numerically computed asymmetry $\left| \eta_B \right|$ and the analytic approximation $\left| \eta_B^{\rm{approx.}}\right|$ from eq.~(\ref{anal}) for example points of the LSS (top two) and ISS (bottom two). All points include all relevant radiative corrections as defined in eq.~(\ref{eqn:inv-masscorrections}) where for these example parameters we have set all Wilson coefficients at lowest order to $1/16\pi^2$.} 
\label{table:approx-etaB}
\end{table}

Finally, the small mass differences between the heavy sterile states generating the resonant enhancement could also lead to coherent oscillations between the SNs. The dynamics of the coherent oscillations will alter the evolution of the lepton asymmetry and could potentially significantly impact the net asymmetry generated for some region of parameter space. To properly account for their effects would require a flavour-covariant set of transport equations, as opposed to the semi-classical Boltzmann equations we employ. We will therefore estimate the impact of coherent sterile neutrino oscillations on the final baryon asymmetry by employing an analytic estimate derived specifically for resonant scenarios of leptogenesis~\cite{Dev:2014laa},
\begin{eqnarray}
\label{asymm-osc}
\eta_B^{\text{osc}} \simeq -\frac{28}{51} \frac{1}{27} \frac{3}{2} \sum\limits_{\alpha,i\neq j} \frac{1}{z_c}\frac{1}{K_{\ell \ell}} (\mathcal{A}_{ij}^{\alpha} \sqrt{x_{ij}} + \mathcal{B}_{ij}^{\alpha}) \frac{2 (m_i^2 - m_j^2) m_N \Gamma_N}{(m_i^2 - m_j^2)^2 + \frac{4 m_N^2 \Gamma_N^2 \text{det}\left[ \Re \left\{ \mathbf{h}^{\dagger}\mathbf{h} \right\} \right]}{(h^{\dagger} h)_{ii}(h^{\dagger} h)_{jj}}},
\end{eqnarray}
where $K_{\ell \ell} = (m_N (h h^\dagger)_{\ell \ell})/(8 \pi H(z=1))$ and $\Gamma_N$ is the average of the sterile neutrino decay widths.
Equation~\ref{asymm-osc} has a similar behaviour to the asymmetry due to standard mixing effects given in eq.~(\ref{cpasymmetry2}) and~\cref{leptotobaryo}. In particular, the appearance of $(\mathcal{A}_{ij}^{\alpha} \sqrt{x_{ij}} + \mathcal{B}_{ij}^{\alpha})$ in both expressions implies that the sign of the asymmetries generated by both of these processes are the same. Therefore for the resonant scenario the effect of including oscillations will only increase the overall asymmetry. The masses of the heavy SNs being larger than the electroweak-phase transition temperature, in combination with the strong washout nature of the parameter space we consider, prevents a `freeze-in' scenario due to oscillations, as in traditional ARS leptogenesis~\cite{Akhmedov:1998qx}. Therefore the asymmetry generated from coherent oscillations should be independent of the initial conditions of the heavy SNs.

\subsection{Inverse seesaw (ISS)}
\label{subsec:ISS}

For the ISS ($\mathcal{Y}_L^{\vphantom{\dagger}}=0$,) the only LNV parameters present are the two small Majorana masses
\begin{equation}
M_{\nu}=\begin{pmatrix}
m_\nu^{\text{loop}} & m_{D} & 0 \\
m_{D}^{T} & \mu_N^{\vphantom{\dagger}} & M_{R}^{\vphantom{T}} \\
0 & M_{R}^{T} & \mu_S^{\vphantom{\dagger}} \end{pmatrix}.
\end{equation}
From eq.~(\ref{eqn:inv-masscorrections}) only $\mu_N^{\vphantom{\dagger}}$ receives radiative corrections from $SO(3)_{N_R}$ breaking terms proportional to the Yukawa spurion $\mathcal{Y}_D^{\vphantom{\dagger}}$. \

We separately consider three scenarios for the ISS: first where no corrections are introduced, second where corrections are introduced at the next lowest order, labelled $\mathcal{N}_1^{\vphantom{\dagger}}$ and, third, corrections up to next-to-leading order, labelled $\mathcal{N}_2^{\vphantom{\dagger}}$:
\begin{equation}
\tilde{\mu}_{N} = \mu_N^{\vphantom{\dagger}} \left( \id_3 + \mathcal{N}_1^{\vphantom{\dagger}} + \mathcal{N}_2^{\vphantom{\dagger}}\right)
\end{equation}
where
\begin{eqnarray}
\label{eqn:ISSspurions}
\mathcal{N}_1^{\vphantom{\dagger}} \,& = &\, n_1^{\vphantom{T}} \left( \mathcal{Y}_D^\dagger \mathcal{Y}_D^{\vphantom{\dagger}} + (\mathcal{Y}_D^\dagger \mathcal{Y}_D^{\vphantom{\dagger}})^T\right),\nonumber\\[5pt]
\mathcal{N}_2^{\vphantom{\dagger}}\, & = & \,n_2^{(1)} \left( \mathcal{Y}_D^\dagger \mathcal{Y}_D^{\vphantom{\dagger}} \mathcal{Y}_D^\dagger \mathcal{Y}_D^{\vphantom{\dagger}} + (\mathcal{Y}_D^\dagger \mathcal{Y}_D^{\vphantom{\dagger}}\mathcal{Y}_D^\dagger \mathcal{Y}_D^{\vphantom{\dagger}})^T\right) + n_2^{(2)} \left( \mathcal{Y}_D^\dagger \mathcal{Y}_D^{\vphantom{\dagger}} (\mathcal{Y}_D^\dagger \mathcal{Y}_D^{\vphantom{\dagger}})^T \right) \nonumber\\
&\mbox{}&+ \,n_2^{(3)} \left( (\mathcal{Y}_D^\dagger \mathcal{Y}_D^{\vphantom{\dagger}})^T \mathcal{Y}_D^\dagger \mathcal{Y}_D^{\vphantom{\dagger}}\right) + n_2^{(4)} \left( \mathcal{Y}_D^{\dagger} \mathcal{Y}_e^{\vphantom{\dagger}} \mathcal{Y}_e^{\dagger} \mathcal{Y}_D^{\vphantom{\dagger}} + (\mathcal{Y}_D^{\dagger} \mathcal{Y}_e^{\vphantom{\dagger}} \mathcal{Y}_e^{\dagger} \mathcal{Y}_D^{\vphantom{\dagger}})^T \right)
\end{eqnarray}
and we fix $n_2^{(i)} = (n_1^{\vphantom{T}})^2$ for computational convenience. Due to the perturbative regime in which we operate, it will always be the case that $||\mathcal{N}_i^{\vphantom{T}}||\ll 1$. All terms are formally flavour invariant and serve to break the flavour degeneracy amongst the heavy SNs when these spurions acquire non-zero VEVs. 

For the ISS as $m_L = 0_{3 \times 3}$ the combination in~\cref{eqn:mddmd} simplifies to
\begin{eqnarray}
\label{eqn:ISS-mddmd}
h^{\dagger} h \,\simeq \, \frac{2}{v^2} \begin{pmatrix}
\hphantom{ii}\left( \frac{1}{\sqrt{2}} - \frac{1}{4\sqrt{2}}\left(\frac{\mu_S^{\vphantom{T}} - \mu_N^{\vphantom{T}}}{2 m_R} \right)\right)^2 m_D^{\dagger} m_D^{\vphantom{\dagger}} &\,\,\, - i \,\,\left(\frac{1}{2} - \frac{1}{32}\left(\frac{\mu_S^{\vphantom{T}} - \mu_N^{\vphantom{T}}}{2 m_R}\right)^2\right) \,\,\,m_D^{\dagger} m_D^{\vphantom{\dagger}}  \\[15pt]
i \,\,\left(\frac{1}{2} - \frac{1}{32}\left(\frac{\mu_S^{\vphantom{T}} - \mu_N^{\vphantom{T}}}{2 m_R}\right)^2\right)\,\,\,\, m_D^{\dagger} m_D^{\vphantom{\dagger}} & \,\hphantom{-ii}\left( \frac{1}{\sqrt{2}} + \frac{1}{4\sqrt{2}}\left(\frac{\mu_S^{\vphantom{T}} - \mu_N^{\vphantom{T}}}{2 m_R} \right)\right)^2 m_D^{\dagger} m_D^{\vphantom{\dagger}} \end{pmatrix}\nonumber\\
\end{eqnarray}
where we are now dealing with unprimed matrices. Due to the fact that $M_R \propto \mu_S^{\vphantom{T}} \propto \id_3$, in the case where only $\mathcal{N}_1^{\vphantom{\dagger}}$ is included the corrected Majorana mass $\mu_N^{\vphantom{T}}$ can be diagonalised without affecting the other entries in the $2 \times 2 $ sub-block, with
\begin{equation}
\begin{pmatrix}
0 & \,m_D^{\vphantom{T}} \,& 0\\[5pt]
m_D^T \,&\,\, \mu_N^{\vphantom{\dagger}}\id_3 + \delta \mu_N^{\vphantom{\dagger}} \,\,&\, m_R \id_3\\[5pt]
0 &\, m_R \id_3\, & \mu_S^{\vphantom{\dagger}}\id_3 \end{pmatrix} \rightarrow \begin{pmatrix}
0 & \,m_D^{\vphantom{T}}  \mathcal{O} \,& 0\\[5pt]
\mathcal{O}^T m_D^T \,&\,\, \hat{\tilde{\mu}}_{N} \,\,&\, m_R \id_3\\[5pt]
0 &\, m_R \id_3\, & \mu_S^{\vphantom{T}} \end{pmatrix}
\end{equation}
under the rotation $N_R \rightarrow \mathcal{O} N_R$ and $S_L \rightarrow \mathcal{O} S_L$. \textit{Similarly} to the minimal type-I scenario, if only $\mathcal{N}_1^{\vphantom{T}}$ is included then $\Re\big\{m_D^{\dagger} m_D^{\vphantom{\dagger}}\big\} \propto \mathcal{N}_1^{\vphantom{T}} = \delta \mu_N^{\vphantom{T}}$ and the combinations of $m_D^{\dagger} m_D^{\vphantom{\dagger}}$ appearing in eq.~(\ref{eqn:ISS-mddmd}) will have no real off-diagonal components. We emphasize that this only occurs for the ISS because $m_L = 0_{3 \times 3}$. \textit{Differently} to the type-I scenario however the off-diagonal blocks of eq.~(\ref{eqn:ISS-mddmd}) have an additional complex phase such that the imaginary components\footnote{This is because $\mathcal{O}^T \Re \left\{ \mathcal{Y}_D^{\dagger} \mathcal{Y}_D^{\vphantom{\dagger}} \right\} \mathcal{O} = \text{diag}(\dots)$ due to the form of the corrections but $\mathcal{O}^T \Im \left\{ \mathcal{Y}_D^{\dagger} \mathcal{Y}_D^{\vphantom{\dagger}} \right\} \mathcal{O} \neq \text{diag}(\dots)$ and therefore off-diagonal imaginary components are generated which become real due to the additional phase in eq.~(\ref{eqn:ISS-mddmd}).} of $\mathcal{O}^T m_D^{\dagger} m_D^{\vphantom{\dagger}} \mathcal{O}$ become real off-diagonal entries in $h^{\dagger} h$. This therefore allows for non-zero generation of lepton asymmetry unlike the type-I scenario.

As only the off-diagonal blocks of $h^\dagger h$ contain the necessary terms, only diagrams between two SNs of different mass splittings, i.e.\ from $\Delta m_{i,j}^{\text{OS}}$, will contribute to asymmetry.  The real components of the $(1,1)$ and $(2,2)$ blocks are diagonal, on the other hand, meaning that SNs with the same sign mass splitting $\Delta m_{i,j}^{\text{SS}}$ will not contribute to $\varepsilon_{\alpha}^{i}$. Once $\mathcal{N}_2^{\vphantom{T}}$ is included however, the flavour mis-alignment between $\delta \mu_N^{\vphantom{T}}$ and $m_D^{\dagger} m_D^{\vphantom{\dagger}}$ will generate real off-diagonal components in $m_D^{\dagger} m_D^{\vphantom{\dagger}}$ and therefore all four blocks of eq.~(\ref{eqn:ISS-mddmd}) will contribute to $\varepsilon_{\alpha}^{i}$.

\begin{figure}[t]
\centering
{
  \includegraphics[width=0.45\linewidth]{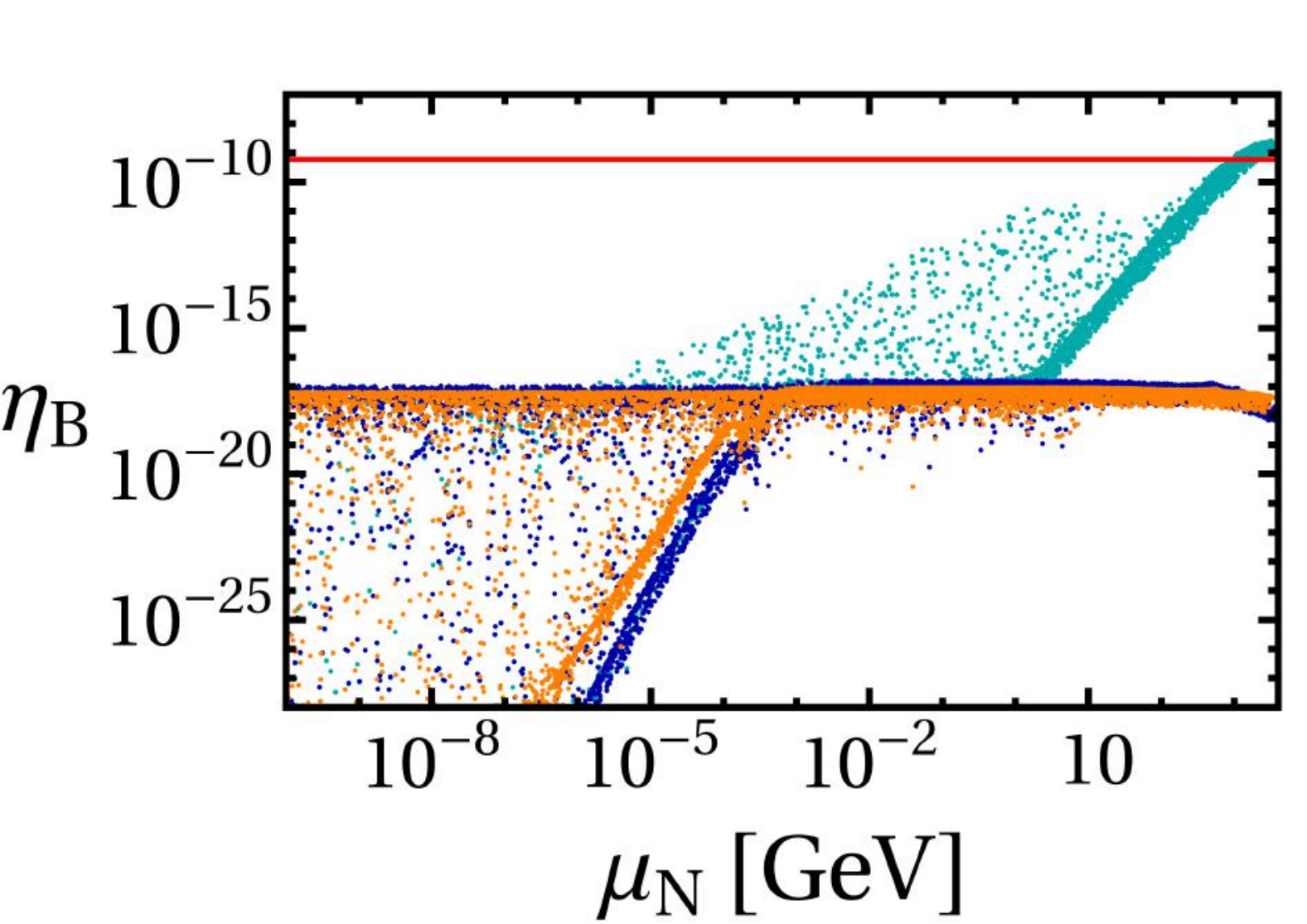}
}
{
  \includegraphics[width=0.45\linewidth]{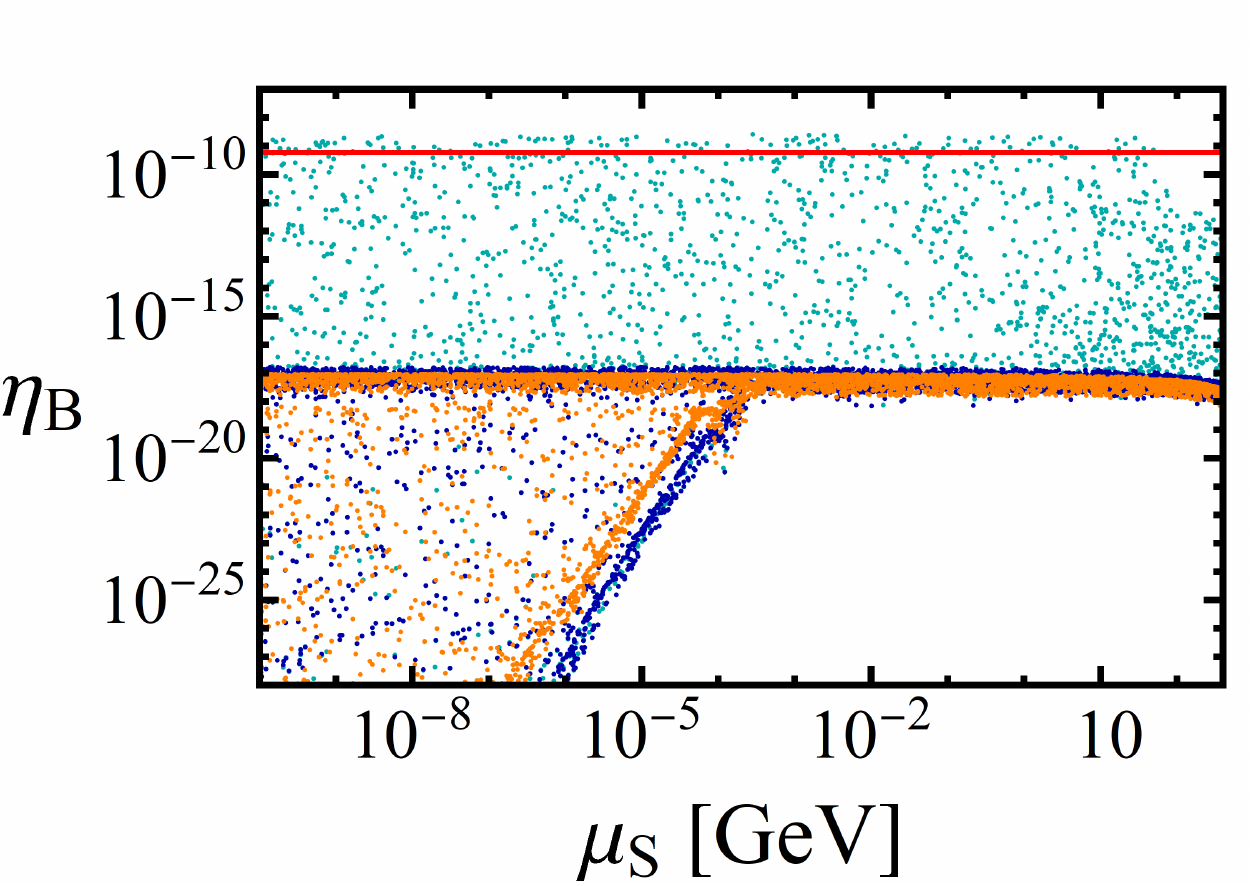}
}
\caption{Plot of the asymmetry generated in the ISS for the individual cases where no radiative corrections are considered (orange), where only $\mathcal{N}_1^{\protect\vphantom{T}}$ is considered (blue) and when both $\mathcal{N}_1^{\protect\vphantom{T}}$ and $\mathcal{N}_2^{\protect\vphantom{T}}$ are included (cyan). This was varied with $\mu_N^{\protect\vphantom{T}}$ \textbf{(left figure)} and $\mu_S^{\protect\vphantom{T}}$ \textbf{(right figure)}. A resonant enhancement in the asymmetry occurs only if the next-to-leading order contributions are included. The red horizontal line indicates the asymmetry required to fit observations.}
\label{figure:allspur-asymm}
\end{figure}

In~\cref{figure:allspur-asymm} we plot the baryon asymmetry numerically calculated as a function of both $\mu_N^{\vphantom{T}}$ and $\mu_S^{\vphantom{T}}$. All three scenarios are simultaneously plotted. Both in the case where radiative effects are ignored as well as when only $\mathcal{N}_1^{\vphantom{\dagger}}$ is included, similar behaviour occurs. The inclusion of $\mathcal{N}_1^{\vphantom{T}}$ breaks the mass degeneracy between all six SNs, as opposed to without its inclusion where two groups of identical mass SNs form. However, as discussed above, due to flavour alignment only diagrams involving opposite mass split SNs contribute to the asymmetry. Therefore the mass difference is generated exclusively from
\begin{equation}
\Delta m_{i,j}^{\text{OS}} = \left( m_{N_\pm} \right)_i - \left( m_{N_\mp} \right) _j \simeq   \mu_N^{\vphantom{T}} + \mu_S^{\vphantom{T}} 
\end{equation}
with
\begin{equation}
m_{N_\pm} \simeq m_R \pm \frac{1}{2}\mu_N^{\vphantom{T}} \pm \frac{1}{2}\mu_S^{\vphantom{T}}.
\end{equation}
While the spurion corrections are present we have $\mu_N^{\vphantom{T}} \gg \mathcal{N}_1$ and therefore they are not the dominant source for the mass splittings in this situation. In this regime resonant leptogenesis is not feasible~\cite{Dolan:2018qpy,Agashe:2018oyk,Agashe:2018cuf}. The inclusion of $\mathcal{N}_2^{\vphantom{T}}$ generates non-zero, off-diagonal entries in the $(1,1)$ and $(2,2)$ entries of eq.~(\ref{eqn:ISS-mddmd}) such that new contributions turn on arising from the mass splitting
\begin{equation}
\Delta m_{i,j}^{\text{SS}} = \left( m_{N_\pm} \right)_i - \left( m_{N_\pm} \right) _j \simeq \mu_N^{\vphantom{T}} \left(\mathcal{Y}_D^{\dagger} \mathcal{Y}_D^{\vphantom{\dagger}} \right)_{ii}.
\end{equation}
We emphasis that even when only $\mathcal{N}_1^{\vphantom{T}}$ is included, the mass degeneracy between the same sign mass split SNs is still broken $\Delta m_{i,j}^{\text{SS}} \neq 0$. No resonant enhancement occurs between two such SNs in this case not because of a mass degeneracy between them (as is the case when no corrections are included) but because of the flavour alignment issue described above. 

\begin{figure}[t]
\centering
{
  \includegraphics[width=0.45\linewidth]{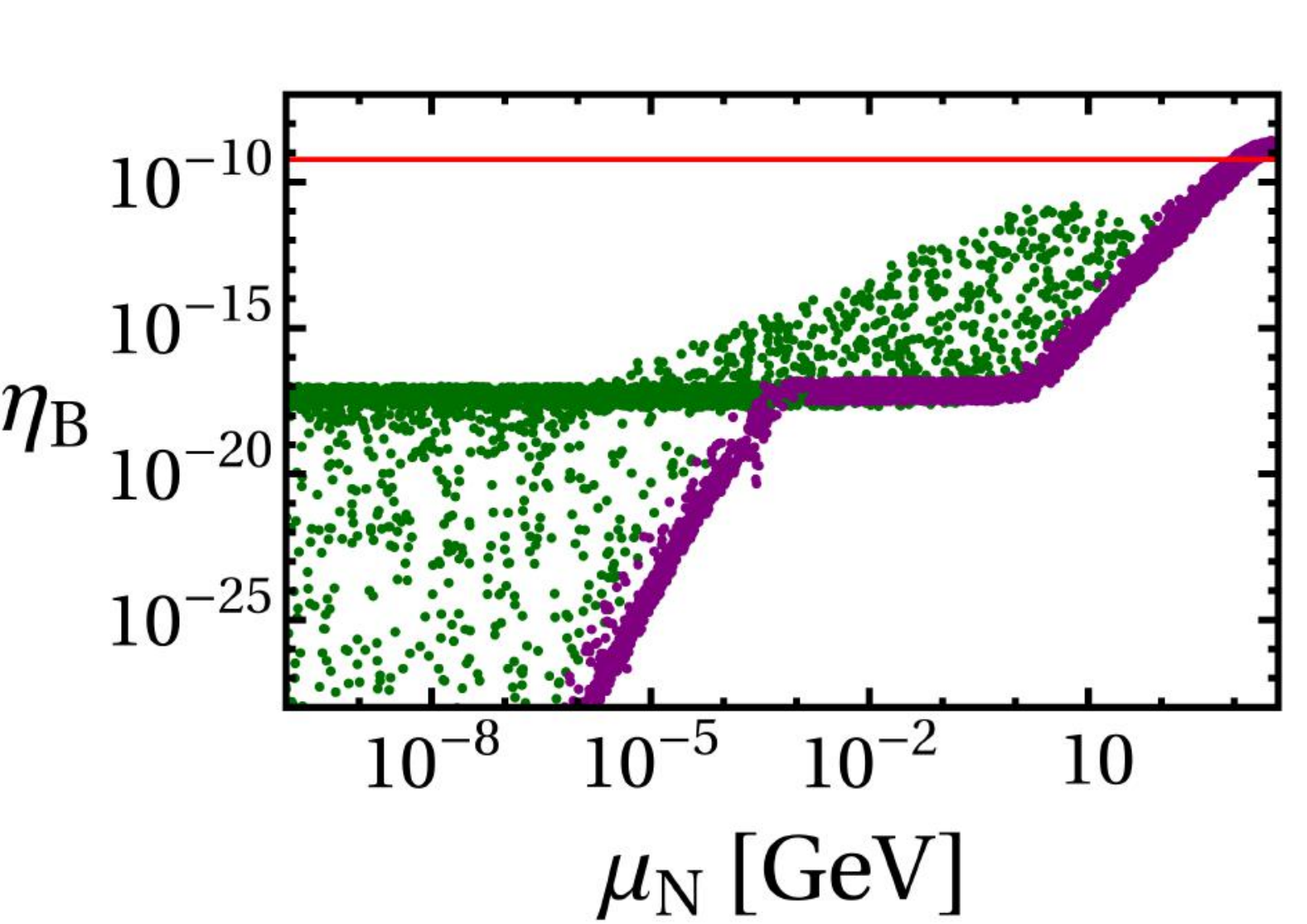}
}
{
  \includegraphics[width=0.45\linewidth]{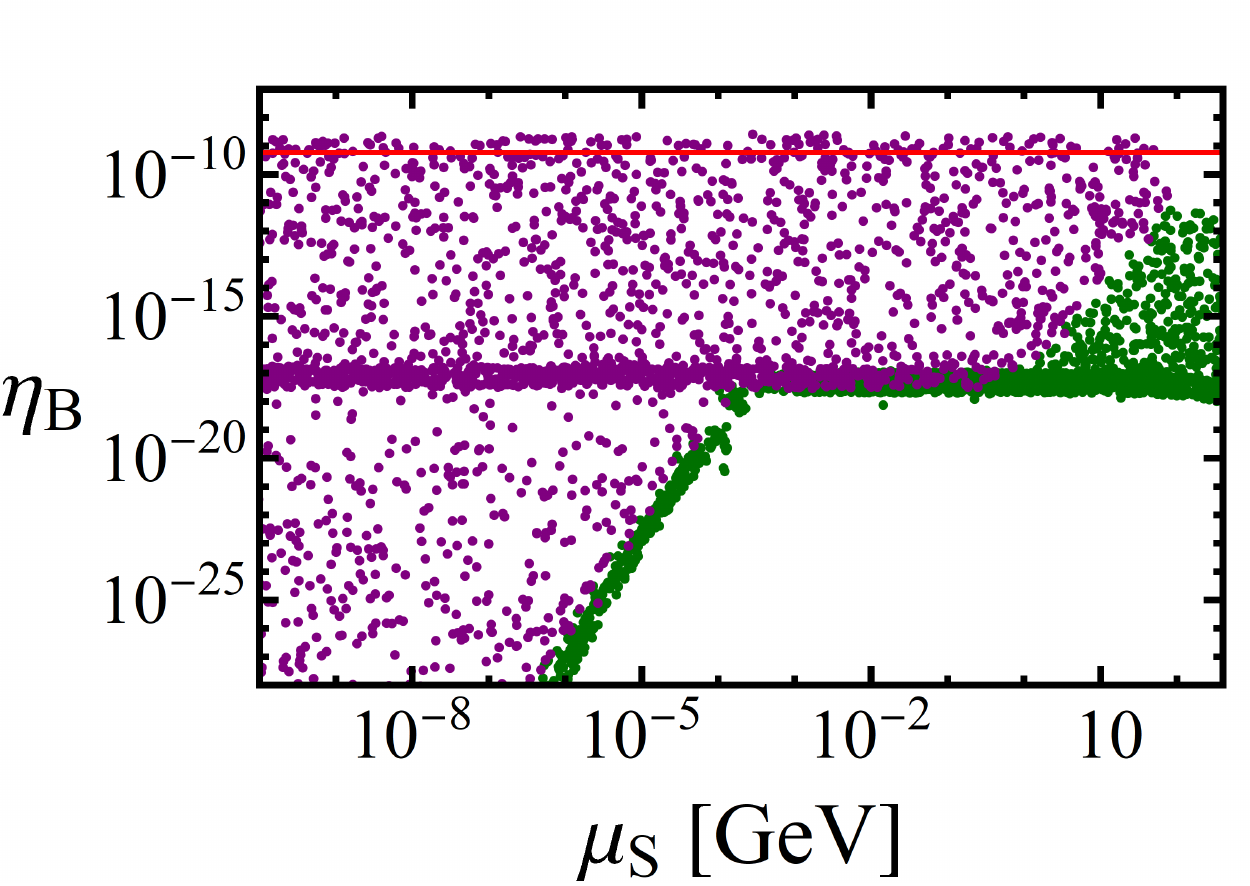}
}
{
  \includegraphics[width=0.45\linewidth]{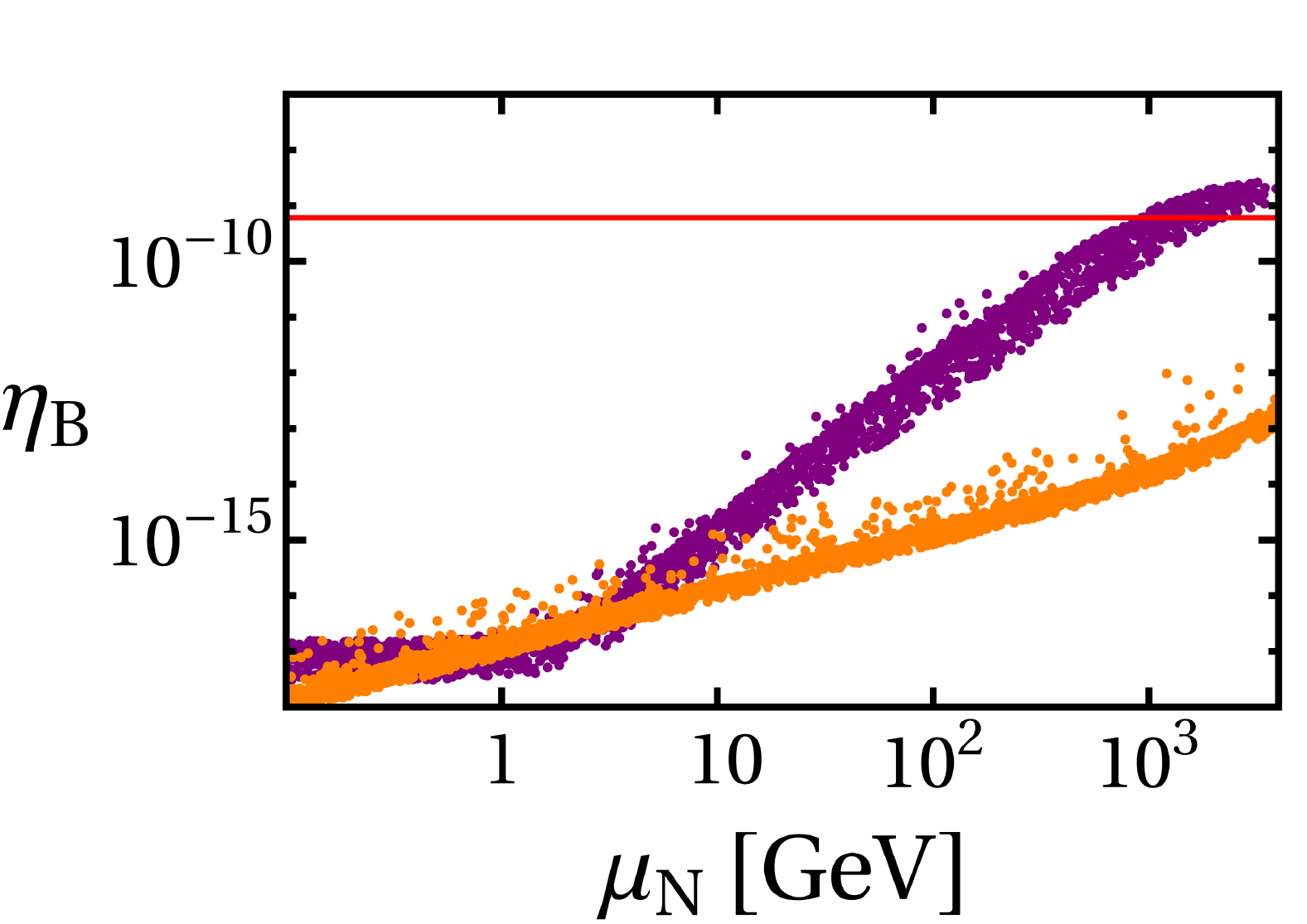}
}
\caption{Plot of the asymmetry generated in the ISS when both $\mathcal{N}_1^{\protect\vphantom{T}}$ and $\mathcal{N}_2^{\protect\vphantom{T}}$ are included, as a function of $\mu_N^{\protect\vphantom{T}}$ \textbf{(top-left figure)} and $\mu_S^{\protect\vphantom{T}}$ \textbf{(top-right figure)}. Points in purple correspond to the regime where $\mu_N^{\protect\vphantom{T}} > \mu_S^{\protect\vphantom{T}}$, while points in green to $\mu_S^{\protect\vphantom{T}} > \mu_N^{\protect\vphantom{T}}$. A resonant enhancement of sufficient size is generated for \textit{large} values of $\mu_N^{\protect\vphantom{T}}$ as long as the hierarchy $\mu_N^{\protect\vphantom{T}} > \mu_S^{\protect\vphantom{T}}$ is satisfied. In order to match active neutrino data, $\mu_S^{\protect\vphantom{T}}$ which generates a tree-level contribution must be less than $\sim 100$ GeV, whereas $\mu_N^{\protect\vphantom{T}}$, which generates it at loop level, is allowed to be larger. The \textbf{bottom} figure compares the asymmetry generated from standard decay to oscillations (in orange) in the relevant regime $\mu_N^{\protect\vphantom{T}} > \mu_S^{\protect\vphantom{T}}$. While for some region of parameter space the two processes are of similar order, in the resonant regime the asymmetry due to oscillations is roughly three orders of magnitude smaller and is therefore a sub-dominant effect.}
\label{figure:spur2-asymm}
\end{figure}

\begin{figure}[t]
\centering
{
  \includegraphics[width=0.35\linewidth]{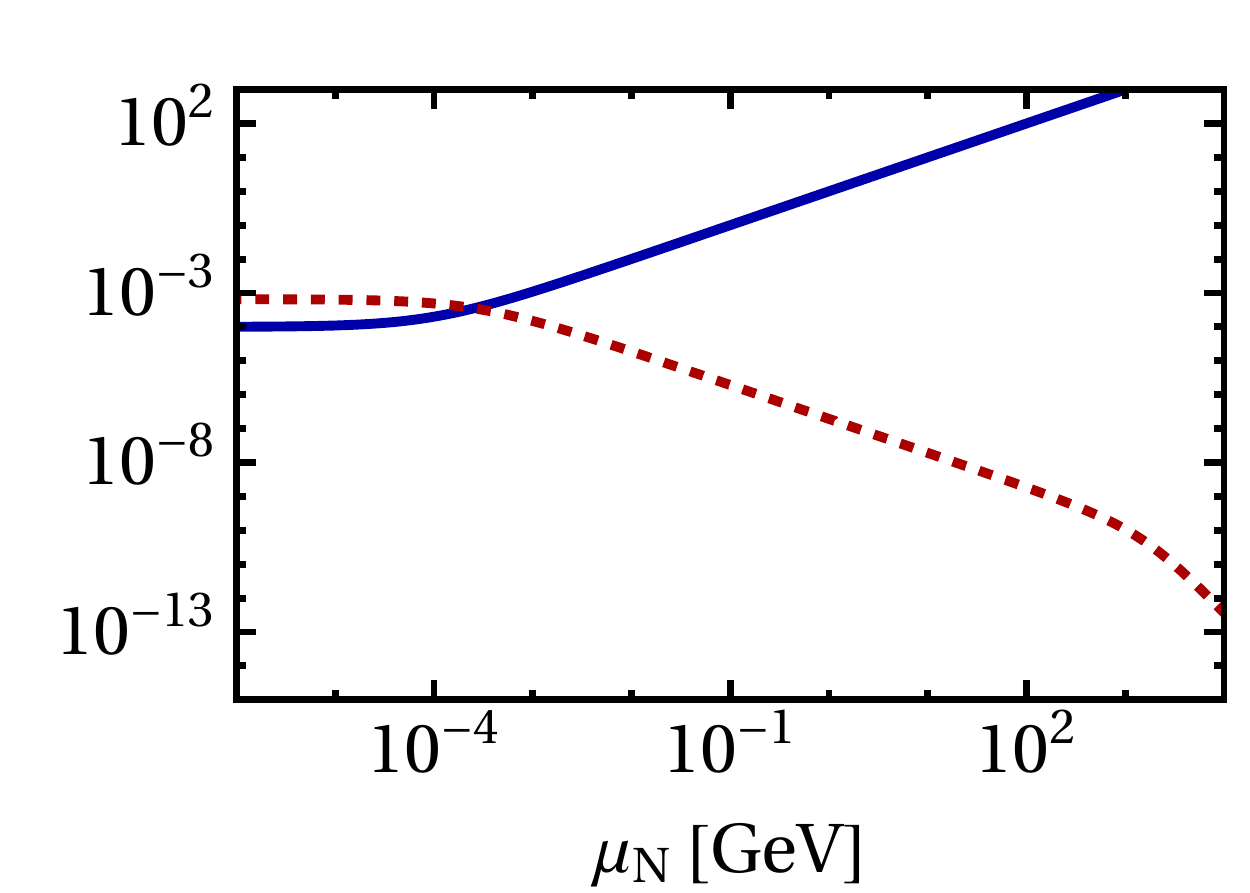}\hfill
}
{
  \hspace*{0.1mm}\includegraphics[width=0.27\linewidth]{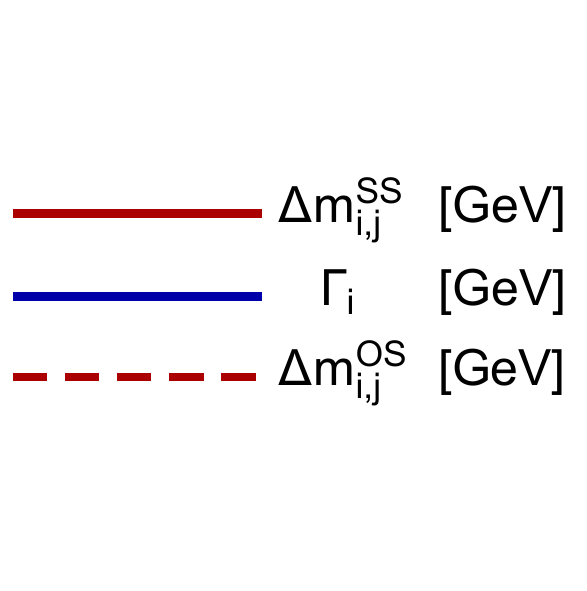}\hfill
}
{
  \includegraphics[width=0.35\linewidth]{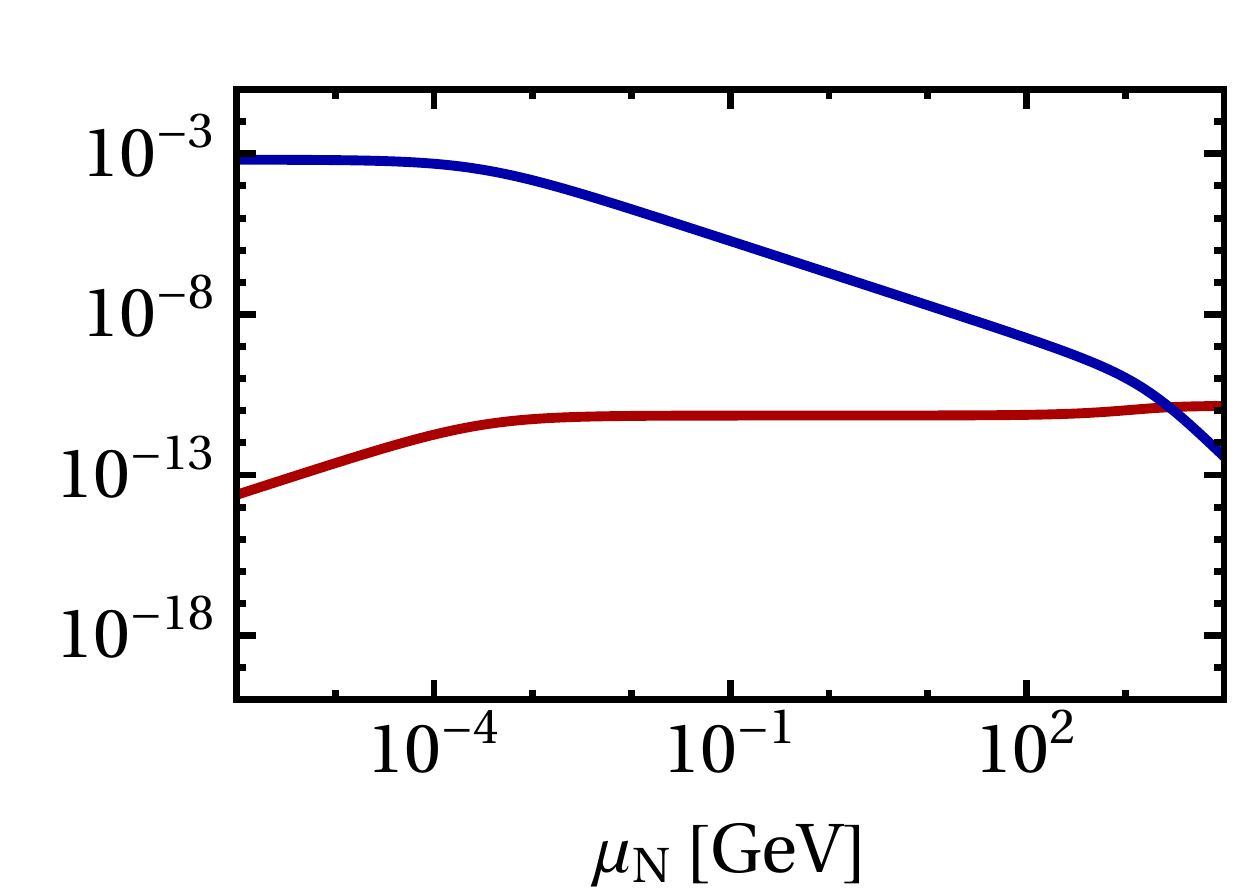}
}
\caption{Plot of the decay width $\Gamma_i \simeq \frac{m_i}{8\pi}\sum_{l} h_{li}^* h_{li}^{\protect\vphantom{*}}$ (blue) along with the mass splitting $\Delta m_{i,j}$ (red) as a function of $\mu_N^{\protect\vphantom{T}}$ for $\mu_S^{\protect\vphantom{T}} = 10^{-4}$ GeV and all radiative effects included. Here the mass splitting among opposite splitting SNs $\Delta m_{i,j}^{\text{OS}}$ \textbf{(left)} is compared to those with same sign splitting $\Delta m_{i,j}^{\text{SS}}$ \textbf{(right)} as defined in~\cref{eqn:ISS-OSsplitting,eqn:ISS-SSsplitting}. While in both plots a region of resonance occurs, in the case on the left it occurs in a region of extremely strong washout c.f.~\cref{figure:washout}. The inclusion of radiative effects allows for a resonance to occur between SNs with same-sign mass splittings in a region of minimised washout such that asymmetry generation can occur. In this region heavy SNs of opposite sign mass splitting grow in mass difference and their contribution becomes irrelevant.}
\label{figure:rescond-ISS}
\end{figure}

Figure~\ref{figure:spur2-asymm} plots the asymmetry when $\mathcal{N}_2^{\vphantom{T}}$ is included and a clear resonant enhancement occurs for large values of $\mu_N^{\vphantom{T}}$. While the mass splitting is proportional to the decay width, it is also scaled by the Majorana mass $\mu_N^{\vphantom{T}}$. For small LNV the mass splitting will be much less than the decay width but for large values
\begin{equation}
\mu_N^{\vphantom{T}} \left(\mathcal{Y}_D^{\dagger} \mathcal{Y}_D^{\vphantom{\dagger}} \right)_{ii} \rightarrow \frac{m_{N_i}}{8\pi} \left(\mathcal{Y}_D^{\dagger} \mathcal{Y}_D^{\vphantom{\dagger}} \right)_{ii} \simeq \Gamma_i
\end{equation}
forcing a resonant enhancement to occur. In~\cref{figure:spur2-asymm} we distinguish between two different hierarchical cases for the Majorana masses: $\mu_N^{\vphantom{T}} > \mu_S^{\vphantom{T}}$ in purple and $\mu_S^{\vphantom{T}} > \mu_N^{\vphantom{T}}$ in green. The asymmetry in the resonant regime is independent of $\mu_S^{\vphantom{T}}$ as in our setup this mass does not receive any radiative corrections. If a stricter setup was considered such that it did receive corrections (if, for example, we took an $SO(3)_{N_R} \times SO(3)_{S_L} \rightarrow SO(3)_{N_R + S_L}$ flavour symmetry) $\mu_S^{\vphantom{T}}$ would require equivalently large masses in order to be placed in the required resonant regime. However, such large values of $\mu_S^{\vphantom{T}}$ are outside the regime of validity required for the ISS and would spoil the guarantee of the parametrisation in~\cref{eqn:ISScasas} to generate the required active neutrino masses. This follows as unlike $\mu_N^{\vphantom{T}}$, $\mu_S^{\vphantom{T}}$ does not have a loop suppressed contribution to the active neutrino masses. Therefore we find that $\mu_S^{\vphantom{T}}$ cannot contribute to the resonance required for MLFV-ISS.

Additionally, the effects of coherent oscillations are estimated (in orange) in the bottom plot of~\cref{figure:spur2-asymm}, where $\mathcal{N}_2^{\vphantom{T}}$ is included specifically for the regime $\mu_N^{\vphantom{T}} > \mu_S^{\vphantom{T}}$ required for successful resonant leptogenesis. Here we find, for the resonant region of parameter space, that the effects of coherent oscillations are roughly three orders of magnitude smaller than that of standard thermal decay. For similar reasons as presented above for resonant leptogenesis, without the inclusion of the second order spurion corrections $\mathcal{N}_2^{\vphantom{T}}$, the asymmetry generated by coherent oscillations is highly suppressed due to the same flavour related cancellation effects. This is evident as the Yukawa structure of eq.~\ref{asymm-osc} is the same as eq.~\ref{cpasymmetry2} and therefore the same arguments apply.

Summarising, we find two criteria for successful MLFV-ISS resonant leptogenesis: (1) large values of the Majorana mass $\mu_N^{\vphantom{T}}$ such that the mass splitting moves on resonance, and (2) the inclusion of spurion effects up to lext-to-leading order in order to break flavour alignment and prevent only opposite mass splitting SNs from contributing to the asymmetry. This is illustrated in~\cref{figure:rescond-ISS} where the decay width and the mass splittings are plotted with the spurion corrections included. For large values of $\mu_N^{\vphantom{T}}$ the opposite mass splitting SNs move further away from resonance whilst the same sign mass splitting SNs move onto resonance. This effect occurs both with $\mathcal{N}_1^{\vphantom{T}}$ and $\mathcal{N}_2^{\vphantom{T}}$ but, as discussed above, if only $\mathcal{N}_1^{\vphantom{T}}$ is included then same sign mass splitting SNs combinations do not generate any flavoured asymmetry.

\begin{figure}[t]
\label{plot:inv-varymv1}
\centering
{
  \includegraphics[width=0.45\linewidth]{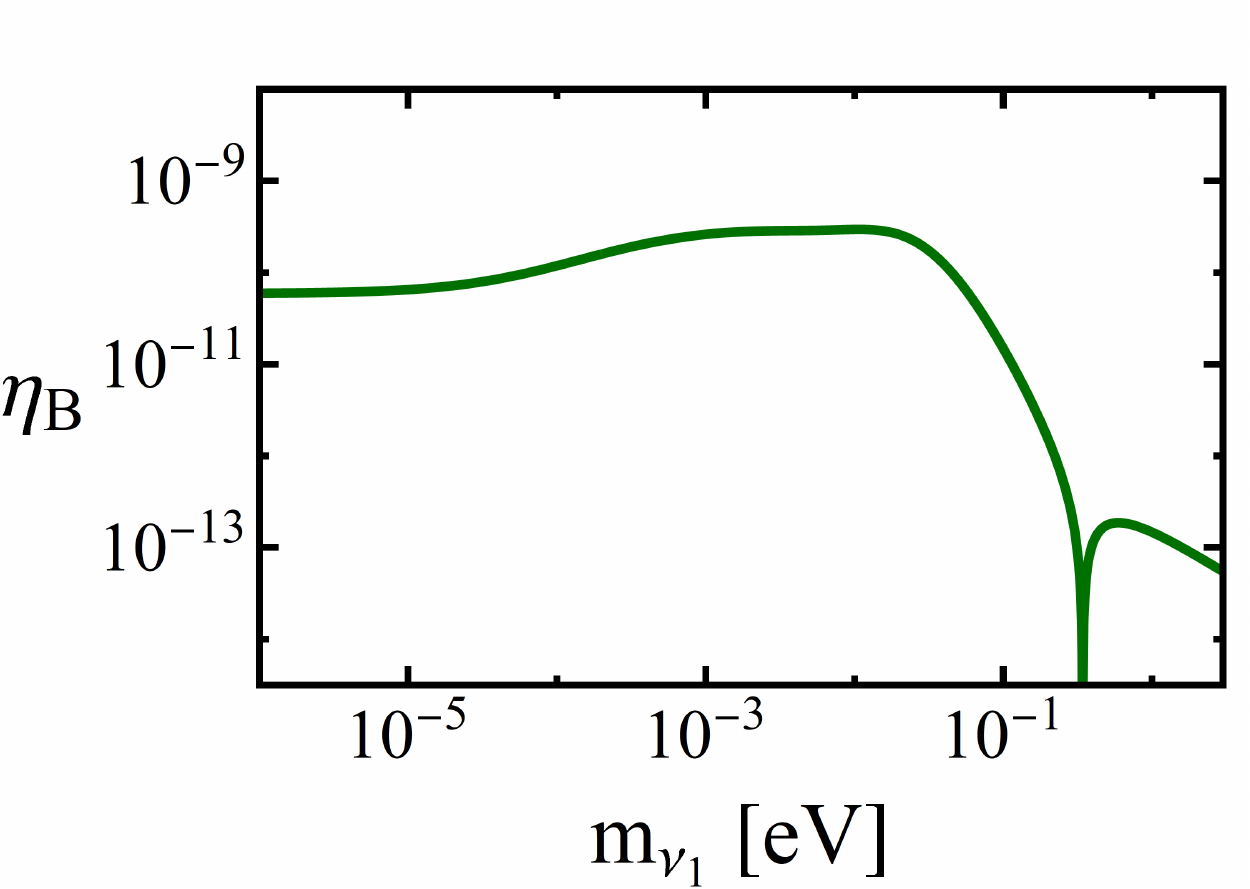}
}
{
  \includegraphics[width=0.45\linewidth]{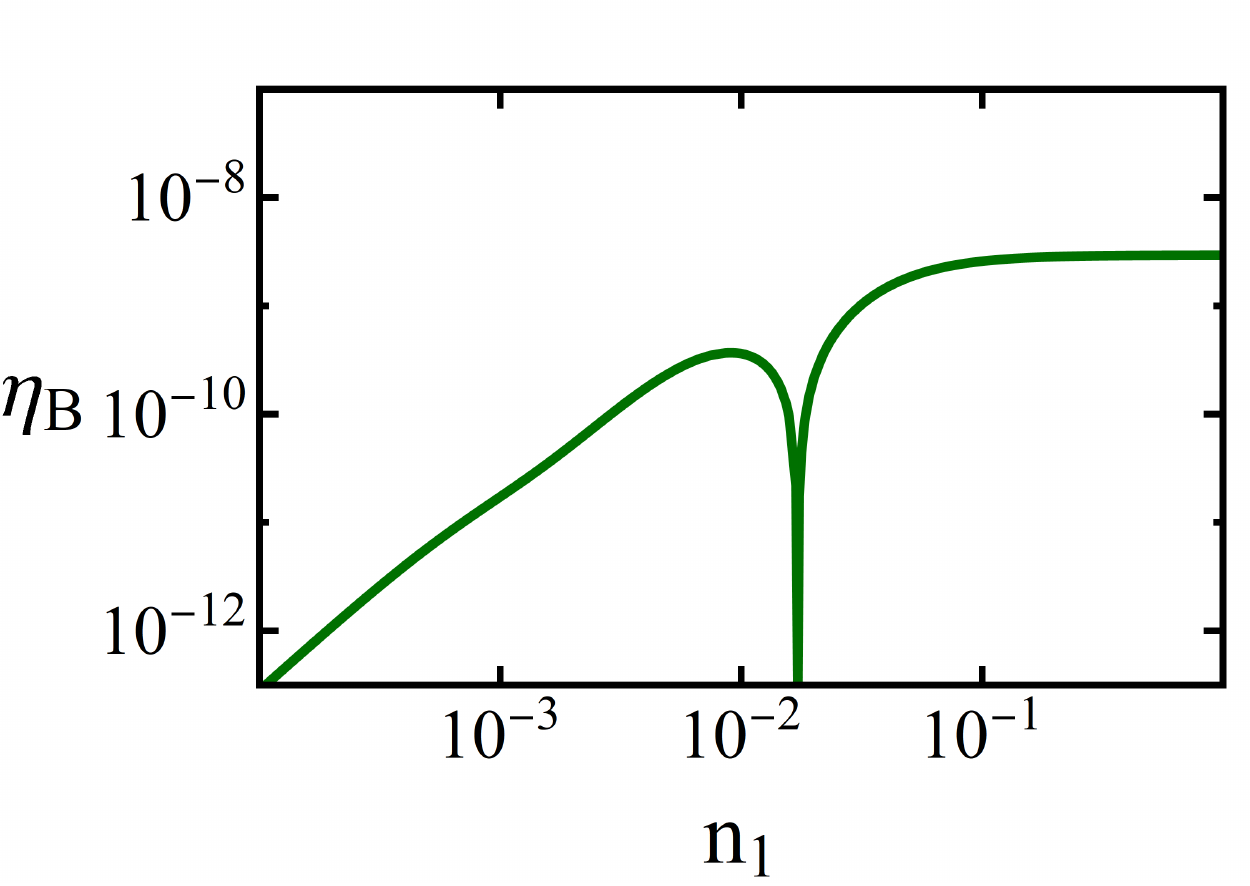}
}
\caption{Variation in the baryon asymmetry as a function of the lightest active neutrino mass $m_{\nu_1}$ \textbf{(left)} and varying the Wilson coefficient $n_1$ \textbf{(right)} for the case where all radiative spurion effects are included. In this scan we fixed $\theta_2^{\text{c}} = 0.7$, $\mu_S^{\protect\vphantom{T}} \simeq 10^{-1}$ GeV and $\mu_N \simeq 900$ GeV and in the left plot set $n_1 = 1/16\pi^2$ and $m_{\nu_1} = 0.01$ eV in the right plot. As the lightest neutrino mass is reduced (becomes hierarchical) the asymmetry freezes out. There is a slight preference for light neutrino masses of $\mathcal{O}(10^{-3}-10^{-2})$ eV. Larger values for $m_{\nu_1}$ lead to a degeneracy amongst the light neutrino masses and decreases the mass splitting generated by the inclusion of $\mathcal{N}_i^{\protect\vphantom{\dagger}}$ decreasing the resonant enhancement. Increasing the size of the Wilson coefficient allows the mass splitting to be closer to resonance allowing for one or two orders of magnitude increased asymmetry.}
\label{figure:assym-varymv1}
\end{figure}

These conclusions where drawn for fixed values of certain parameters. Most importantly the Wilson coefficients were fixed such that $c_1^{\vphantom{T}} = 1/16\pi^2$, which we chose under the assumption that the high-scale dynamics arise from radiative effects. Increasing the size of the Wilson coefficients will increase the overall size of $\mathcal{N}_2^{\vphantom{\dagger}}$ from eq.~(\ref{eqn:ISSspurions}), allowing for smaller values of $\mu_N^{\vphantom{T}}$ by one or two orders of magnitude. Similarly, we fixed the lightest neutrino mass $m_{\nu_1}$ in our scans. Smaller values of the lightest neutrino mass will increase the hierarchy within $m_D^{\dagger} m_D^{\vphantom{\dagger}}$, which impacts the level of resonance through~\cref{eqn:ISScasas}. To illustrate this behaviour, in~\cref{figure:assym-varymv1} the asymmetry is plotted as a function of these two parameters where we have fixed all other parameters to a benchmark point within the resonant region. From this figure we conclude that there is a preference for smaller values of neutrino mass $m_{\nu_1} \ll 0.1$ eV and if larger values of the Wilson coefficients were allowed, a larger region of parameter space would go on resonance.

\begin{figure}[t]
\centering
{
  \includegraphics[width=0.45\linewidth]{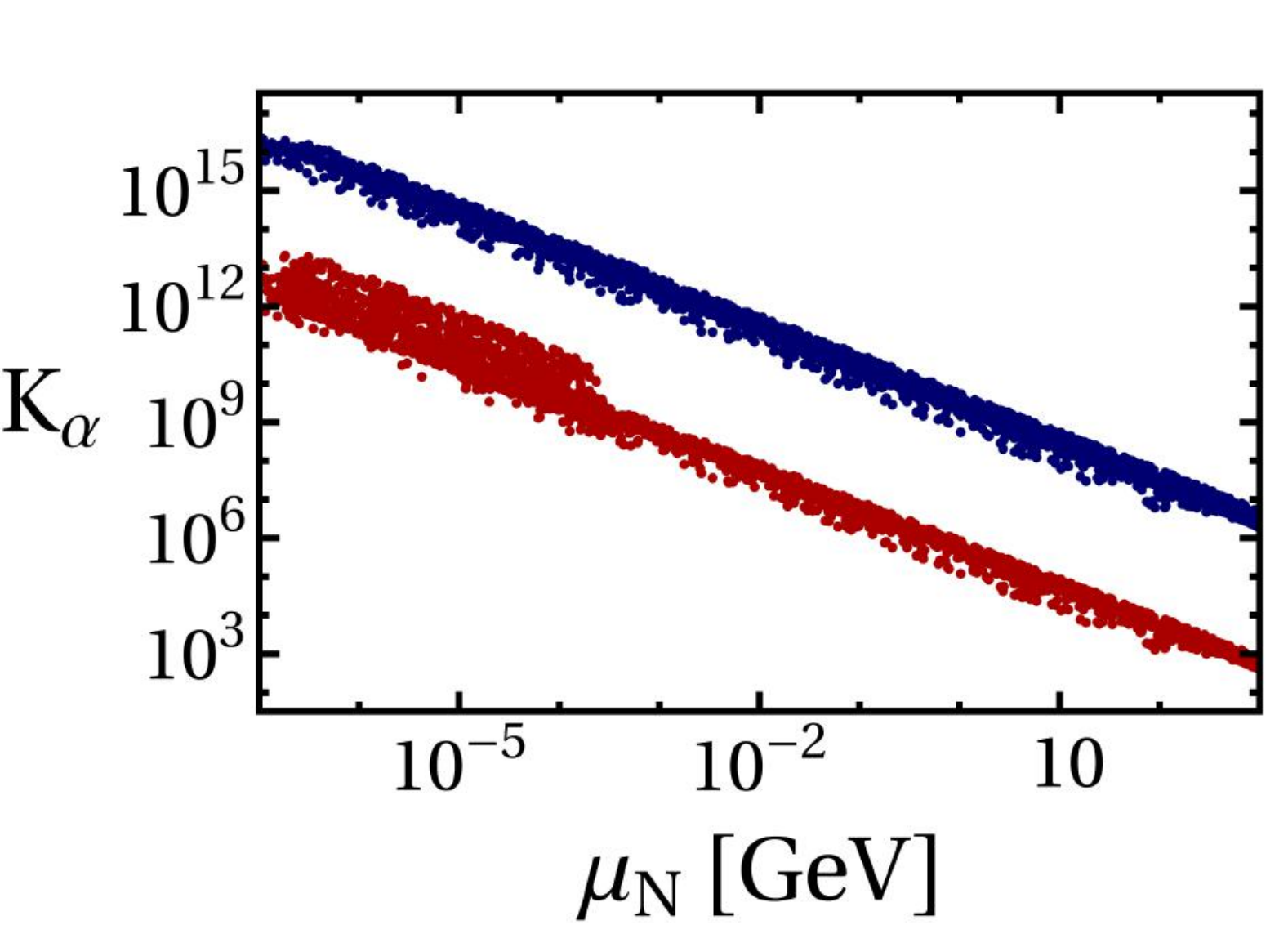}
}
{
  \includegraphics[width=0.45\linewidth]{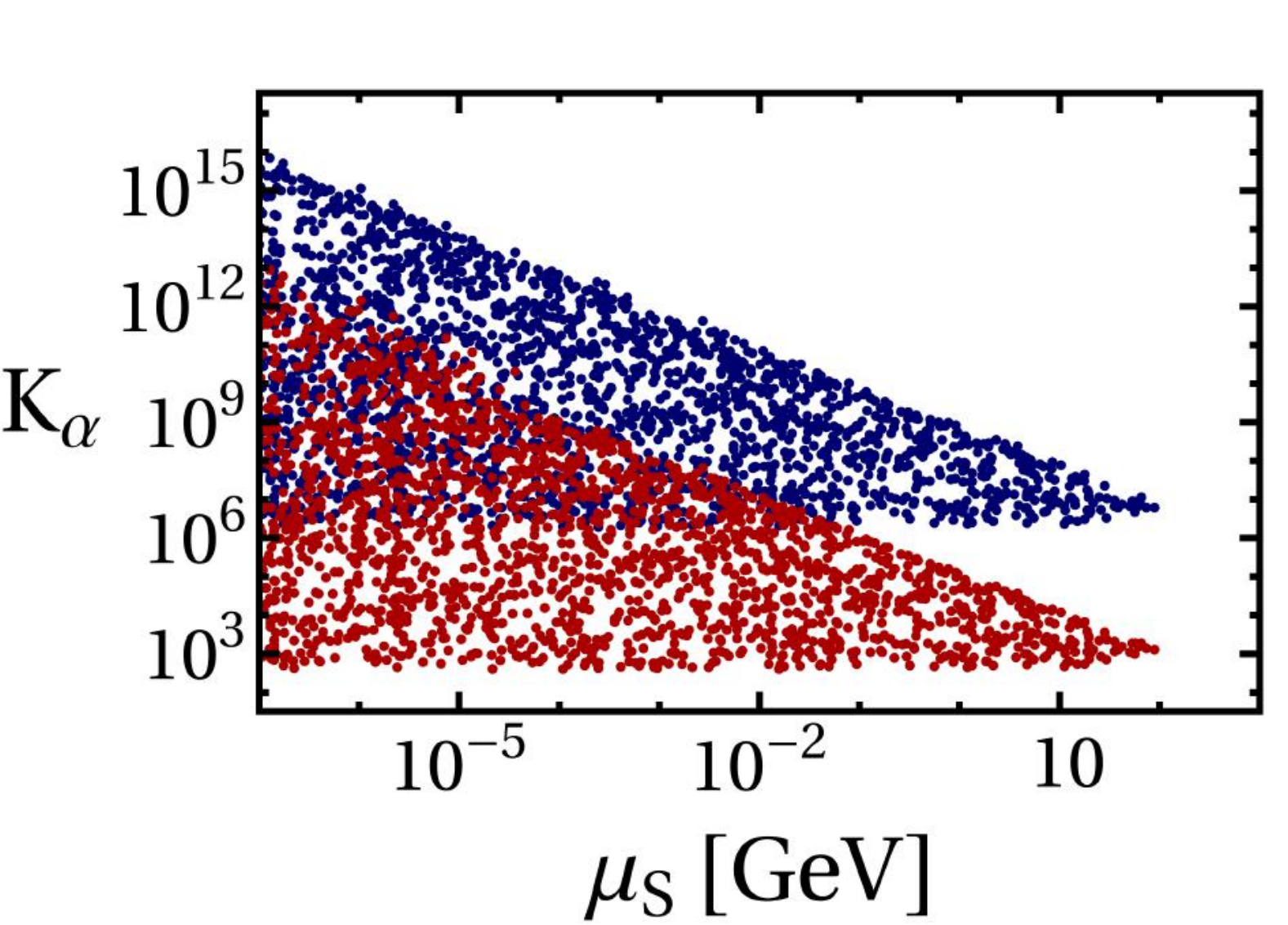}
}
\caption{Plot of the washout as a function of $\mu_N^{\protect\vphantom{T}}$ \textbf{(left)} and $\mu_S^{\protect\vphantom{T}}$ \textbf{(right)} in the scenario where $\mu_N^{\protect\vphantom{T}} > \mu_S^{\protect\vphantom{T}}$ relevant for resonant leptogenesis. In blue the na\"{\i}ve washout is plotted while in red the effective washout which is defined in eq.~(\ref{washout}) relevant for situations with approximate lepton number conservation. As can be seen, na\"{\i}vely the washout is overestimated by several orders of magnitude. In regions of large $\mu_N^{\protect\vphantom{T}}$ the effective washout is low enough (but still within the strong regime) for resonant leptogenesis to be feasible.}
\label{figure:washout}
\end{figure}

Asymmetry generation is sufficient for large Majorana masses not simply because of the enhancement in the mass splitting. Due to the relationship between the Yukawa couplings required to satisfy active neutrino mixing data and the input parameters from~\cref{eqn:ISScasas}, larger values of the Majorana masses (for fixed sterile Dirac mass $m_R$) results in a overall \textit{decrease} in the couplings required to generate the same light neutrino masses which leads to less efficient washout of any generated asymmetry. Figure~\ref{figure:washout} plots the washout as a function of the two Majorana masses for the hierarchy $\mu_N^{\vphantom{T}} > \mu_S^{\vphantom{T}}$. Both the na\"{\i}ve washout $K_\alpha^{\vphantom{T}} \propto \Gamma_i m_i / \text{H}(m_i)$ and the effective washout $K_\alpha^{\text{eff}}$ defined in eq.~(\ref{washout}) (and relevant for scenarios with \textit{small} LN violating parameters~\cite{Blanchet:2009kk,Deppisch:2010fr,Borah:2017qdu} due to the sensitivity to $2 \leftrightarrow 2$ scatterings) are plotted together. 

While the na\"{\i}ve washout grossly overestimates the efficiency of washout, it is clear that for large values of the LNV parameters palatable values of washout (albeit still very much in the strong washout regime) of $\mathcal{O}(10^3)$ or below are possible and therefore resonant leptogenesis is roughly feasible only in this specific region.

\begin{figure}[t]
\centering
{
  \includegraphics[width=0.45\linewidth]{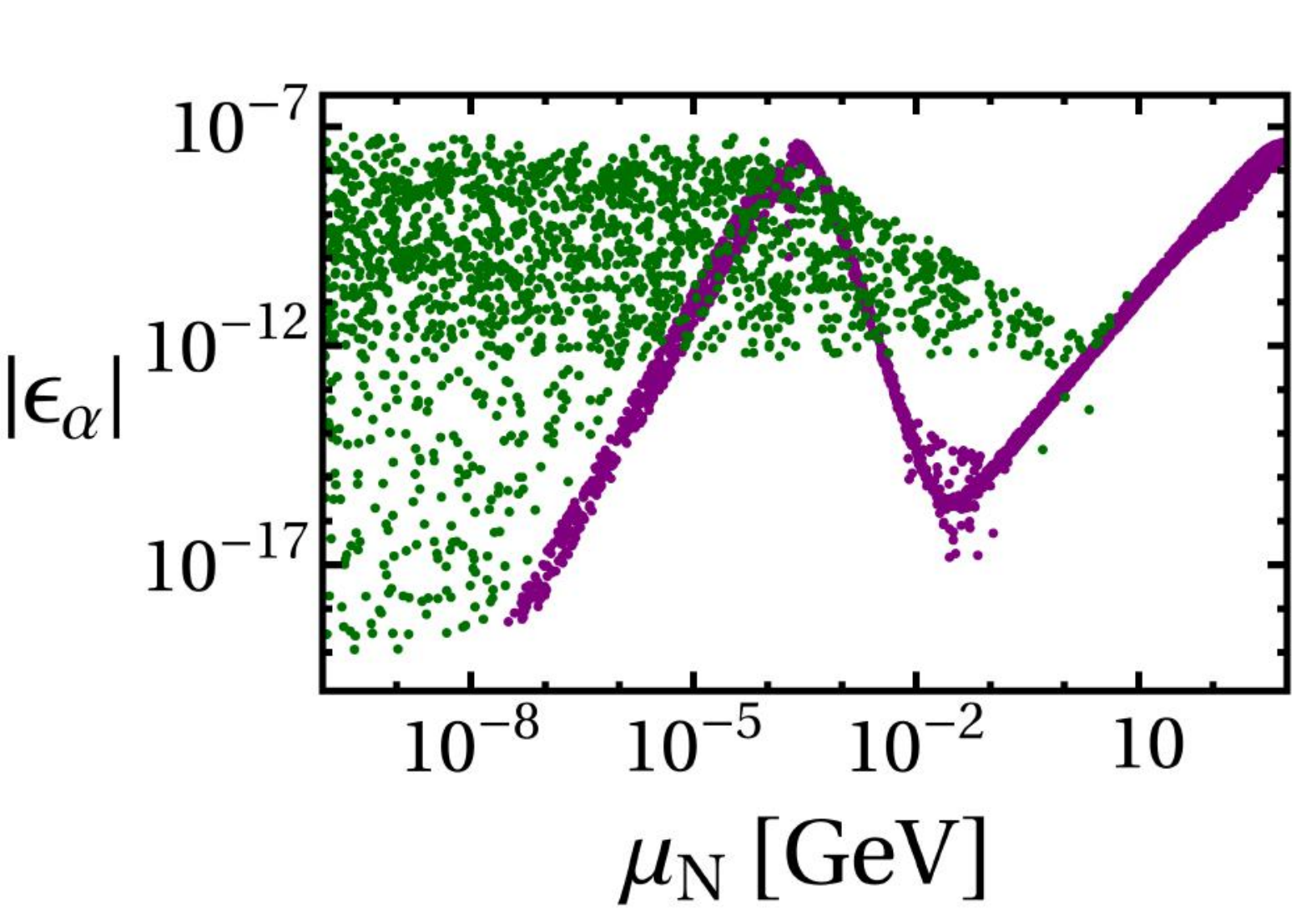}
}
{
  \includegraphics[width=0.45\linewidth]{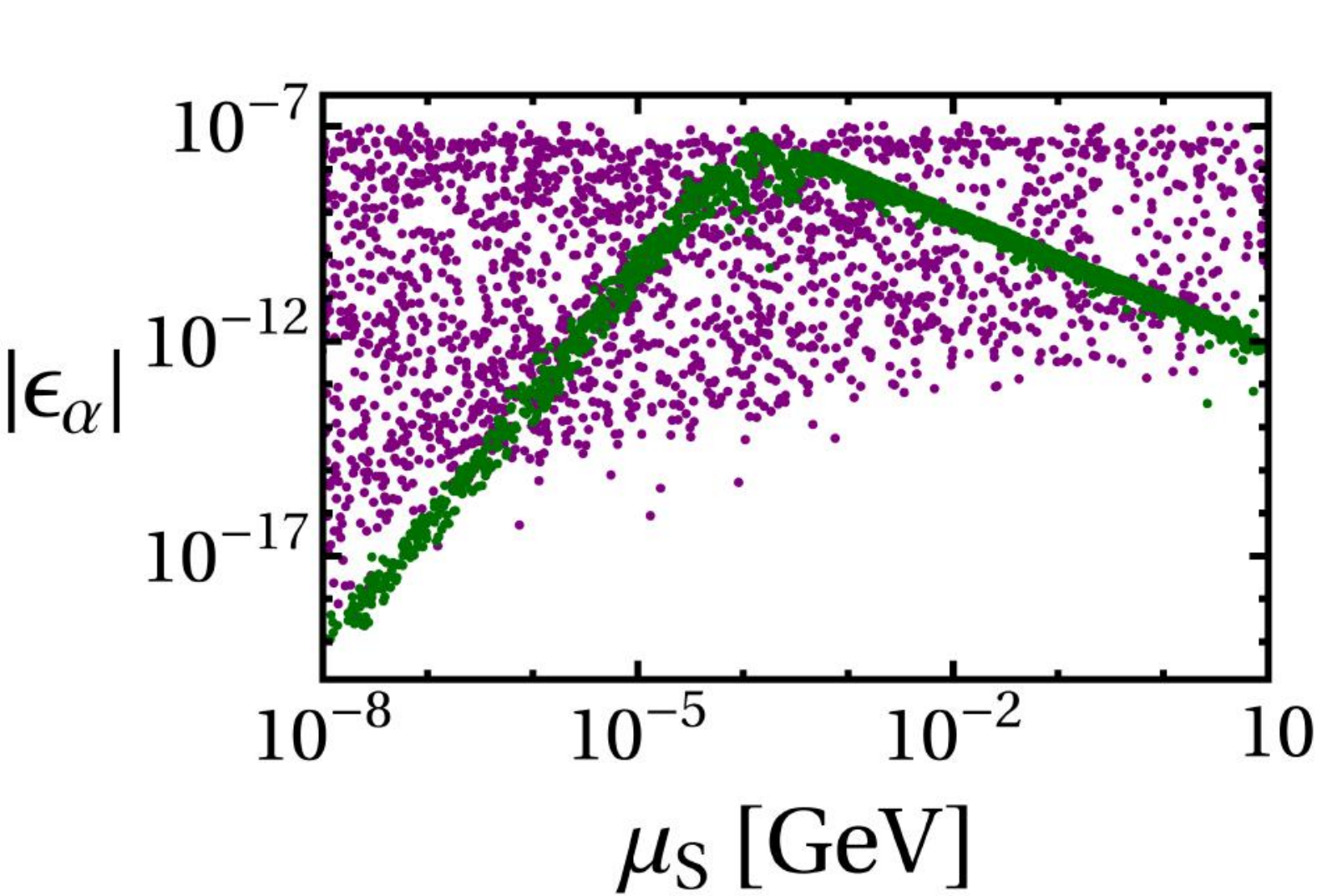}
}
\caption{Plot of the CP asymmetry into a specific flavour $\epsilon_\alpha^{\protect\vphantom{T}} = \sum_i \epsilon_{\alpha}^{i}$ generated in the ISS when both $\mathcal{N}_1^{\protect\vphantom{T}}$ and $\mathcal{N}_2^{\protect\vphantom{T}}$ are included, as a function of $\mu_N^{\protect\vphantom{T}}$ \textbf{(left figure)} and $\mu_S^{\protect\vphantom{T}}$ \textbf{(right figure)}. Points in purple correspond to the regime where $\mu_N^{\protect\vphantom{T}} > \mu_S^{\protect\vphantom{T}}$, while points in green to $\mu_S^{\protect\vphantom{T}} > \mu_N^{\protect\vphantom{T}}$. At around $10^{-4}$ GeV a natural resonance occurs generated by $\Delta m_{i,j}^{\text{OS}}$ in agreement with~\cite{Dolan:2018qpy}. Once radiative effects are included an additional resonance occurs for large values of $\mu_N^{\protect\vphantom{T}}$ and in the regime $\mu_N^{\protect\vphantom{T}} > \mu_S^{\protect\vphantom{T}}$.}
\label{figure:spur2-eps}
\end{figure}

\begin{figure}[t]
\centering
{
  \includegraphics[width=0.45\linewidth]{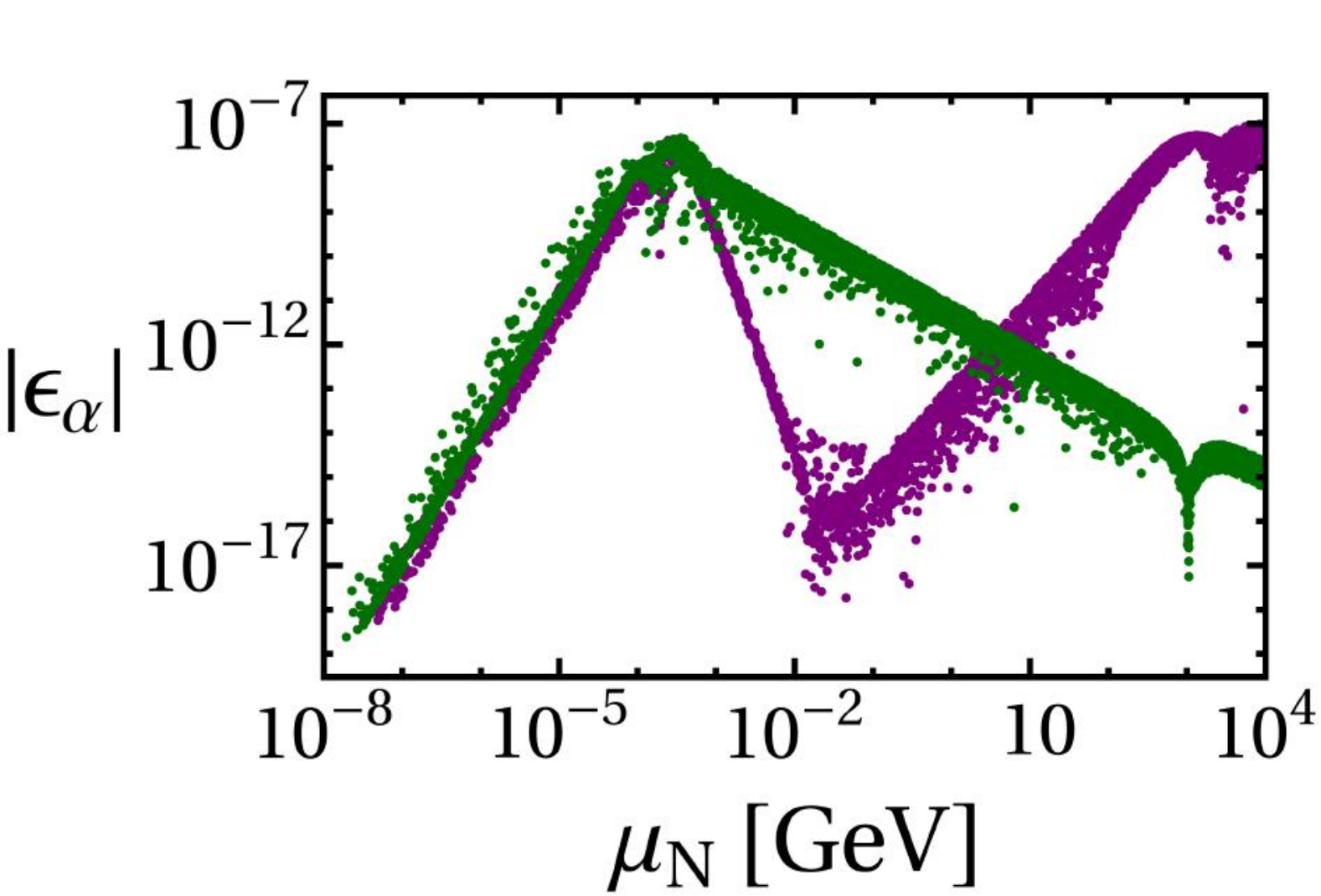}
}
{
  \includegraphics[width=0.45\linewidth]{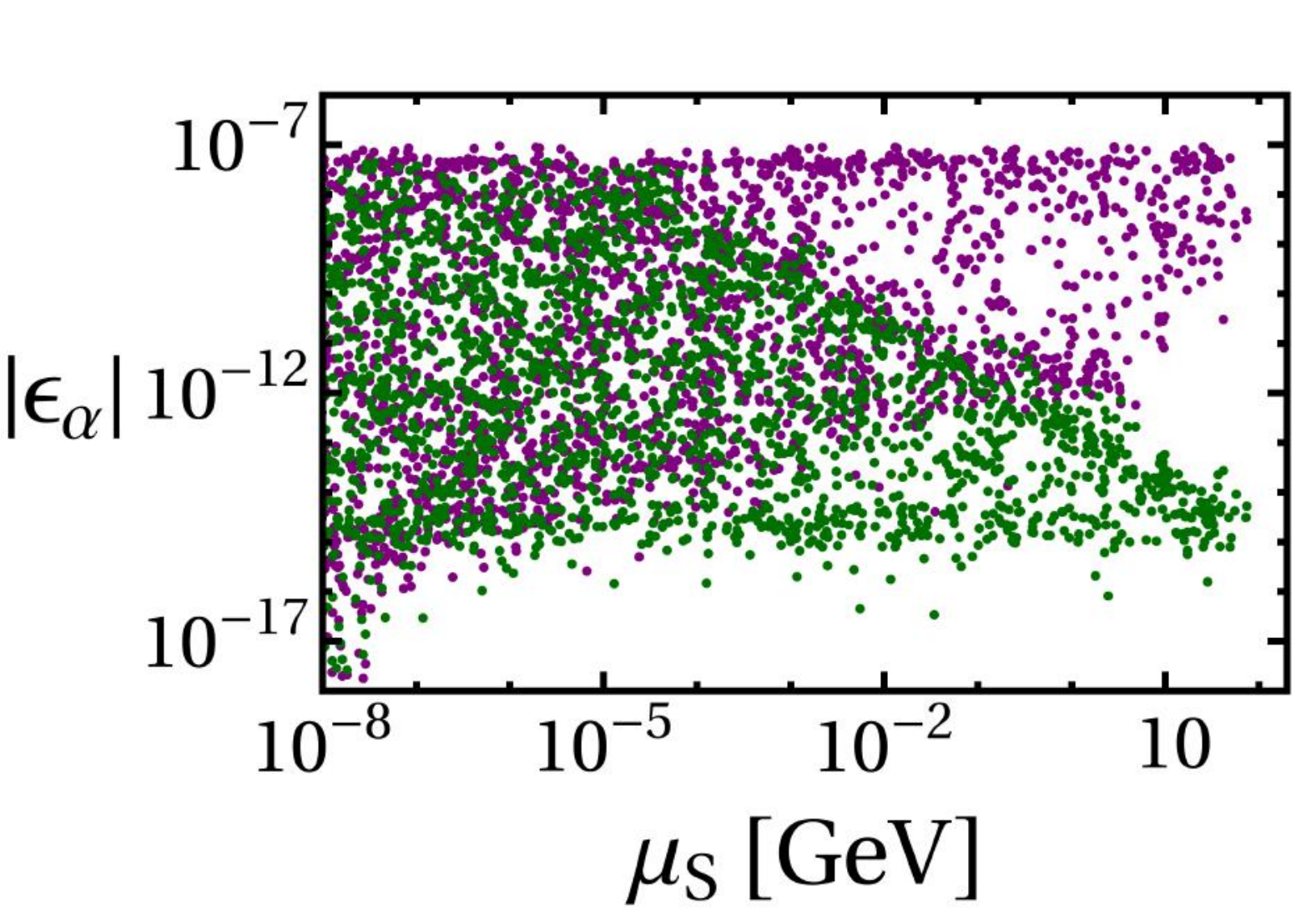}
}
\caption{Plot of the CP asymmetry into a specific flavour $\epsilon_\alpha^{\protect\vphantom{T}} = \sum_i \epsilon_{\alpha}^{i}$ generated in the ISS with no radiative effects included (green) and with $\mathcal{N}_1^{\protect\vphantom{T}}$ and $\mathcal{N}_2^{\protect\vphantom{T}}$ included (purple). This is varied with $\mu_N^{\protect\vphantom{T}}$ \textbf{(left figure)} and $\mu_S^{\protect\vphantom{T}}$ \textbf{(right figure)}. A resonant enhancement in the asymmetry occurs only if the next-to-leading order contributions are included.}
\label{figure:allspur-eps}
\end{figure}

Figures.~\ref{figure:spur2-eps} and~\ref{figure:allspur-eps} plot the CP-asymmetry parameter to a specific lepton flavour\footnote{Due to the anarchic nature of our scenario there is no preference for a specific lepton flavour.} $\alpha$ as a function of the Majorana masses. A `natural' resonance occurs in the region of $\mu_i \simeq 10^{-4}$ GeV induced by $\Delta m_{i,j}^{\text{OS}}$ independent of the radiative spurion contributions, in agreement with~\cite{Dolan:2018qpy}. However, this region is accompanied by a much larger effective washout and cannot accommodate sufficient asymmetry generation. In contrast only when both $\mathcal{N}_1^{\vphantom{\dagger}}$ and $\mathcal{N}_2^{\vphantom{\dagger}}$ are included does a clear second resonance peak form for large values of $\mu_N^{\vphantom{T}}$. Larger values of $\mu_N^{\vphantom{T}}$ would cause $\mu_{\text{eff}} \gg M_R$ and the approximations involved in deriving eq.~(\ref{eqn:fullissmass}) and~(\ref{eqn:ISScasas}) would break down.

\begin{figure}[t]
\centering
{
  \includegraphics[width=0.45\linewidth]{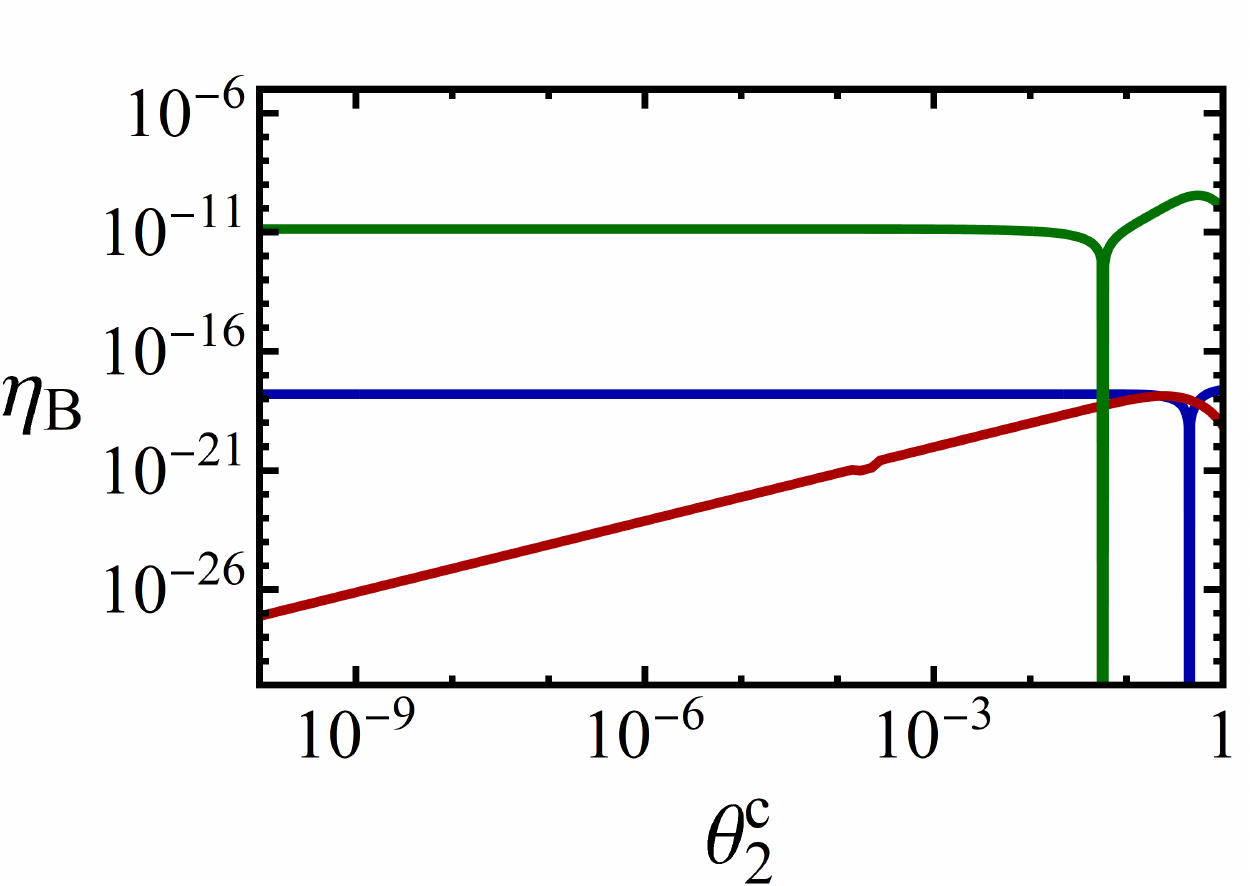}
}
{
  \includegraphics[width=0.45\linewidth]{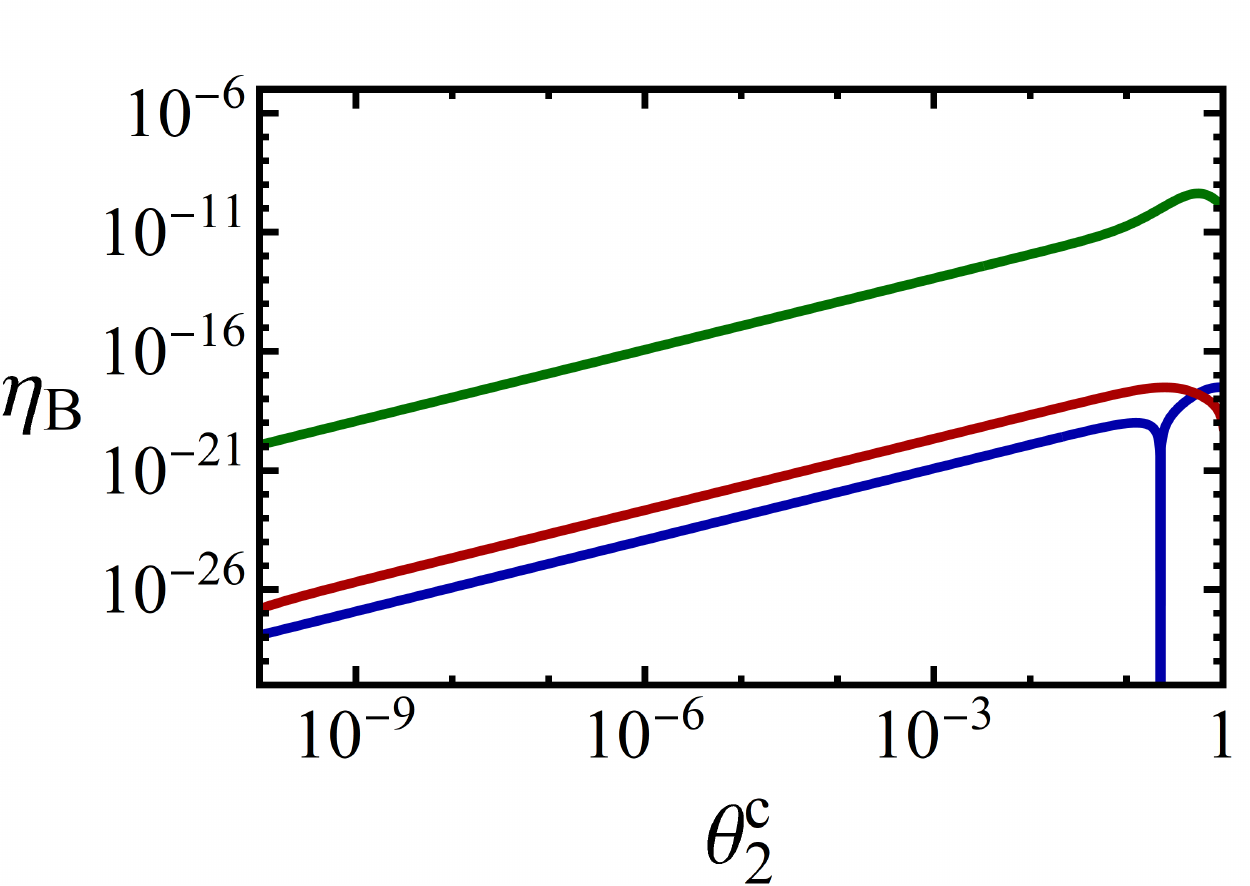}
}
\caption{Plot of the baryon asymmetry as a function of the complex angle $\theta_2^{\text{c}}$ for $\delta_{CP}^{\protect\vphantom{T}}=3\pi/2$ \textbf{(left)} and $\delta_{CP}^{\protect\vphantom{T}} = 0$ \textbf{(right)}. Here we have fixed $\mu_N^{\protect\vphantom{T}} \simeq 950$ GeV and $\mu_S^{\protect\vphantom{T}} \simeq 10$ GeV to be on resonance. In blue no radiative effects are included, in red only to leading order and in green next-to-leading order effects are included. The behaviour of the asymmetry as a function of the complex angle is clear. At small values it freezes out to that provided by the Dirac phase $\delta_{CP}^{\protect\vphantom{T}}$ however if only the leading radiative effects are included (red), the asymmetry tends to zero with decreasing $\theta_2^{\text{c}}$ even for non-zero $\delta_{CP}^{\protect\vphantom{T}}$. This effect is similar to what occurs in the type-I MLFV scenario~\cite{Hambye:2004jf,Cirigliano:2006nu,Cirigliano:2007hb}. Asymmetry generation is possible both with low-energy phases as well as complex entries of the $R$ matrix where we loosely find the criterion $0.1 \lesssim \theta_2^{\text{c}} \lesssim 1$ for sufficient asymmetry generation when $\delta_{CP}^{\protect\vphantom{T}} = 0$.  }
\label{figure:theta-varyY}
\end{figure}

Finally in~\cref{figure:theta-varyY} the asymmetry is varied against the complex angle $\theta_2^{\text{c}}$ for all three scenarios for a fixed benchmark point on resonance. In the case where no radiative effects are included, there is a slight dependence on the size of $\theta_2^{\text{c}}$ which can vary the asymmetry generated by one or two orders of magnitude. For small values of $\theta_2^{\text{c}}$ the asymmetry is generated by the CP-violating phase $\delta_{CP}$ within the low-energy mixing matrix. Similar behaviour occurs when $\mathcal{N}_2^{\protect\vphantom{T}}$ is included, with an overall preference for the range $0.1 < \theta_2^{\text{c}} < 1$. In the case where only $\mathcal{N}_1^{\protect\vphantom{T}}$ is included, we confirm its strong dependence on the angle $\theta_2^{\text{c}}$ similarly to the type-I scenario~\cite{Cirigliano:2006nu}. However, we note that in our case the asymmetry is not zero exactly unless $\theta_2^{\text{c}} = 0$. Asymmetry generation can only occur for non-zero values of $\theta_2^{\text{c}}$ independent of any low-energy CP violating phases as phases from $U_{\text{PMNS}}$ cancel in the term $m_D^{\dagger} m_D^{\vphantom{\dagger}}$ in combination with the flavour alignment described above. Leptogenesis is viable both through the Dirac phase as well as high scale CPV but we note that within MLFV-ISS the region in which it is possible is very narrow.

\subsection{$\ell_i \rightarrow \ell_j \gamma$ and MLFV-ISS	 }

Here we briefly discuss the consequences of MLFV-ISS on low-energy cLFV processes, specifically the impact that the introduction of CPV has on predictions for $\ell_i \rightarrow \ell_j \gamma$. We also assess whether a future measurement of the region important for resonant leptogenesis will be possible. Similar to the MFV hypothesis of the quark sector, models of MLFV predict relationships between the rates for different charged lepton flavour decays.

In the SM effective field theory framework the process $\ell_i \rightarrow \ell_j \gamma$ arises from dimension six effective operators that contain $\left( \overline{\ell} \, \Gamma\, e_{R} \right)$~\cite{Cirigliano:2005ck,Cirigliano:2006su,Dinh:2017smk} where family indices have been suppressed.  These operators are not invariant under the flavour symmetry defined for MLFV in eq.~(\ref{eqn:ISSMLFVsymm}). Insertion of spurion combinations transforming as $\bm{(3,\,\overline{3},\,1,\,1)}$ make these terms formally invariant. At lowest order this is simply the spurion $\mathcal{Y}_e^{\vphantom{T}}$. However, as we work in a basis where this matrix is diagonal, it does not contribute to flavour-violating processes. The lowest order spurion combination that makes the relevant operators flavour invariant and contains non-diagonal entries is $\Delta \mathcal{Y}_e^{\vphantom{T}}$, where
\begin{equation}
\label{eqn:ISS-cLFV}
\Delta = \mathcal{Y}_D^{\vphantom{T}} \mathcal{Y}_D^{\dagger}.
\end{equation}  
The combination transforms as $\left( \Delta \mathcal{Y}_e^{\vphantom{T}} \right) \rightarrow \mathcal{U}_{\ell_L}^{\vphantom{T}} \left( \Delta \mathcal{Y}_e^{\vphantom{T}} \right) \, \mathcal{U}_{e_R}^{\dagger}$ as required.

A simple expression for the branching ratio is obtained~\cite{Cirigliano:2005ck} in the limit $m_{\ell_j} \ll m_{\ell_i}$,
\begin{equation}
B_{\ell_i \rightarrow \ell_j \gamma} = 384 \pi^2 e^2\frac{v^4}{4 \Lambda_{\text{LFV}}^4} \left| \Delta_{ij} \right|^2 \left| C \right|^2
\end{equation}
where $C$ accounts for a combination of Wilson coefficients of the relevant operators. As is conventional when considering cLFV in MLFV, we set $|C|^2=1$. Note that in a specific UV-complete model based on MLFV these parameters will be fixed by the high-scale dynamics and may have a different magnitude.

In~\cref{eqn:ISS-cLFV} the only parameter which carries family indices is the spurion combination $\Delta_{ij}$. Therefore useful predictive parameters for MLFV are ratios of branching ratios for different lepton flavour decays,
\begin{equation}
R_{(i,j)[k,l]} \equiv \frac{B_{\ell_i \rightarrow \ell_j \gamma}}{B_{\ell_k \rightarrow \ell_l \gamma}} = \frac{\left| \Delta_{ij} \right|^2}{\left| \Delta_{kl} \right|^2}.
\end{equation}
In these ratios, the unknown scale of LFV ($\Lambda_{\text{LFV}}$) cancels out. While it can be identified with $M_R$ in the ISS model, it is also possible that it could arise as the result of some unknown dynamics not directly related to the seesaw mechanism.

Many detailed explorations of cLFV in MLFV have been made for the minimal type-I seesaw scenario~\cite{Cirigliano:2005ck,Cirigliano:2006su,Branco:2006hz,Dinh:2017smk,Coy:2018bxr}, the results of which should also hold in the ISS. Here we briefly consider the impact of the complex angle $\theta_2^{\text{c}}$ and the Dirac phase $\delta_{CP}$ on the predictions for specific ratios of cLFV observables. In particular we compare the CP-conserving scenario (assumed in the simplest version of MLFV) with the CP-violating scenario and the size of deviation their inclusion introduces. As we have demonstrated that leptogenesis is viable with both the low-energy phase $\delta_{CP}^{\vphantom{T}}$ and with the inclusion of the CP-violating angle $\theta_2^{\text{c}}$, we analyse the two limiting cases where CPV arises solely from one of these angles.

\begin{figure}[t]
\centering
{
  \includegraphics[width=0.45\linewidth]{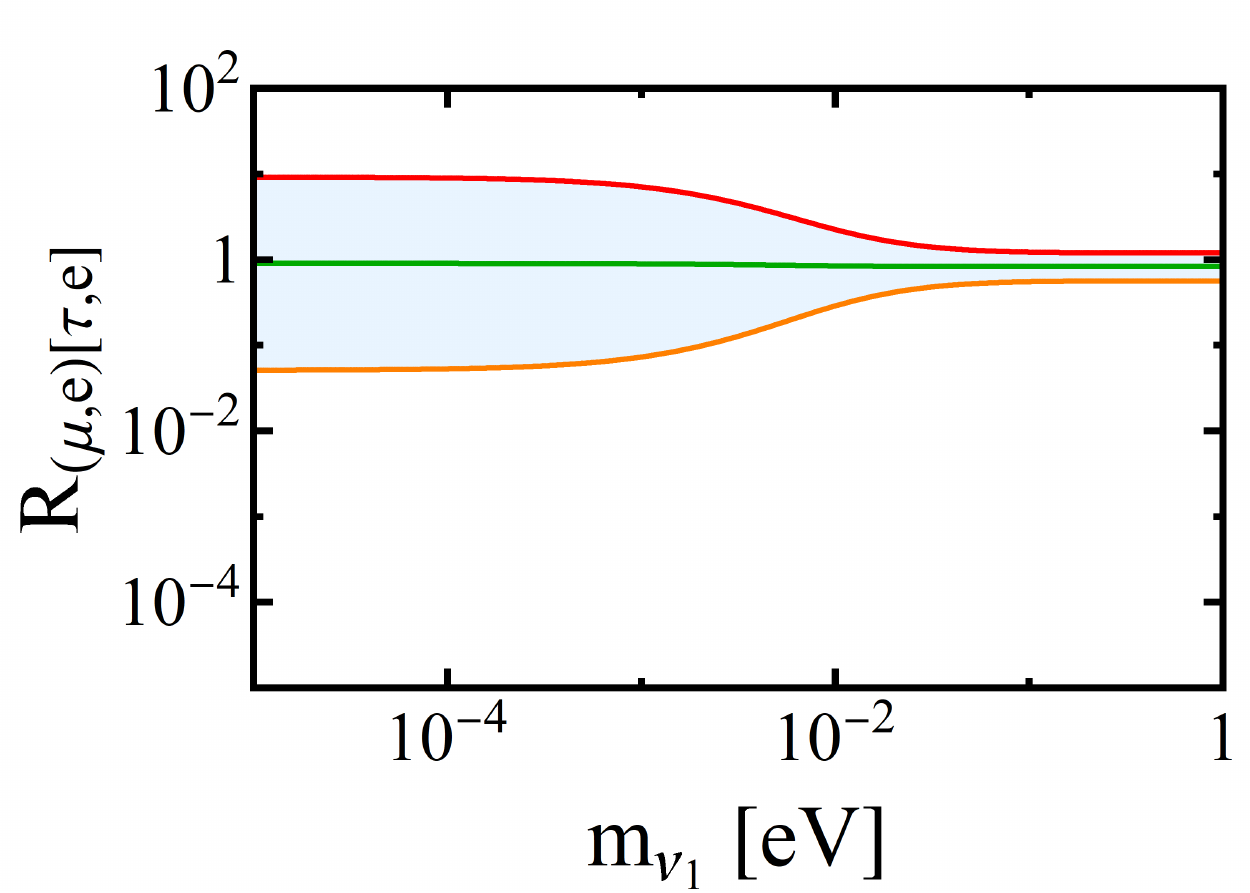}
}
{
  \includegraphics[width=0.45\linewidth]{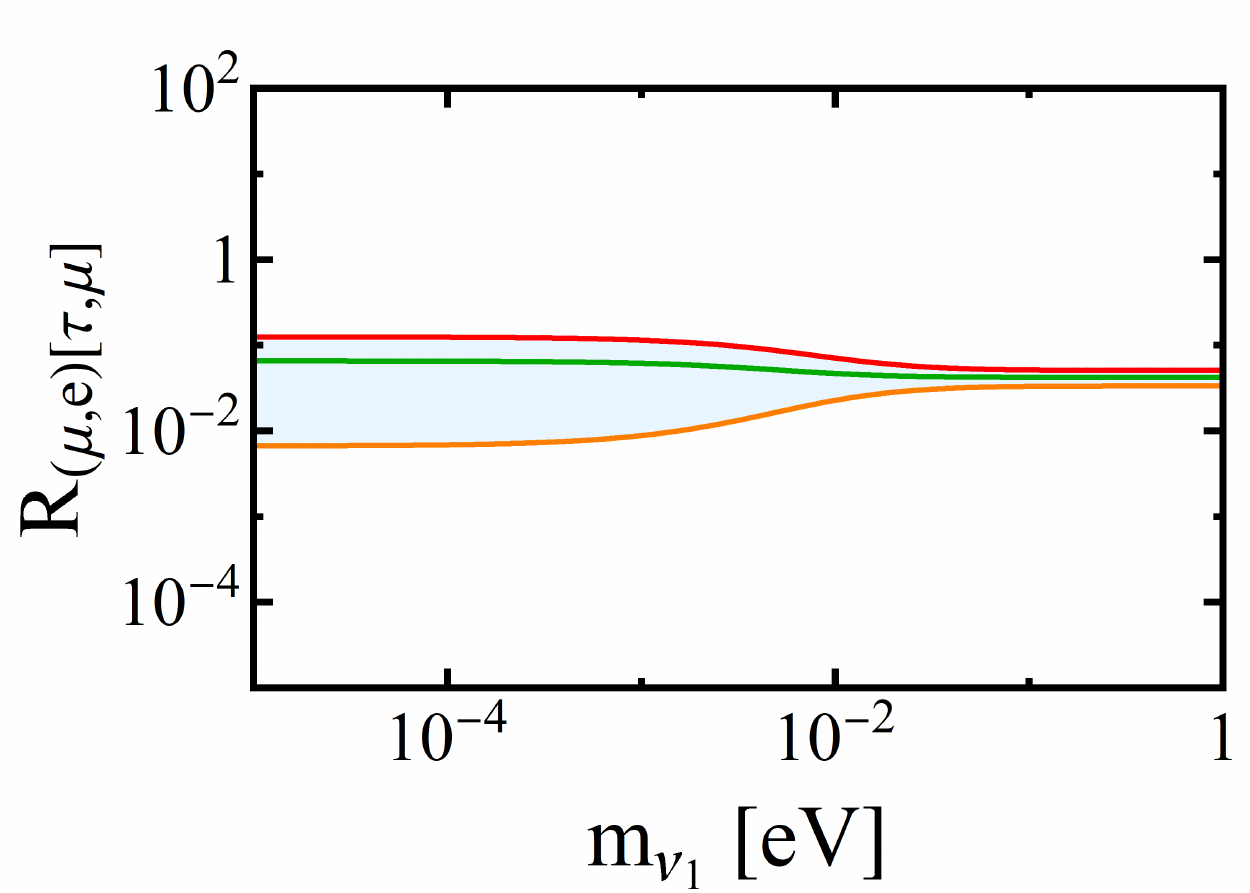}
}
{
  \includegraphics[width=0.45\linewidth]{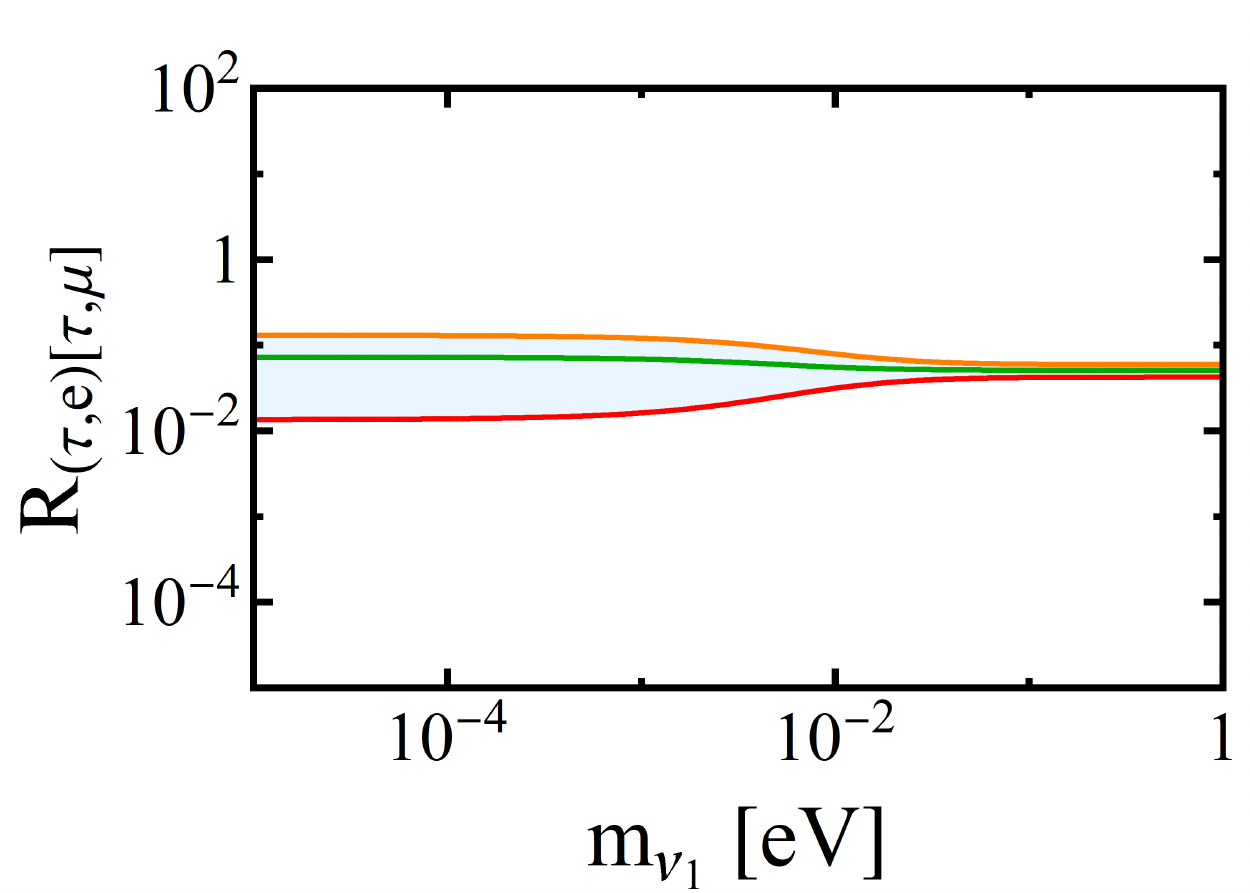}
}
\caption{Plot of the ratios $R_{(i,j)[k,l]}$ for various combinations of flavour initial and final states as a function of the lightest neutrino mass $m_{\nu_1}$. Here the complex angle $\theta_2^{\text{c}}$ has been switched off but the low-energy phase $\delta_{CP}$ is varied. Lines in red and orange correspond to the CP-conserving cases $\delta_{CP} = 0 \text{ and } \delta_{CP} = \pi$ respectively. In green is the maximally violating case of $\delta_{CP} = \pi/2 \text{ or } \delta_{CP}=3\pi/2$. The shaded blue region corresponds to $\delta_{CP} = (0,2\pi) \backslash \{\pi/2,\,\pi,\,3\pi/2\}$. All results are in full agreement with~\cite{Dinh:2017smk} for the CP-conserving cases.}
\label{figure:R-varydel}
\end{figure}

Figure~\ref{figure:R-varydel} plots three ratios of cLFV observables as a function of the lightest neutrino mass $m_{\nu_1}$ for $\theta_2^{\text{c}} = 0$ while varying $\delta_{CP}^{\vphantom{T}}$. The red and orange lines correspond to the CP-conserving scenarios $\delta_{CP} = 0 \text{ or } \pi$, which allows for full reconstruction of $\mathcal{Y}_D^{\vphantom{T}}$ from low energy observables. The blue shaded region corresponds to all other values of $\delta_{CP}^{\vphantom{T}}$ where the green line specifically corresponds to the maximally CP-violating cases $\delta_{CP} = \pi/2 \text{ or } 3\pi/2$. The inclusion of CP-violating phases does not spoil the generic MLFV prediction that $R_{(\mu,e)[\tau,\mu]} \ll 1$ and $R_{(\tau,e)[\tau,\mu]} \ll 1$ in both the hierarchical and degenerate scenarios for the light neutrinos. A more potentially measurable effect occurs for the ratio $R_{(\mu,e)[\tau,e]}$, specifically for a hierarchical spectrum of light neutrinos. The CP-conserving case predicts either $R_{(\mu,e)[\tau,e]} > 1$ or $R_{(\mu,e)[\tau,e]}<1$ depending on the choice of $\delta_{CP}^{\vphantom{T}}$, while the maximally CP-violating scenario predicts $R_{(\mu,e)[\tau,e]} \simeq 1$.

\begin{figure}[t]
\centering
{
  \includegraphics[width=0.45\linewidth]{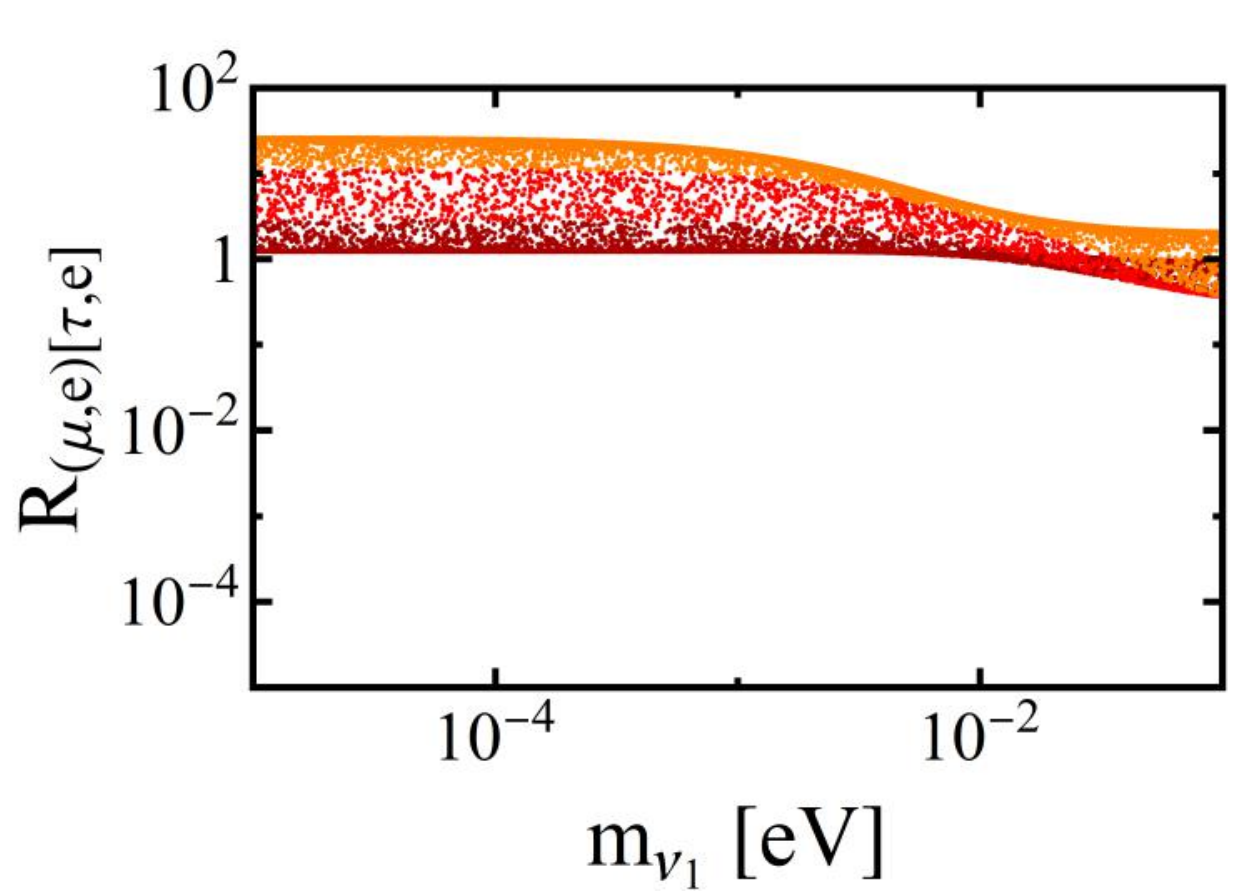}
}
{
  \includegraphics[width=0.45\linewidth]{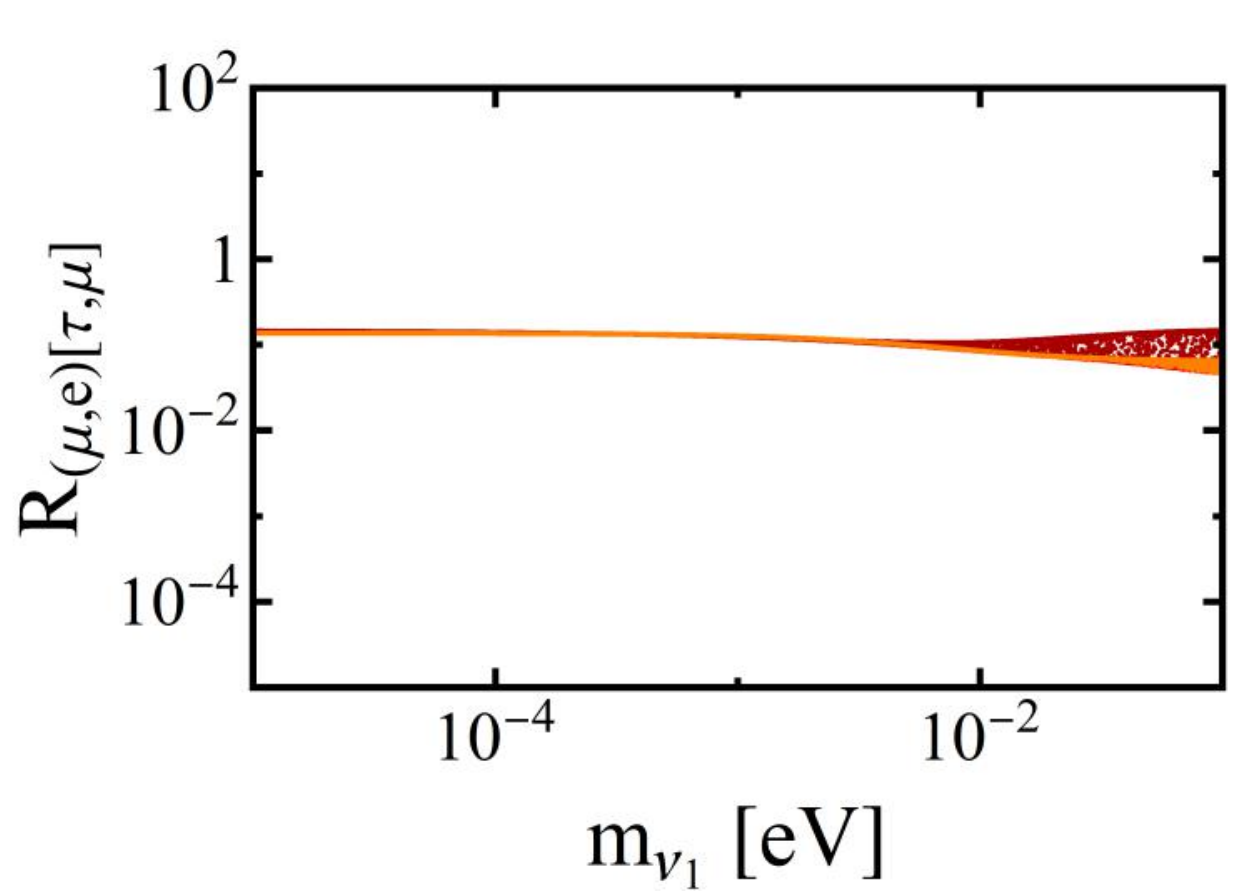}
}
{
  \includegraphics[width=0.45\linewidth]{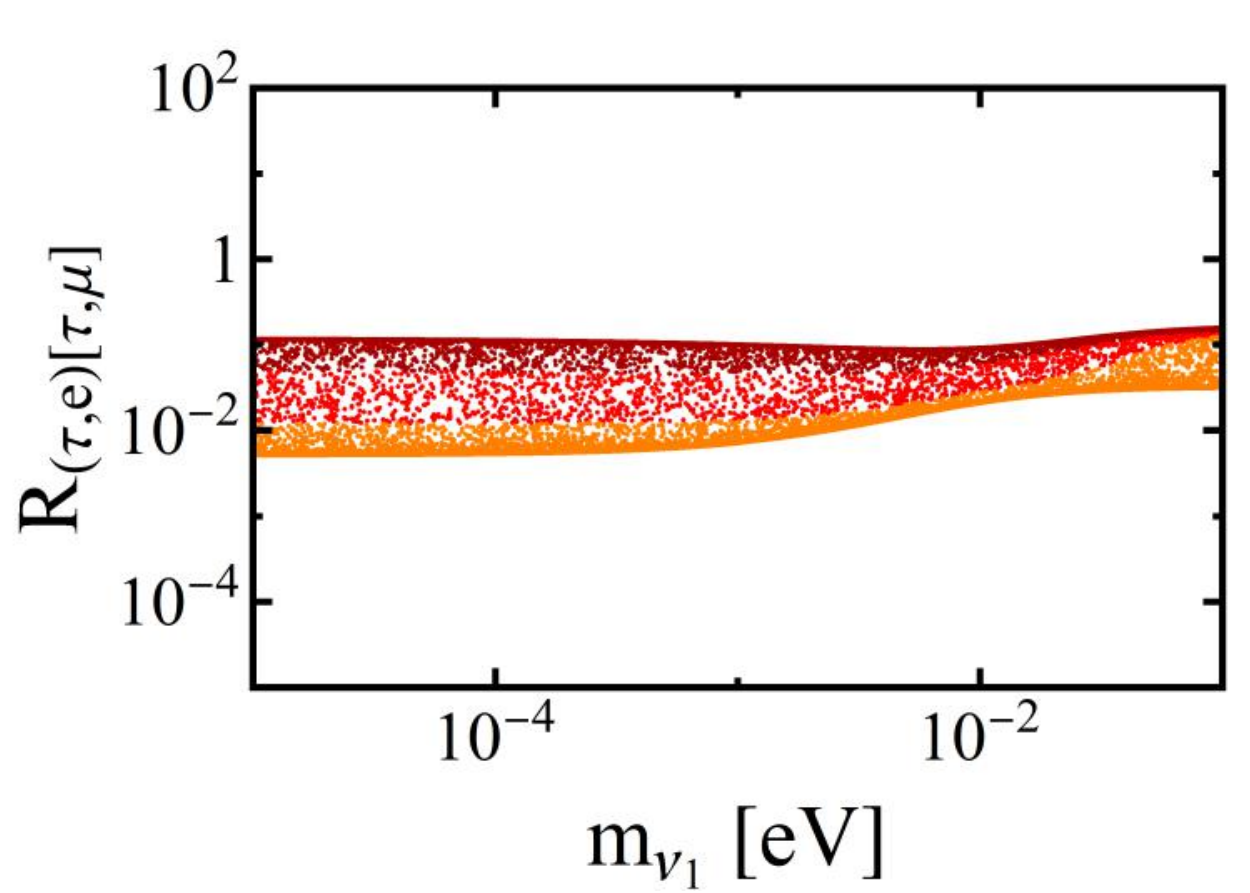}
}
\caption{Plot of the ratios $R_{(i,j)[k,l]}$ for various combinations of flavour initial and final states as a function of the lightest neutrino mass $m_{\nu_1}$. Here the low-energy phase $\delta_{CP}^{\protect\vphantom{T}}$ has been switched off but the complex angle $\theta_2^{\text{c}}$ is varied. We find the same generic predictions as in the CP-conserving case of $R_{(\mu,e)[\tau,e]} > 1$ \textbf{(top-left)}  $R_{(\mu,e)[\tau,\mu]} < 1$ \textbf{(top-right)} and  $R_{(\tau,e)[\tau,\mu]} < 1$ \textbf{(bottom)}. For the ratios $R_{(\mu,e)[\tau,e]}$ and $R_{(\tau,e)[\tau,\mu]}$ the introduction of CP-violation brings the ratios closer to one and this difference is most apparent with a hierarchical spectrum of light neutrinos. In orange $\theta_2^{\text{c}} < 0.1$, in red $0.1 < \theta_2^{\text{c}} < 0.3$ and in burgundy $\theta_2^{\text{c}} > 0.3$.}
\label{figure:R-varyY}
\end{figure}

Similar plots are presented in~\cref{figure:R-varyY} where we plot the ratios as a function of the lightest neutrino mass $m_{\nu_1}$ for $\delta_{CP}^{\vphantom{T}} = 0 $ and varying $\theta_2^{\text{c}}$. Once again a hierarchical spectrum of light neutrinos is required in order for CP-violating phases to have their most significant effect. The inclusion of $\theta_2^{\text{c}}$ does not spoil the generic predictions from the CP-conserving cases presented in~\cref{figure:R-varydel}, however it is clear in the case of $R_{(\mu,e)[\tau,e]}$ and $R_{(\tau,e)[\tau,\mu]}$ that the inclusion of CP-violation (CP-conservation) moves the ratios closer (further) from one. Therefore future measurement of this ratio could constrain scenarios of MLFV that include CP violation.

\begin{figure}[t]
\centering
{
  \includegraphics[width=0.45\linewidth]{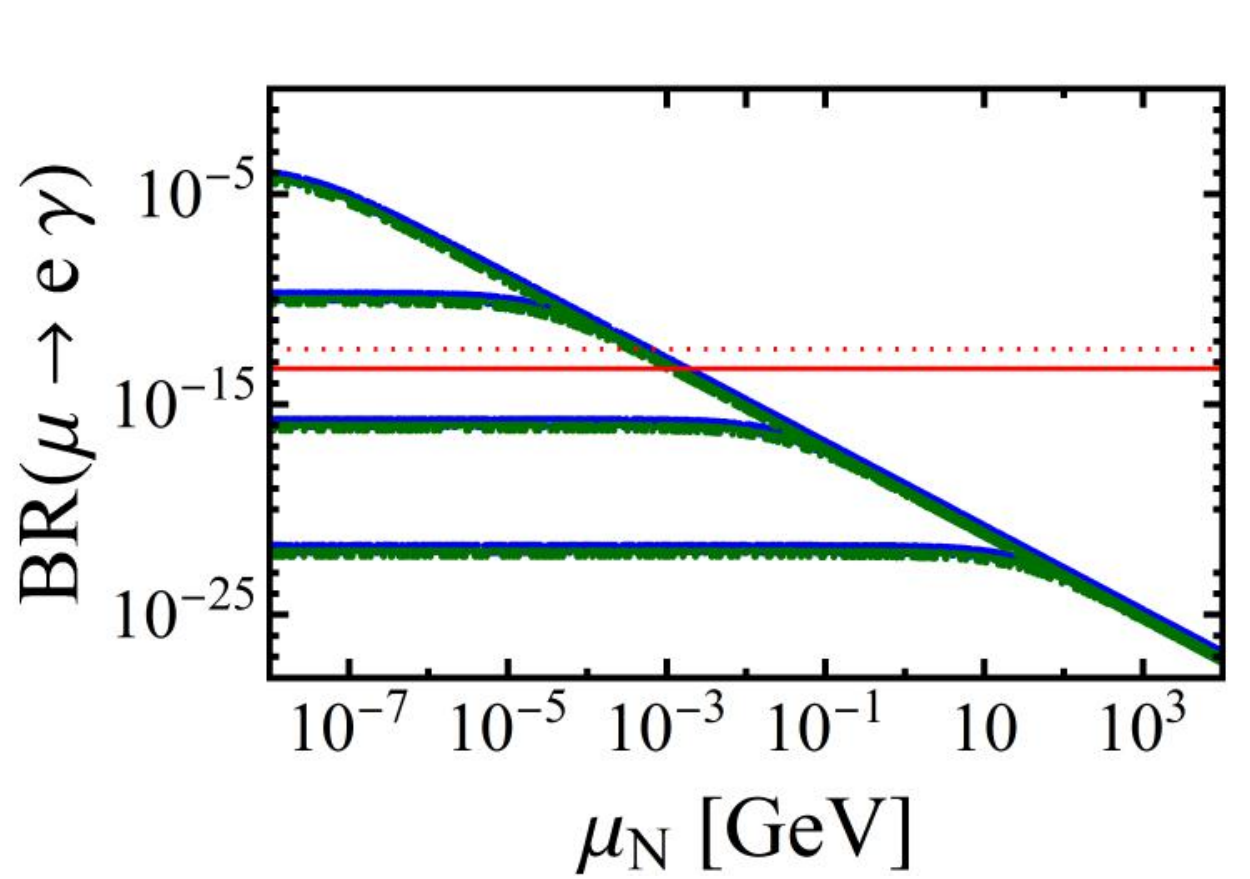}
}
{
  \includegraphics[width=0.45\linewidth]{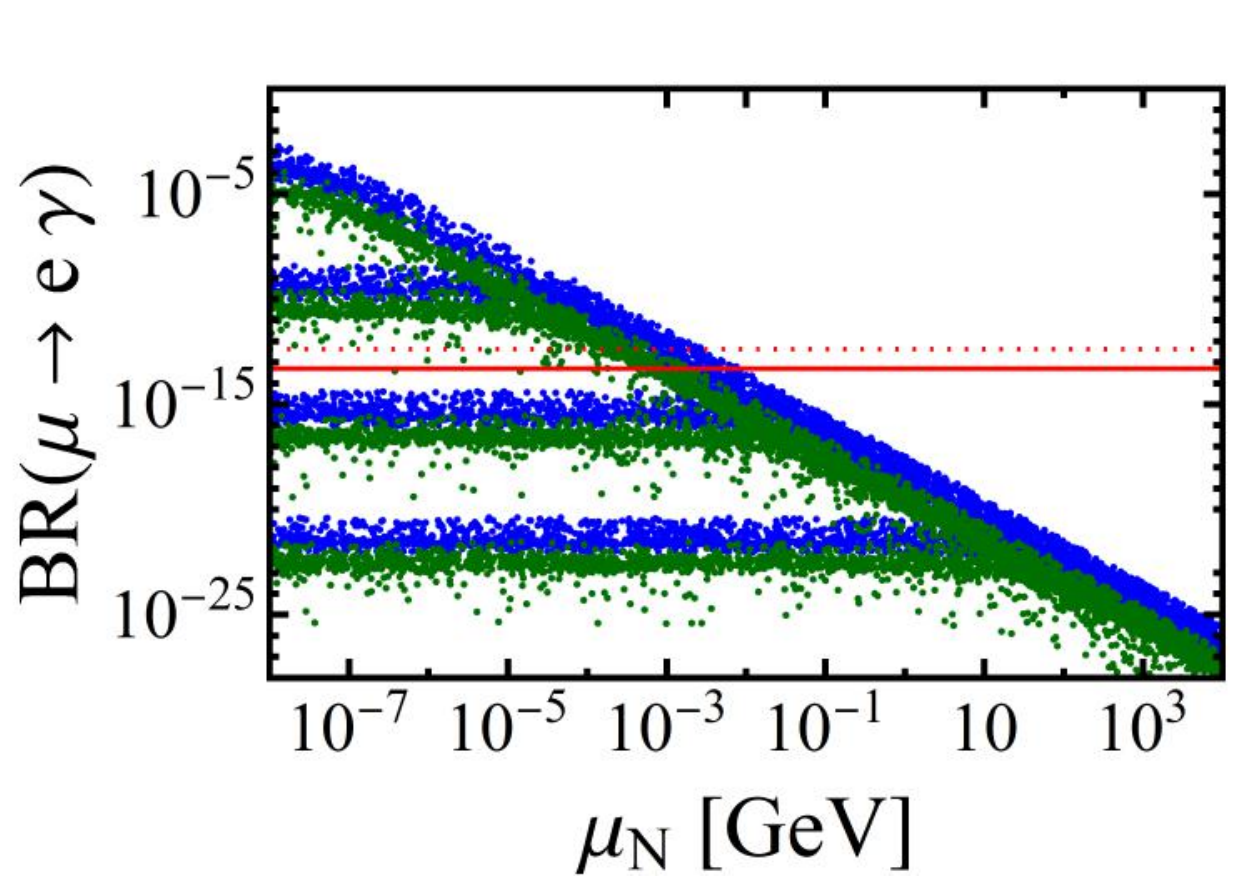}
}
\caption{Plot of the branching ratio BR($\mu \rightarrow e \gamma$) for $\theta_2^{\text{c}} = 0$ \textbf{(left)} and $0.1 < \theta_2^{\text{c}} < 1$ \textbf{(right)} as a function of $\mu_N^{\protect\vphantom{T}}$. Points in blue correspond to the choice $\delta_{CP}^{\protect\vphantom{T}} = 0$ whereas points in green correspond to $\delta_{CP}^{\protect\vphantom{T}} = 3\pi/2$. In these plots the other LNV parameter $\mu_S^{\protect\vphantom{T}}$ has been fixed to the values $10^{-8},\,10^{-5},\,10^{-2}\text{ and }10$ GeV. The ratio plateaus whenever $\mu_N^{\protect\vphantom{T}} < \mu_S^{\protect\vphantom{T}}$. The red dotted (solid) line corresponds to the current (future) sensitivity of MEG and MEG-II~\cite{TheMEG:2016wtm,Cattaneo:2017psr}, respectively, for this decay mode. Clearly, small values of the LNV parameters are experimentally accessible, while large values (necessary for MLFV-ISS resonant leptogenesis) correspond to a suppressed signal. The inclusion of low-scale CPV through the Dirac phase has no effect on the prediction, while CPV from $\theta_2^{\text{c}}$ can alter the prediction by approximately an order of magnitude.}
\label{figure:meg-ISS}
\end{figure}

Finally in~\cref{figure:meg-ISS} the branching ratio BR($\mu \rightarrow e \gamma$) is plotted as a function of the LNV parameter $\mu_N^{\vphantom{T}}$. Here the cases where no CPV is present is distinguished from the cases where CPV arises from the low-energy observable $\delta_{CP}^{\vphantom{T}}$ and from the complex angle $\theta_2^{\text{c}}$. The LFV scale has been fixed to $\Lambda_{\text{LFV}}^{\vphantom{T}} = M_R^{\vphantom{T}} = 1\text{ TeV}$ and the LNV parameter $\mu_S^{\vphantom{T}}$ has been fixed to the values described in the figure caption. The horizontal dotted (solid) red line corresponds to the current (future) sensitivity of MEG (MEG-II)~\cite{TheMEG:2016wtm,Cattaneo:2017psr} for this decay process. Currently, MLFV-ISS has been excluded for very low values of $\mu_N^{\vphantom{T}}$ and $\mu_S^{\vphantom{T}}$ for which the scales of LNV $\Lambda_{\text{LNV}} = m_R^2 / \mu_{\text{eff}}$ and LFV $\Lambda_{\text{LFV}} = m_R$ are most disparate. 

Successful MLFV-ISS resonant leptogenesis, however, requires very large values of $\mu_N^{\vphantom{T}}$ in order for sufficiently reduced washout to occur in combination with a resonant enhancement of the asymmetry. This corresponds to a reduction in the hierarchy between the LNV and LFV scale, suppressing the overall size of this cLFV decay. Therefore, while some overall predictions can be made within MLFV-ISS on the relative strength of various combinations of cLFV observables, a measurement of cLFV in near future experiments will be in conflict with the viable parameter space for MLFV-ISS leptogenesis in the absence of some additional mechanism to reduce the overall washout in this region.

\subsection{Linear Seesaw (LSS)}

For the LSS ($\mathcal{Y}_{\mu_S}^{\vphantom{T}} = \mathcal{Y}_{\mu_N}^{\vphantom{T}} = 0$) the LNV parameter arises from the $(1,3)$ and $(3,1)$ entry of the full neutrino mass matrix,
\begin{equation}
M_{\nu}=\begin{pmatrix}
m_\nu^{\text{loop}} & m_{D}^{\vphantom{T}} & m_{L}^{\vphantom{T}} \\
m_{D}^{T} & 0 & M_{R}^{\vphantom{T}} \\
m_{L}^{T} & M_{R}^{T} & 0 \end{pmatrix}.
\end{equation}
From eq.~(\ref{eqn:inv-masscorrections}) as long as $\mathcal{Y}_L^{\vphantom{T}} \not\propto \id_3$ there is a radiative contribution to both the $(2,2)$ and $(3,3)$ entry which will break the $SO(3)^2$ invariance, leading to
\begin{eqnarray}
\tilde{\mu}_{N} &=&  \mu_N^{\vphantom{T}} \left( \mathcal{N}_1^{\vphantom{\dagger}} + \mathcal{N}_2^{\vphantom{\dagger}} \right) \nonumber\\[5pt]
\tilde{\mu}_{S} &=&  \,\mu_S^{\vphantom{T}} \left(\mathcal{S}_1^{\vphantom{\dagger}}\, +\, \mathcal{S}_2^{\vphantom{\dagger}}\right),
\end{eqnarray}
where
\begin{eqnarray}
\label{eqn:LSSspurions}
\mathcal{N}_1^{\vphantom{\dagger}} \,& = &\, n_1^{\vphantom{T}} \left( \mathcal{Y}_D^\dagger \mathcal{Y}_D^{\vphantom{\dagger}} + (\mathcal{Y}_D^\dagger \mathcal{Y}_D^{\vphantom{\dagger}})^T\right),\nonumber\\[5pt]
\mathcal{N}_2^{\vphantom{\dagger}}\, & = & \,n_2^{(1)} \left( \mathcal{Y}_D^\dagger \mathcal{Y}_D^{\vphantom{\dagger}} \mathcal{Y}_D^\dagger \mathcal{Y}_D^{\vphantom{\dagger}} + (\mathcal{Y}_D^\dagger \mathcal{Y}_D^{\vphantom{\dagger}}\mathcal{Y}_D^\dagger \mathcal{Y}_D^{\vphantom{\dagger}})^T\right) + n_2^{(2)} \left( \mathcal{Y}_D^\dagger \mathcal{Y}_D^{\vphantom{\dagger}} (\mathcal{Y}_D^\dagger \mathcal{Y}_D^{\vphantom{\dagger}})^T \right) \nonumber\\
&\mbox{}&+ \,n_2^{(3)} \left( (\mathcal{Y}_D^\dagger \mathcal{Y}_D^{\vphantom{\dagger}})^T \mathcal{Y}_D^\dagger \mathcal{Y}_D^{\vphantom{\dagger}}\right) + n_2^{(4)} \left( \mathcal{Y}_D^{\dagger} \mathcal{Y}_e^{\vphantom{\dagger}} \mathcal{Y}_e^{\dagger} \mathcal{Y}_D^{\vphantom{\dagger}} + (\mathcal{Y}_D^{\dagger} \mathcal{Y}_e^{\vphantom{\dagger}} \mathcal{Y}_e^{\dagger} \mathcal{Y}_D^{\vphantom{\dagger}})^T \right)\nonumber\\[5pt]
\mathcal{S}_1^{\vphantom{\dagger}} \,& = &\, s_1^{\vphantom{T}} \left( \mathcal{Y}_L^\dagger \mathcal{Y}_L^{\vphantom{\dagger}} + (\mathcal{Y}_L^\dagger \mathcal{Y}_L^{\vphantom{\dagger}})^T\right),\nonumber\\[5pt]
\mathcal{S}_2^{\vphantom{\dagger}}\, & = & \,s_2^{(1)} \left( \mathcal{Y}_L^\dagger \mathcal{Y}_L^{\vphantom{\dagger}} \mathcal{Y}_L^\dagger \mathcal{Y}_L^{\vphantom{\dagger}} + (\mathcal{Y}_L^\dagger \mathcal{Y}_L^{\vphantom{\dagger}}\mathcal{Y}_L^\dagger \mathcal{Y}_L^{\vphantom{\dagger}})^T\right) + s_2^{(2)} \left( \mathcal{Y}_L^\dagger \mathcal{Y}_L^{\vphantom{\dagger}} (\mathcal{Y}_L^\dagger \mathcal{Y}_L^{\vphantom{\dagger}})^T \right) \nonumber\\
&\mbox{}&+ \,s_2^{(3)} \left( (\mathcal{Y}_L^\dagger \mathcal{Y}_L^{\vphantom{\dagger}})^T \mathcal{Y}_L^\dagger \mathcal{Y}_L^{\vphantom{\dagger}}\right) + s_2^{(4)} \left( \mathcal{Y}_L^{\dagger} \mathcal{Y}_e^{\vphantom{\dagger}} \mathcal{Y}_e^{\dagger} \mathcal{Y}_L^{\vphantom{\dagger}} + (\mathcal{Y}_L^{\dagger} \mathcal{Y}_e^{\vphantom{\dagger}} \mathcal{Y}_e^{\dagger} \mathcal{Y}_L^{\vphantom{\dagger}})^T \right).
\end{eqnarray}
As before, we consider separately the cases where no corrections are included, corrections are included at leading order and corrections are included at next-to-leading order. 

For the ISS scenario the corrections to the Majorana mass $\tilde{\mu}_{N}$ were proportional to the non-zero bare mass $\mu_N^{\vphantom{T}}$ itself and scaled as $\mu_N^{\vphantom{T}}$. For the LSS, by contrast, we operate in a regime where an explicit Majorana mass term in the Lagrangian is forbidden. However, flavour-invariant combinations such as $\mathcal{Y}_D^{\dagger} \mathcal{Y}_D^{\vphantom{T}}$ and $\mathcal{Y}_L^{\dagger} \mathcal{Y}_L^{\vphantom{T}}$ transform in the necessary way to induce mass terms for each SN.

Here the dimensionful parameters $\mu_N^{\vphantom{T}}$ and $\mu_S^{\vphantom{T}}$ cannot be identified with bare Majorana mass terms. Rather, they arise from the unknown UV complete dynamics. As we remain agnostic about these dynamics, yet under the MLFV Ansatz we must include such terms, we choose to fix the values of $\mu_N^{\vphantom{T}}$ and $\mu_S^{\vphantom{T}}$ in some plausible way. For example, it seems reasonable to take $\mu_i^{\vphantom{T}} \simeq v^2/M_{\text{X}}$, which would be true if $\mu_i^{\vphantom{T}}$ arose from some effective coupling with the SM Higgs doublet mediated by a heavy new particle $X$, generating an operator similar to the Weinberg operator. 

We fix these parameters to $\mu_N^{\vphantom{T}} = \mu_S^{\vphantom{T}} = 1\text{ GeV}$ which, in the example given above, would arise if $m_X \simeq \mathcal{O}(10-1000)\text{ TeV}$ depending on the strength of the relevant couplings. We also make the reasonable assumption that this mass scale is independent of any parameters appearing in the matrix $m_L^{\vphantom{T}}$.

\begin{figure}[t]
\centering
{
  \includegraphics[width=0.45\linewidth]{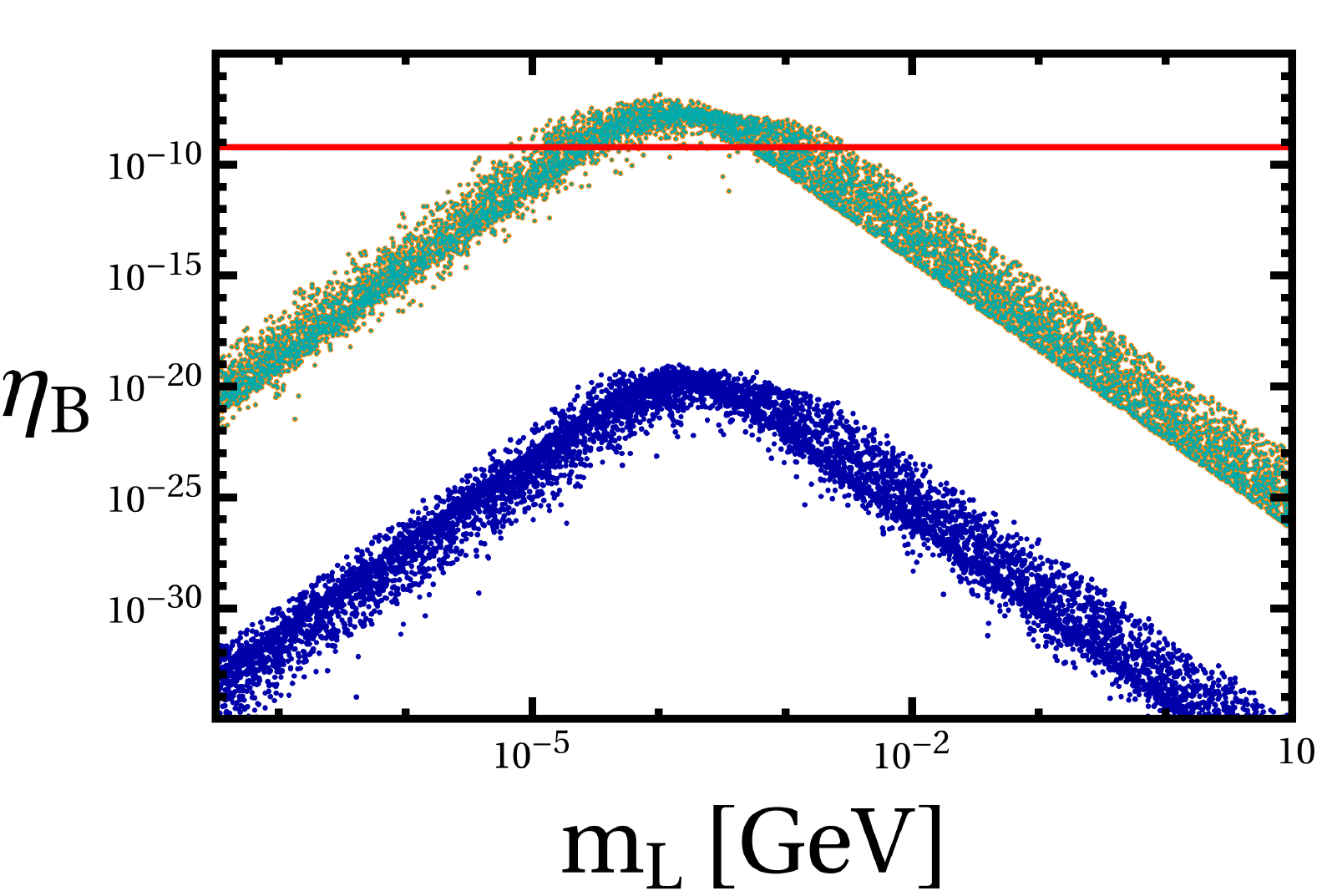}
}
{
  \includegraphics[width=0.45\linewidth]{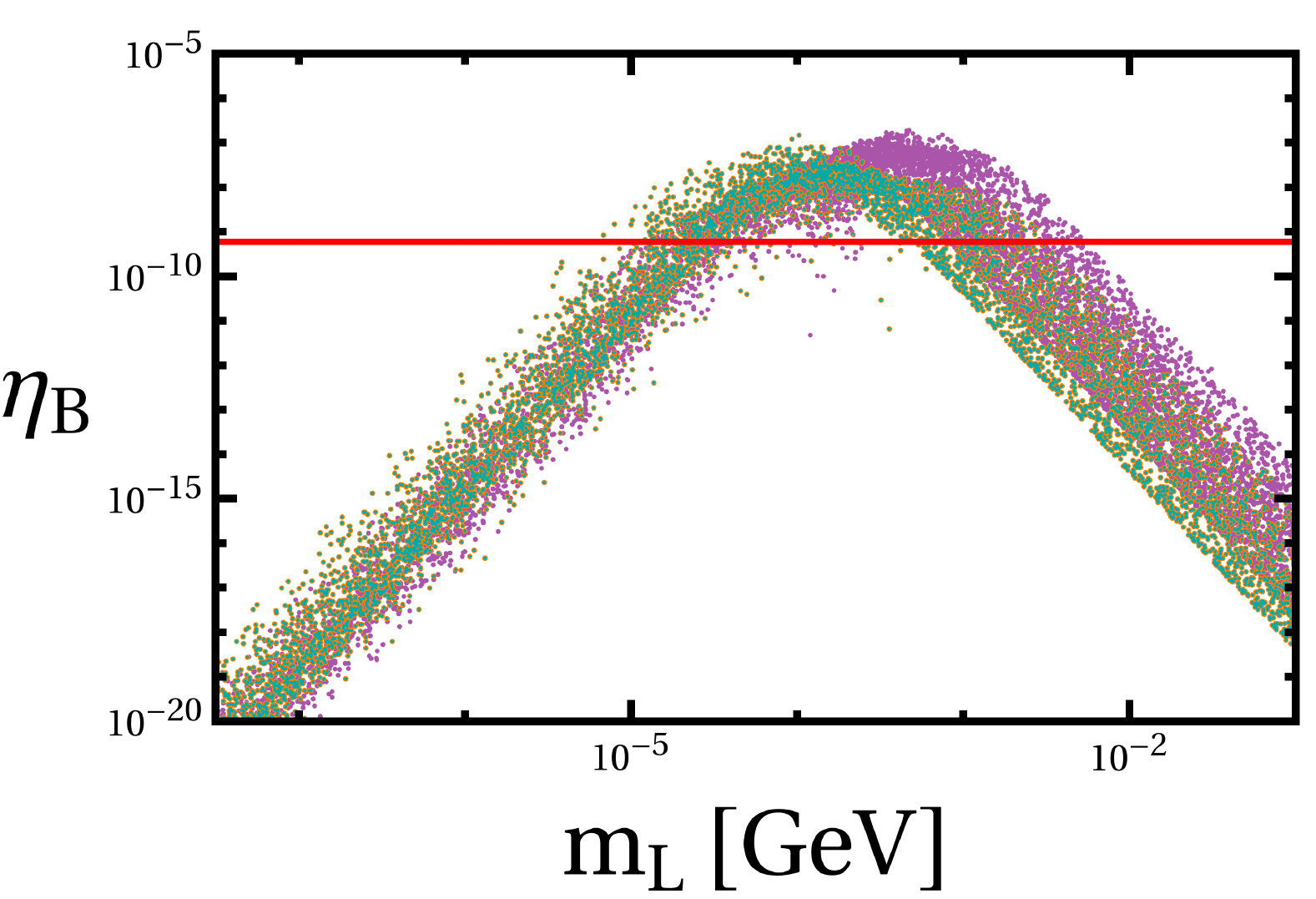}
}
\caption{Plot of the asymmetry generated in the LSS for the three scenarios \textbf{(left)} and comparing the asymmetry generated from mixing to oscillations \textbf{(right)} as a function of $\bm{m_L^{\protect\vphantom{T}}}$, the average of the non-zero entries of the LNV matrix $m_L^{\protect\vphantom{T}}$ in GeV. Points in blue correspond to no radiative corrections, points in orange are when $\mathcal{N}_1^{\protect\vphantom{T}}$ and $\mathcal{S}_1^{\protect\vphantom{T}}$ are included and points in cyan include all terms. Contrary to the case of MLFV-ISS, corrections at leading order are sufficient to generate the necessary asymmetry. The case without any radiative corrections is highly suppressed as the self-energy contributions to $\varepsilon_i^{\alpha}$ are identically zero. However a non-zero asymmetry is generated due to the differences in the decay widths generating slight deviations in the vertex contribution for each SN, c.f.~\cref{eqn:vertexonlyeps}. Points in purple correspond to the estimated asymmetry generated due to oscillation effects between the heavy sterile neutrinos. The asymmetry due to oscillations is predicted to be of the same order as the asymmetry from standard thermal leptogenesis and modifies the predictions for the allowed range of couplings only slightly.}
\label{figure:LSS-asym}
\end{figure}

\begin{figure}[t]
\centering
{
  \includegraphics[width=0.65\linewidth]{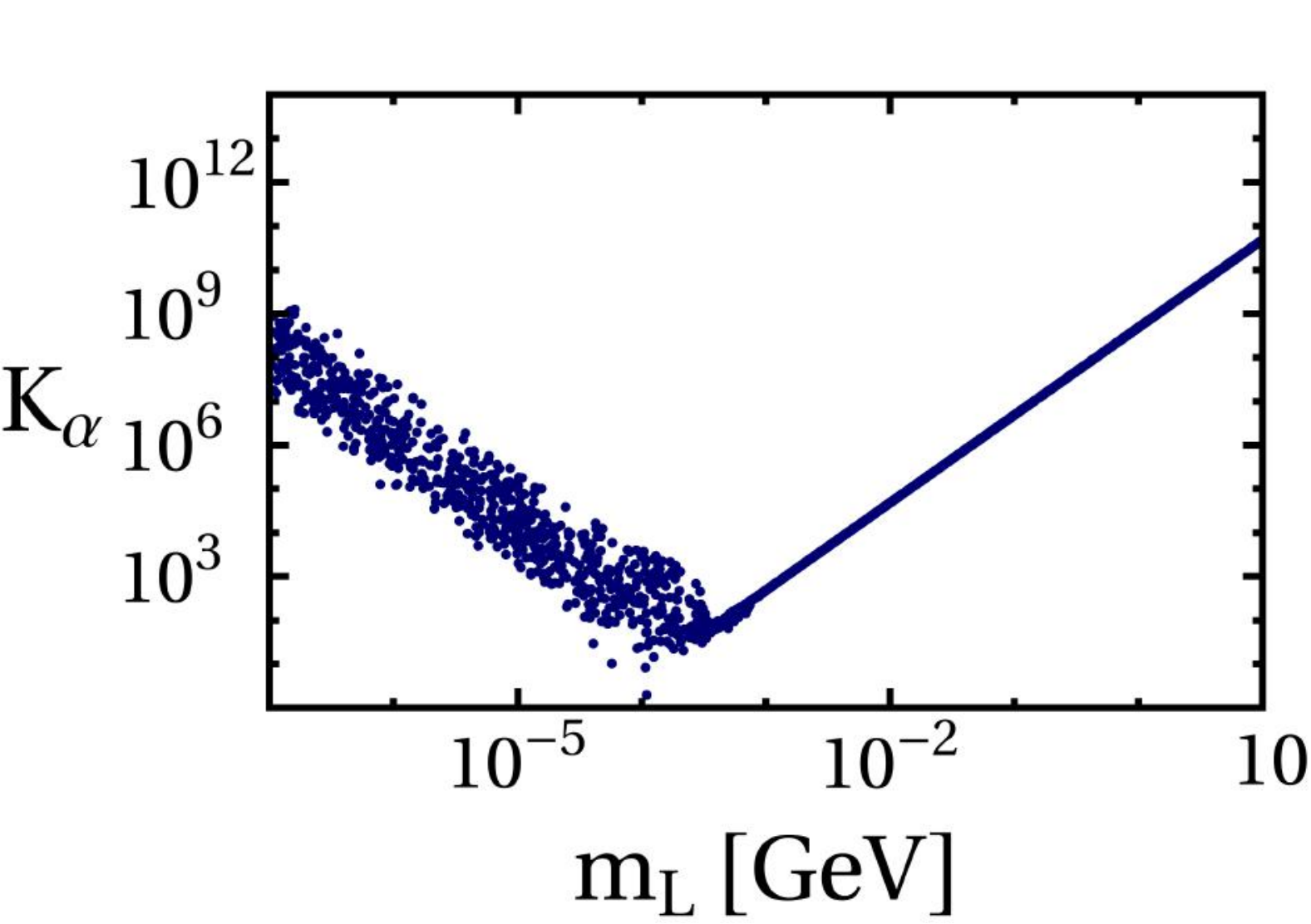}
}
\caption{Plot of the effective washout to a specific lepton flavour $K_{\alpha}$ generated in the LSS as defined in eq.~(\ref{washout}). We find the washout is minimised for $\bm{m_L^{\protect\vphantom{T}}} \simeq \mathcal{O}(10^{-5} - 10^{-3})\text{ GeV}$ in agreement with~\cite{Dolan:2018qpy}. The washout which is related to the Yukawa couplings in the mass basis of the heavy SNs from eq.~(\ref{eqn:mddmd}), is dominantly controlled by the parameters of $m_L^{\protect\vphantom{T}}$ ($m_D^{\protect\vphantom{T}}$) when $\| m_L^{\protect\vphantom{T}} \| > \| m_D^{\protect\vphantom{T}}\|$ ( $\| m_D^{\protect\vphantom{T}}\| > \| m_L^{\protect\vphantom{T}} \|$). For large values of $\bm{m_L^{\protect\vphantom{T}}}$ the washout is not dependent on the size of the complex parameter $a_2^{\text{c}}$ whereas for small values where the parameters of $m_D^{\protect\vphantom{T}}$ dominate, different values of $a_2^{\text{c}}$ can produce different values of washout for the same choices in $m_L^{\protect\vphantom{T}}$.}
\label{figure:wash-lin}
\end{figure}

Figure~\ref{figure:LSS-asym} plots the asymmetry generated as a function of $\bm{m_L^{\vphantom{T}}}$, which we define to be the average of the non-zero entries of $m_L^{\vphantom{T}}$ in the three scenarios considered. Points in blue correspond to the asymmetry without including $\mathcal{N}_i^{\vphantom{T}}$ and $\mathcal{S}_i^{\vphantom{T}}$. As discussed previously near~\cref{eqn:vertexonlyeps}, in this regime (where all the SNs have identical masses) the self-energy contribution to the CP-asymmetry is switched off but a non-zero asymmetry is generated from the vertex contribution and differences in the decay widths of each SN. The most asymmetry (albeit much too small) is generated in the region where the washout is minimised, as shown~\cref{figure:wash-lin}, in agreement with~\cite{Dolan:2018qpy}. 

Points in orange correspond to spurion insertions at lowest order and points in cyan include all relevant terms. Now, due to the inclusion of the radiative Majorana masses, mass splittings occur between the six heavy SNs. The self-energy component of eq.~(\ref{cpasymmetry2}) turns on and increases the overall CP-asymmetry (due to a resonance between the SN masses) by several orders of magnitude. Here, including the next-to-leading order contributions does not change the size of the asymmetry generated. By contrast with with MLFV-ISS, the flavour mis-alignment between the matrix structure of $m_D^{\vphantom{T}}$ and $m_L^{\vphantom{T}}$ generates non-diagonal real entries in~\cref{eqn:mddmd} at lowest order which also allows for larger values of $\varepsilon_i^{\alpha}$.

The right side of~\cref{figure:LSS-asym} estimates the contribution to the asymmetry arising from coherent oscillations with the spurion contributions included. For the entire region of parameter space we considered, the asymmetry generated from coherent oscillations is of similar size to that generated from conventional thermal leptogenesis. The inclusion of oscillation effects will therefore slightly increase the total asymmetry, but we find no region where the effects of coherent oscillations dominate.

\begin{figure}[t]
\centering
{
  \includegraphics[width=0.6\linewidth]{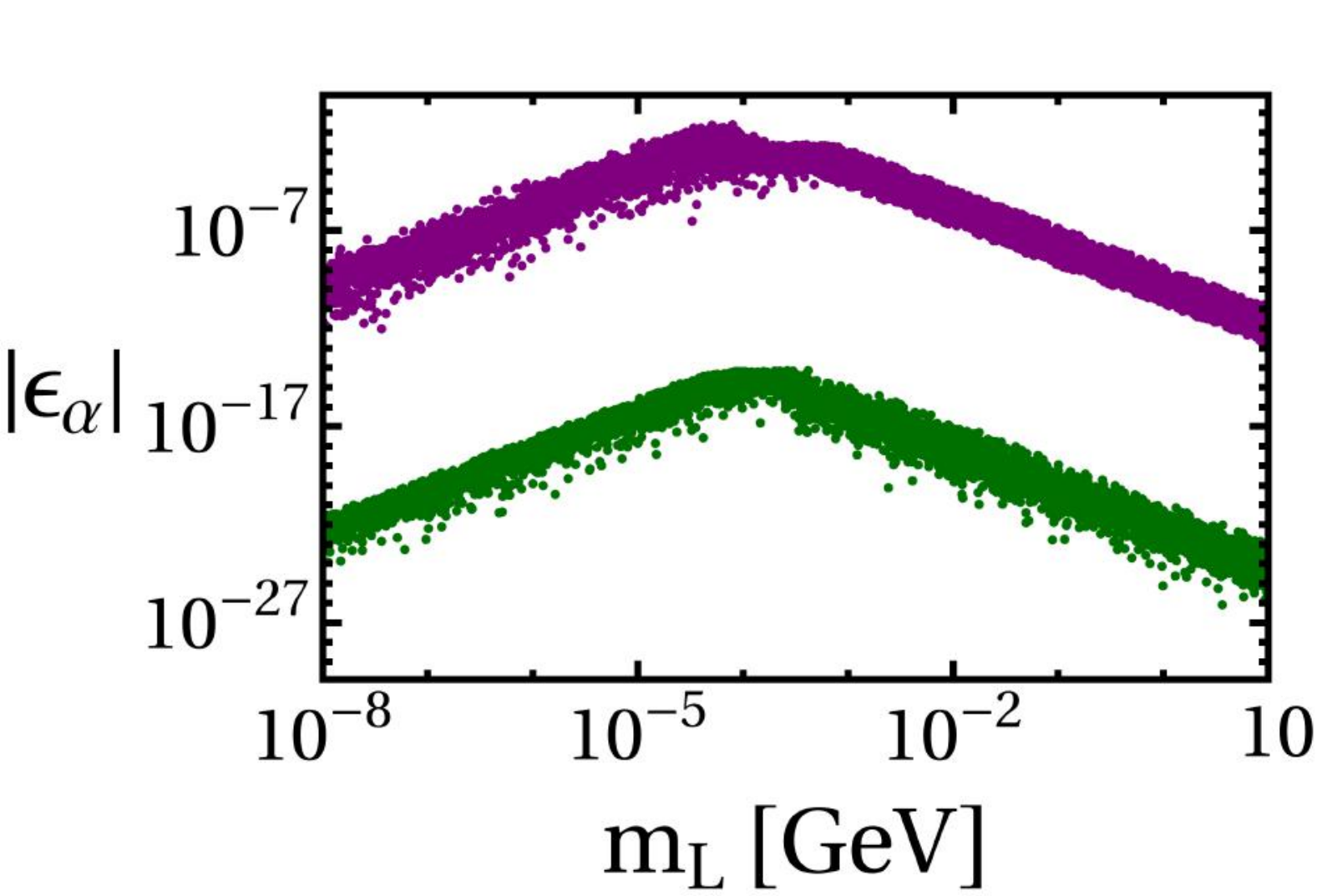}
}
\caption{Plot of the CP-asymmetry to a specific lepton flavour $\alpha$ as a function of $\bm{m_L^{\protect\vphantom{T}}}$. Points in green correspond to when $\mathcal{N}_i^{\protect\vphantom{T}}$ and $\mathcal{S}_i^{\protect\vphantom{T}}$ are not included and points in purple correspond to when they are included (at lowest order and next-to-leading order the points are identical). Clearly a resonance due to their inclusion increases the overall CP-asymmetry by many orders of magnitude. The mass splittings induced in this scenario are directly proportional to the decay widths and therefore for any choice of the parameters within $m_L^{\protect\vphantom{T}}$ the mass splittings will always be in a resonant regime.}
\label{figure:LSS-eps}
\end{figure}

Figure~\ref{figure:LSS-eps} plots the CP-asymmetry into a specific lepton flavour $\alpha$ in the three scenarios, where once again due to the anarchic nature each flavour has the same behaviour and overall size. It is clear that, for all values of $\bm{m_L^{\vphantom{T}}}$, the CP-asymmetry is orders of magnitude larger when the radiative Majorana masses are included. In other words, for the entire region of the parameter space, roughly the same resonant enhancement is occurring. This forced-resonance is occurring as the mass splitting between the heavy SNs is related to the small Majorana parameters,
\begin{eqnarray}
m_{N_j} &\simeq & m_R + \frac{\tilde{\mu}_{N}}{2} + \frac{\tilde{\mu}_{S}}{2} \nonumber\\[5pt]
m_{N_i} &\simeq & m_R - \frac{\tilde{\mu}_{N}}{2} - \frac{\tilde{\mu}_{S}}{2}
\end{eqnarray}
leading to
\begin{equation}
\Delta m_{ij} =  m_{N_i} - m_{N_j} = \tilde{\mu}_{N} + \tilde{\mu}_{S} = \mu_N^{\vphantom{T}}\,n_1\, \left( \mathcal{Y}_D^{\dagger} \mathcal{Y}_D^{\vphantom{\dagger}} \right) + \mu_S^{\vphantom{T}}\,s_1\, \left( \mathcal{Y}_L^{\dagger} \mathcal{Y}_L^{\vphantom{\dagger}} \right)\propto \Gamma_{i,j},
\end{equation}
where we take the parameters $\tilde{\mu}_{N}$ and $\tilde{\mu}_{S}$ to be independent of the parameters of $m_L^{\vphantom{T}}$. Therefore the same level of resonant enhancement occurs for the entire parameter space. Unlike with MLFV-ISS, both $\Delta m_{i,j}^{\text{OS}} \neq 0$ and $\Delta m_{i,j}^{\text{SS}} \neq 0$ occurs at lowest order in the corrections and they both contribute to the asymmetry generated.

\begin{figure}[t]
\centering
{
  \includegraphics[width=0.6\linewidth]{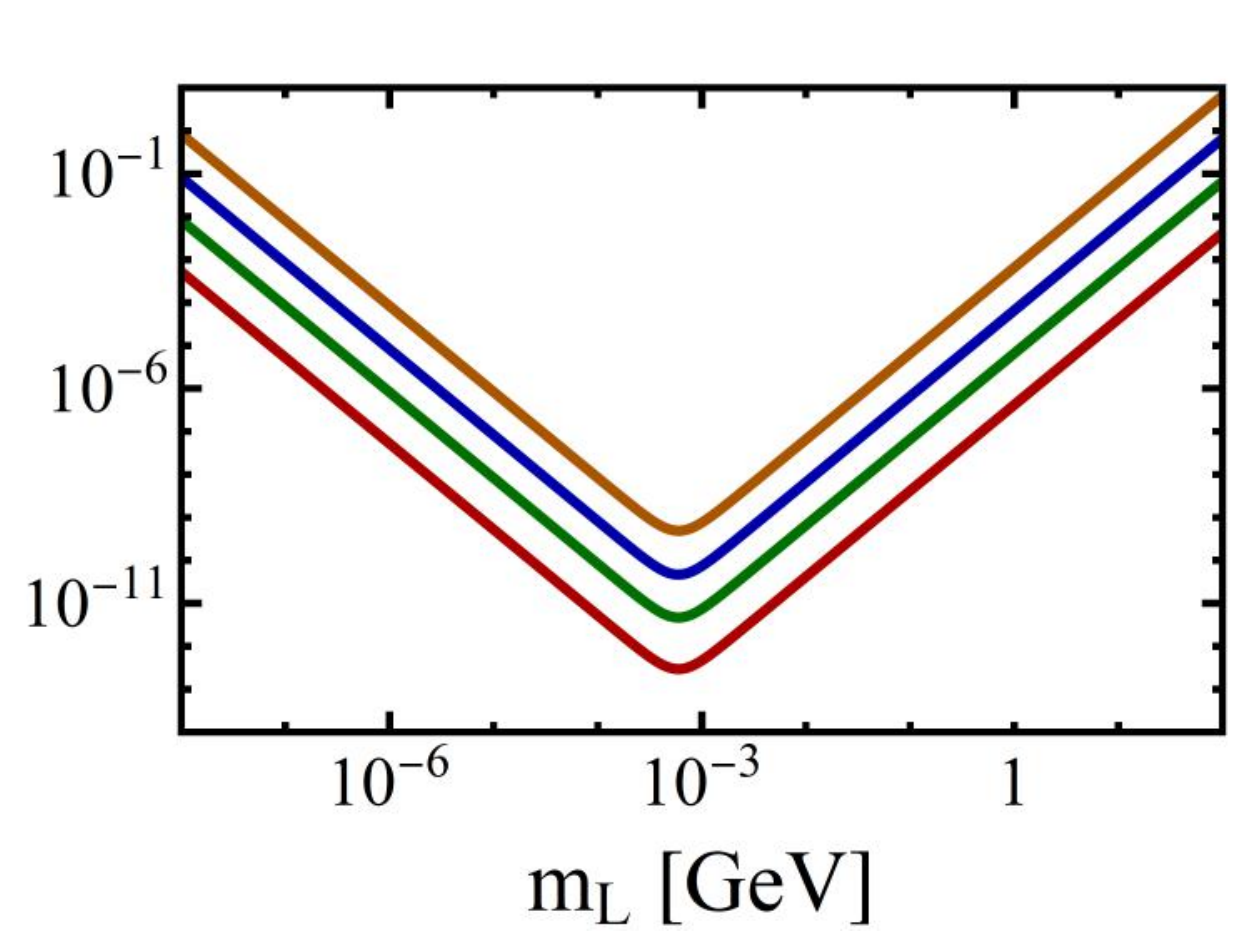}
}
\raisebox{20mm}{
\includegraphics[width=0.35\linewidth]{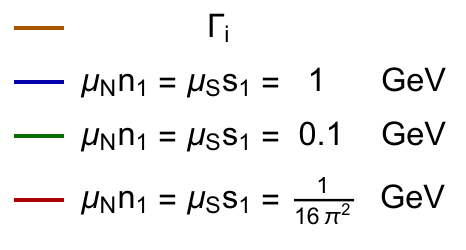}
}
\caption{Plot of the mass splitting $\Delta m_{ij}$ as a function of $\bm{m_L^{\protect\vphantom{T}}}$ in GeV when $\mathcal{N}_1^{\protect\vphantom{T}}$ and $\mathcal{S}_1^{\protect\vphantom{T}}$ are included. In red $\mu_N^{\protect\vphantom{T}} \,n_i = \mu_S^{\protect\vphantom{T}} \,s_i = 1/(4\pi)^2\text{ GeV}$, in green $\mu_N^{\protect\vphantom{T}} n_i = \mu_S^{\protect\vphantom{T}} s_i = 0.1 \text{ GeV}$ and in blue $\mu_N^{\protect\vphantom{T}} n_i = \mu_S^{\protect\vphantom{T}} s_i = 1\text{ GeV}$. This is plotted against the decay width $\Gamma_{i,j}$ in brown. As the combination $\mathcal{N}_1^{\protect\vphantom{T}} + \mathcal{S}_1^{\protect\vphantom{T}} \propto \Gamma_{i,j}$ the mass splitting induced will always be on resonance independent of the value of $m_L^{\protect\vphantom{T}}$. Here there is no distinction between $\Delta m_{i,j}^{\text{SS}}$ and $\Delta m_{i,j}^{\text{OS}}$ and they both behave in a similar way.}
\label{figure:LSS-diff-gamma}
\end{figure}

To illustrate this,~\cref{figure:LSS-diff-gamma} plots the mass splitting $\Delta m_{ij}$ and the decay width $\Gamma_i$ as a function of $m_L^{\vphantom{T}}$. The parameter which most significantly impacts the level of resonance (where the mass splitting overlaps with the decay width) is the combination of $\mu_N^{\vphantom{T}}\, n_i$ and $\mu_S^{\vphantom{T}}\, s_i$. The resonant enhancement is maximised when
\begin{equation}
\Delta m_{i,j} \simeq \Gamma_{i,j} = \frac{\left(h^{\dagger} h\right)}{8\pi} m_{N_i} \simeq 40 \left(h^{\dagger} h\right) \text{ GeV}.
\end{equation}
This places constraints on the overall size of the combination $\mu_N^{\vphantom{T}}\, n_i$ and $\mu_S^{\vphantom{T}} \,s_i$ required in order for enough asymmetry to be generated. While this forced resonance occurs, in order for it to significantly impact the asymmetry, it relies on the radiative Majorana masses generated arising from a scale around $\mathcal{O}(1-1000)\text{ GeV}$.

\begin{figure}[t]
\centering
{
  \includegraphics[width=0.45\linewidth]{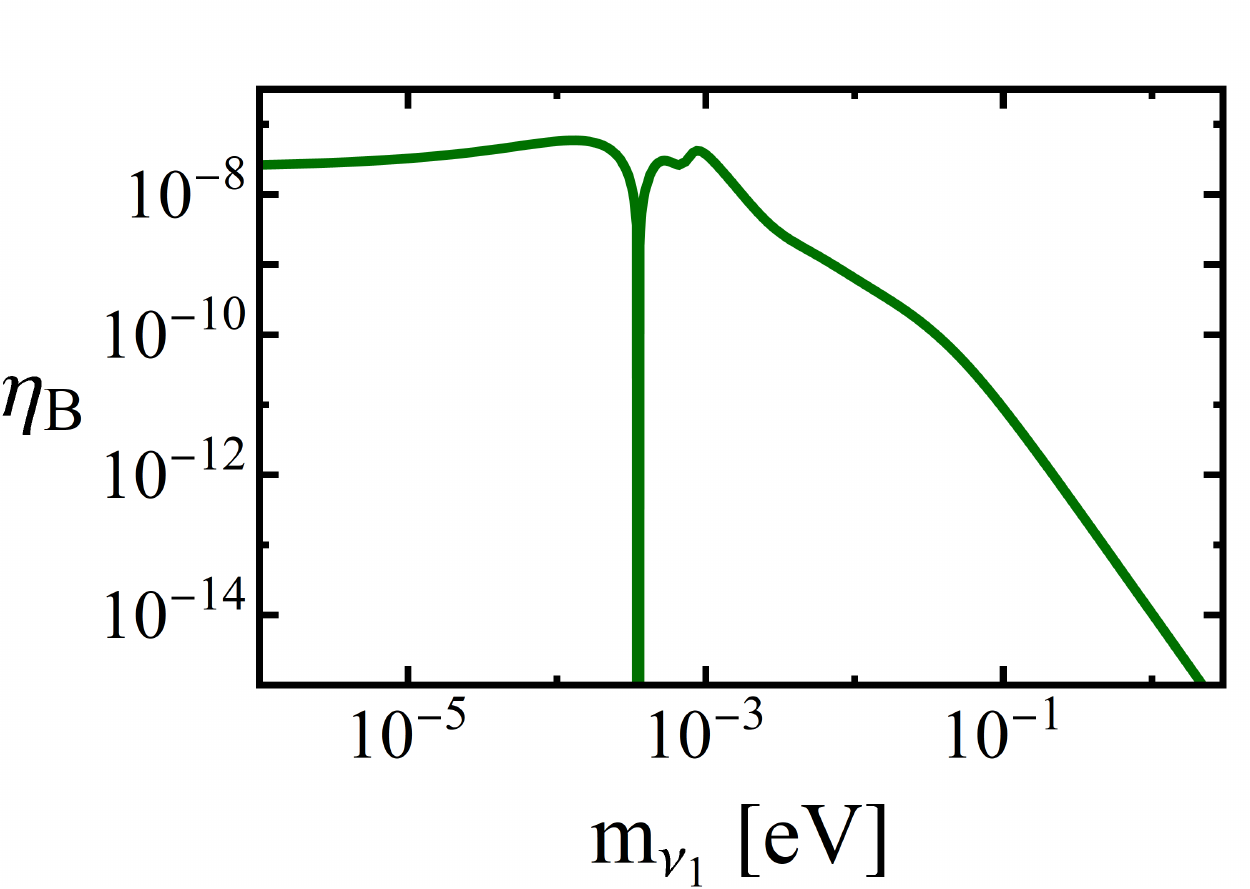}
}
{
  \includegraphics[width=0.45\linewidth]{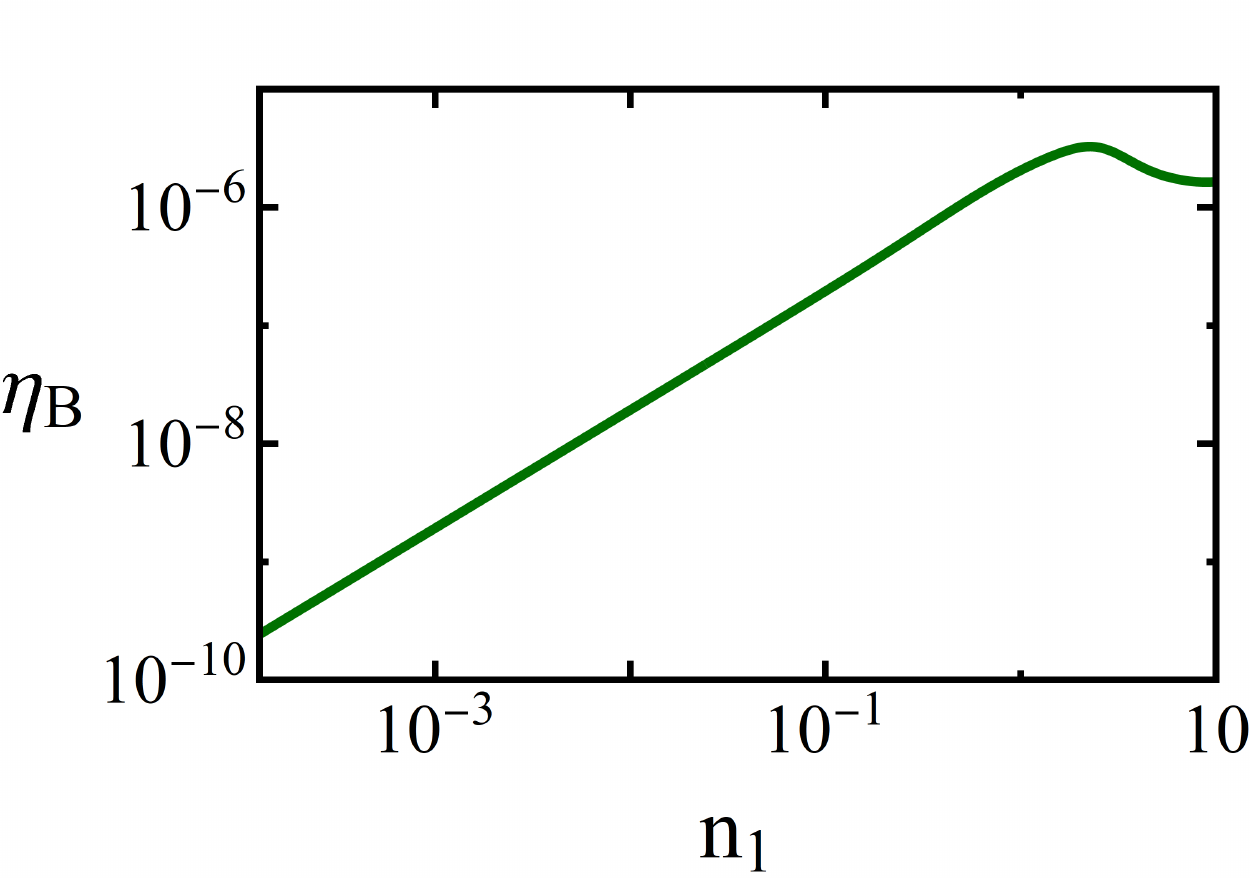}
}
\caption{Variation in the baryon asymmetry as a function of the lightest active neutrino mass $m_{\nu_1}$ \textbf{(left)} and varying the Wilson coefficients $n_i$ and $s_i$ \textbf{(right)} for the case where all radiative spurion effects are included. In this scan we fixed $a_2^{\text{c}} = 0.7$ and in the left plot set $n_i,\,s_i = 1/16\pi^2$ whereas we fixed $m_{\nu_1} = 0.01\text{ eV}$ in the right plot. In both plots similar behaviour compared to MLFV-ISS is found. }
\label{figure:assym-varymv1-lin}
\end{figure}

In~\cref{figure:assym-varymv1-lin} we vary the lightest neutrino mass $m_{\nu_1}$ and the Wilson coefficients for fixed choices of the other parameters as described in the figure caption. We find similar conclusions to MLFV-ISS where large values of the light neutrino mass $m_{\nu_1}$ correspond to smaller asymmetry generation. Masses less than $\mathcal{O}(10^{-3})\text{ eV}$  maximise the asymmetry generated. Larger values for the Wilson coefficient leads to a larger asymmetry and allows for a wider range of values within the parameters of $m_L^{\vphantom{T}}$ to produce the necessary asymmetry along with smaller values for the CPV parameters.

Based on these two scenarios, we conclude that successful MLFV resonant leptogenesis will also occur if the ISS and LSS were operative together. Appropriate choices for the now three LNV parameters based on the two scenarios here will allow for minimised washout with mass splittings related to the heavy SN decay widths for the necessary resonance to occur. However, as resonant leptogenesis is already feasible~\cite{Dolan:2018qpy} in this scenario without the need for radiative resonant leptogenesis, we do not consider this scenario further.

\begin{figure}[t]
\centering
{
  \includegraphics[width=0.45\linewidth]{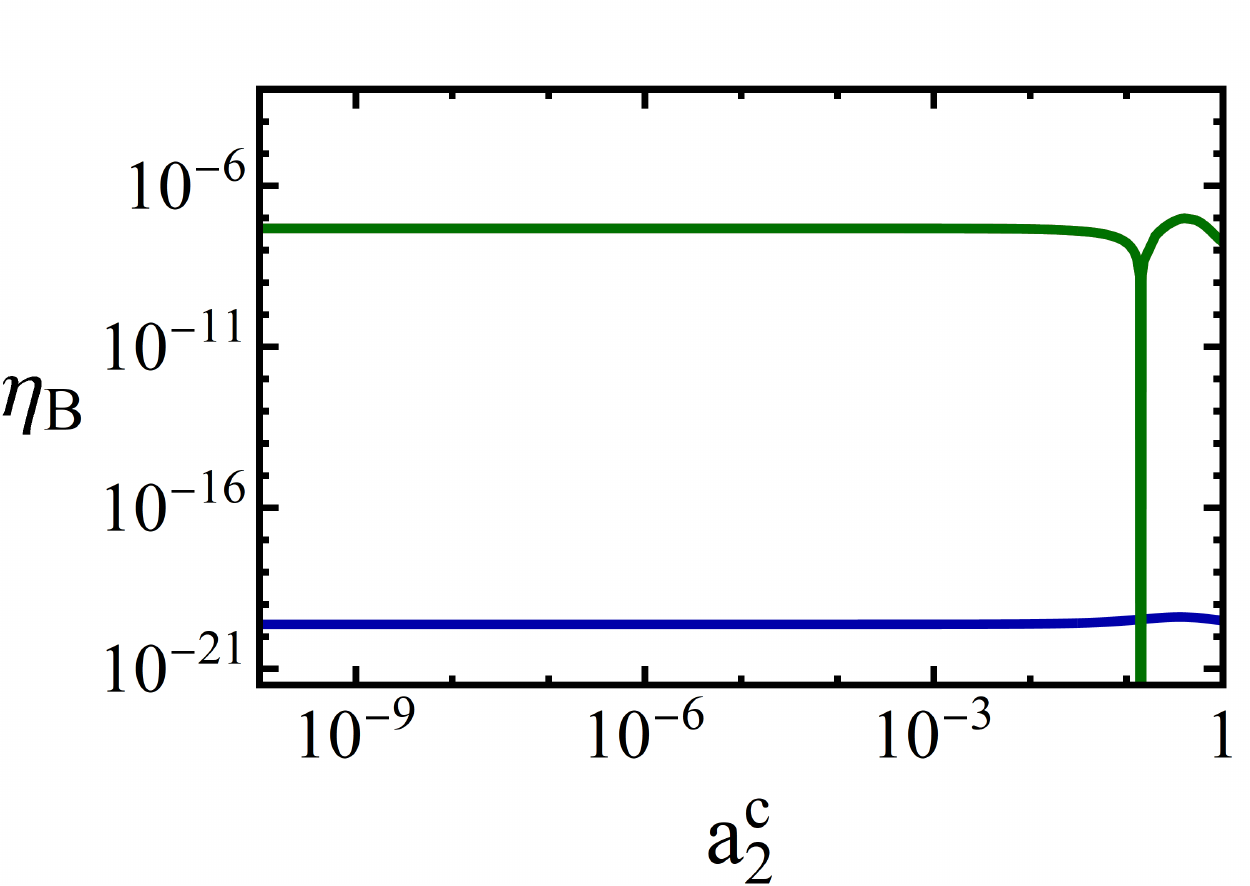}
}
{
  \includegraphics[width=0.45\linewidth]{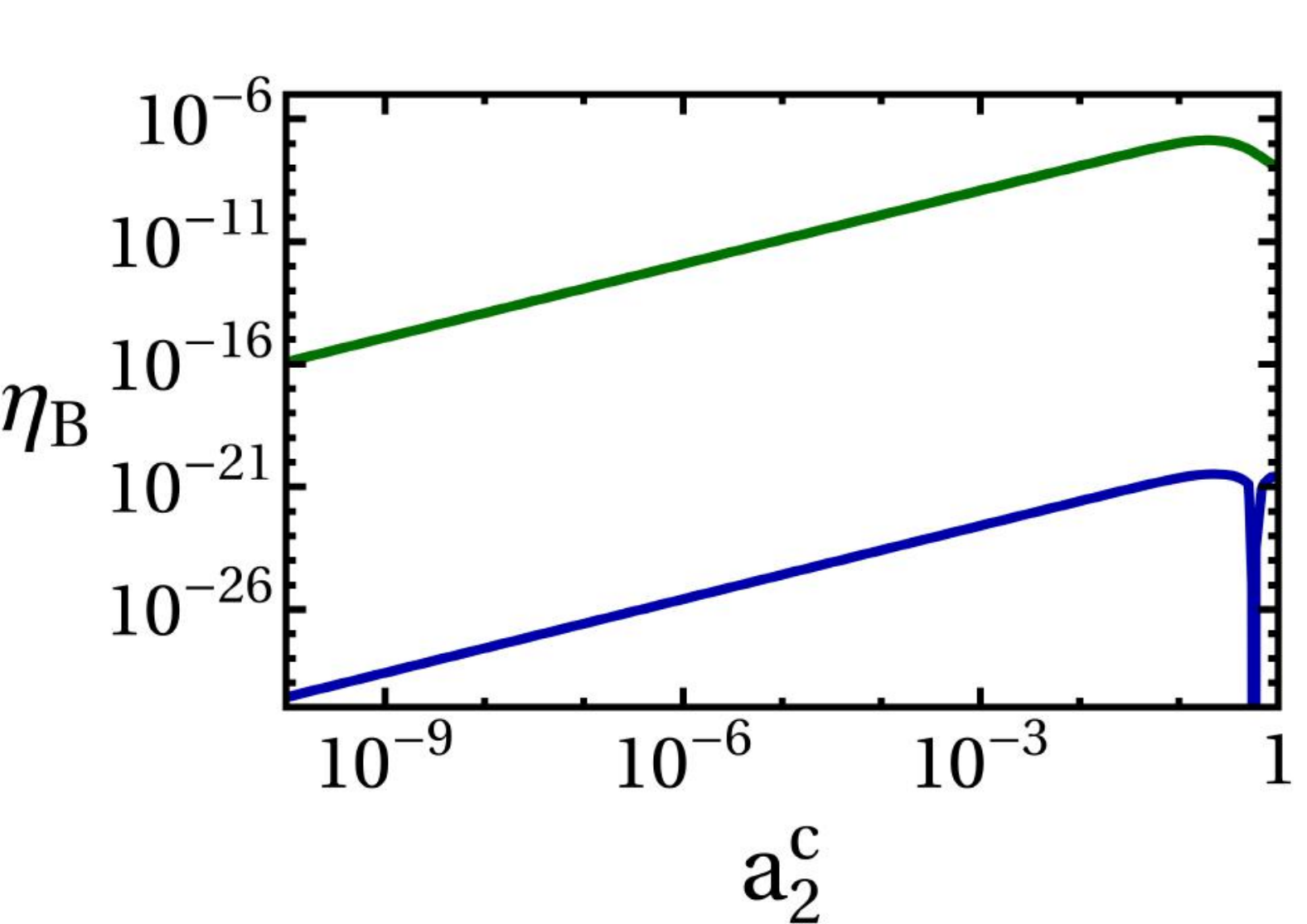}
}
\caption{Plot of the baryon asymmetry as a function of the complex parameter $a_2^{\text{c}}$ for $\delta_{CP}^{\protect\vphantom{T}}=3\pi/2$ \textbf{(let)} and $\delta_{CP}^{\protect\vphantom{T}} = 0$ \textbf{(right)}. Here we have fixed $m_L^{\protect\vphantom{T}} \simeq 10^{-4} \times \text{diag}(1,2,5) \text{ GeV}$ to be in a region with sufficient washout suppression for necessary asymmetry generation. In blue no radiative effects are included and in green is the scenario where spurion effects are included. The behaviour of the asymmetry as a function of the complex angle has consistent behaviour as it is taken to zero. As before asymmetry generation is possible both with low-energy phases as well as complex entries of the $C$ matrix where smaller values of $a_2^{\text{c}}$ are allowed compared to the ISS scenario.}
\label{figure:theta-varyY-lin}
\end{figure}

Finally, in~\cref{figure:theta-varyY-lin} we plot the behaviour of the asymmetry as a function of the complex component of the $C$ matrix when low-energy CPV is included and when it is not. Similarly to the ISS case, asymmetry generation can be predominately due to either the Dirac phase $\delta_{CP}$ or the complex component $a_2^{\text{c}}$ of the $C$ matrix. We find consistent behaviour for the baryon asymmetry as these CPV parameters are taken to zero. It is clear from both~\cref{figure:LSS-asym,figure:theta-varyY-lin} that a larger portion of the parameter space provides the necessary asymmetry generation allowing for smaller sizes of the CPV parameters, decreasing their contribution to flavour-violating processes.

\subsection{$\ell_i \rightarrow \ell_j \gamma$ and MLFV-LSS	 }

Once again we briefly consider the prospects of detection of MLFV-LSS through cLFV processes. Unlike in the MLFV-ISS scenario a much less tuned region of parameter space is required in order to generate the necessary baryon asymmetry which may lead to improved detection prospects.

As before we require insertions of spurions transforming as a $\bm{(3,\overline{3},1,1)}$ in order to make the necessary dimension six effective operators invariant. Off-diagonal terms are required in order for LFV processes to occur. The lowest order combination which satisfies this is once again 
\begin{equation}
\Delta \mathcal{Y}_{e}^{\vphantom{T}} = \mathcal{Y}_D^{\vphantom{T}} \mathcal{Y}_D^{\dagger} \mathcal{Y}_{e}^{\vphantom{T}} + \mathcal{Y}_L^{\vphantom{T}} \mathcal{Y}_L^{\dagger} \mathcal{Y}_{e}^{\vphantom{T}}.
\end{equation}
While the combination $\mathcal{Y}_L^{\vphantom{T}} \mathcal{Y}_L^{\dagger}$ may transform in the correct way, it does not contain the necessary off-diagonal terms in our scan and therefore cLFV will once again be controlled by $\mathcal{Y}_D^{\vphantom{T}} \mathcal{Y}_D^{\dagger}$.

\begin{figure}[t]
\centering
{
  \includegraphics[width=0.45\linewidth]{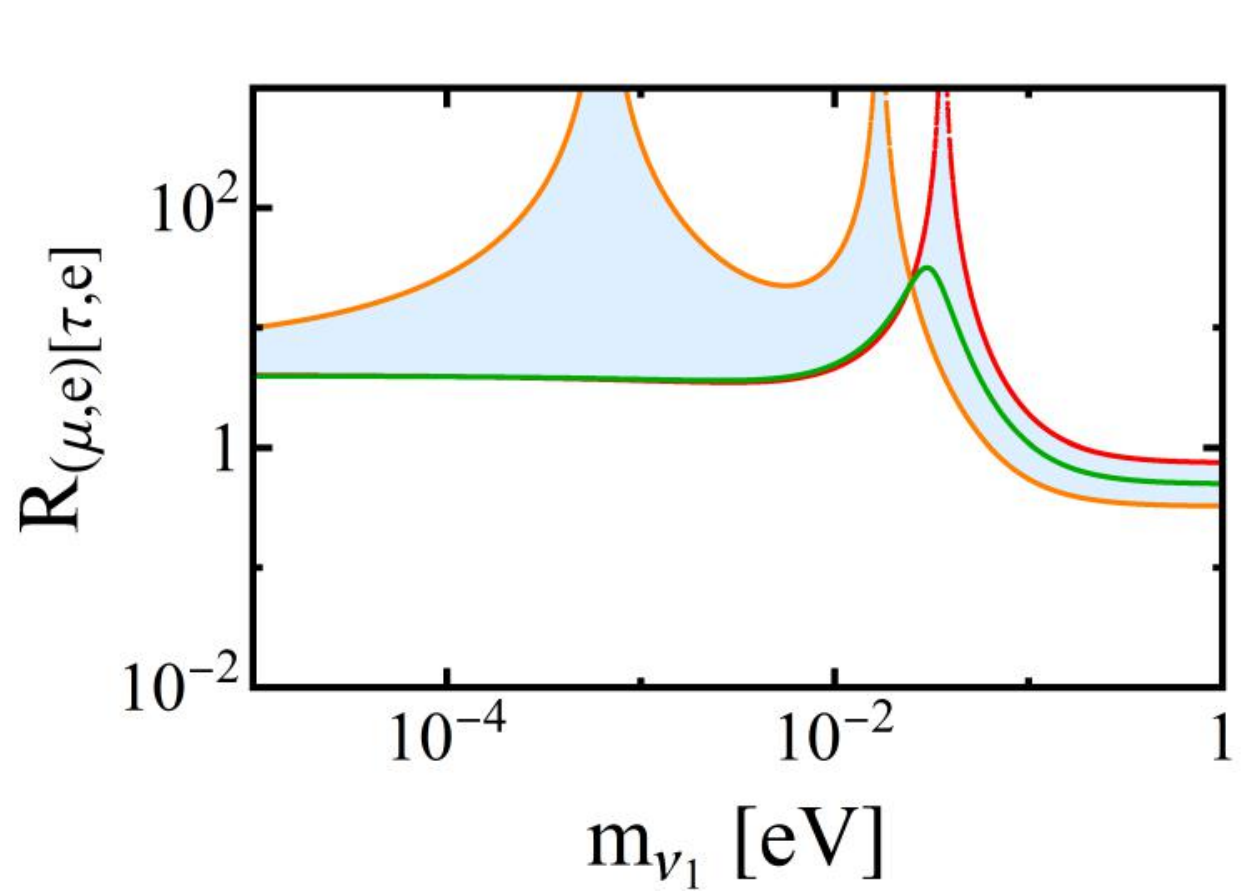}
}
{
  \includegraphics[width=0.45\linewidth]{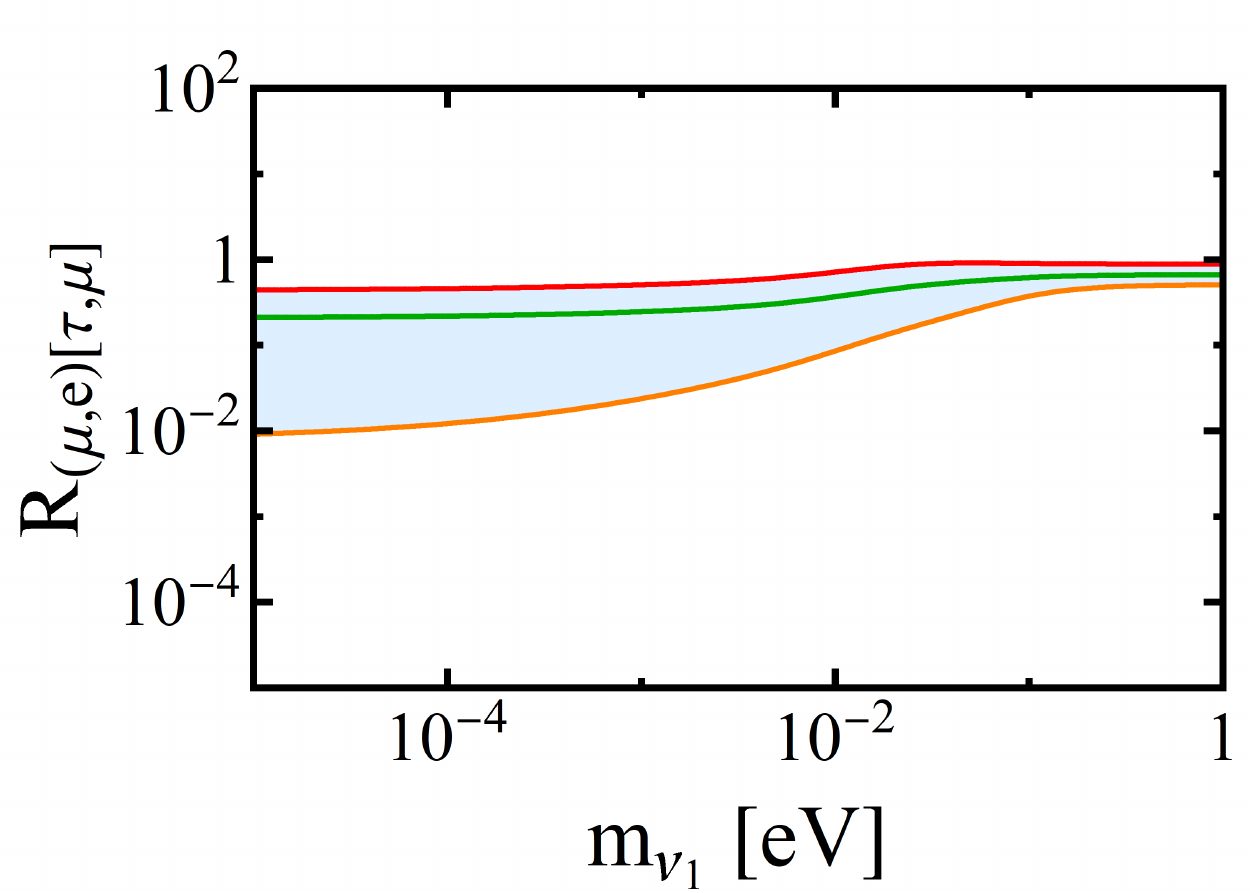}
}
{
  \includegraphics[width=0.45\linewidth]{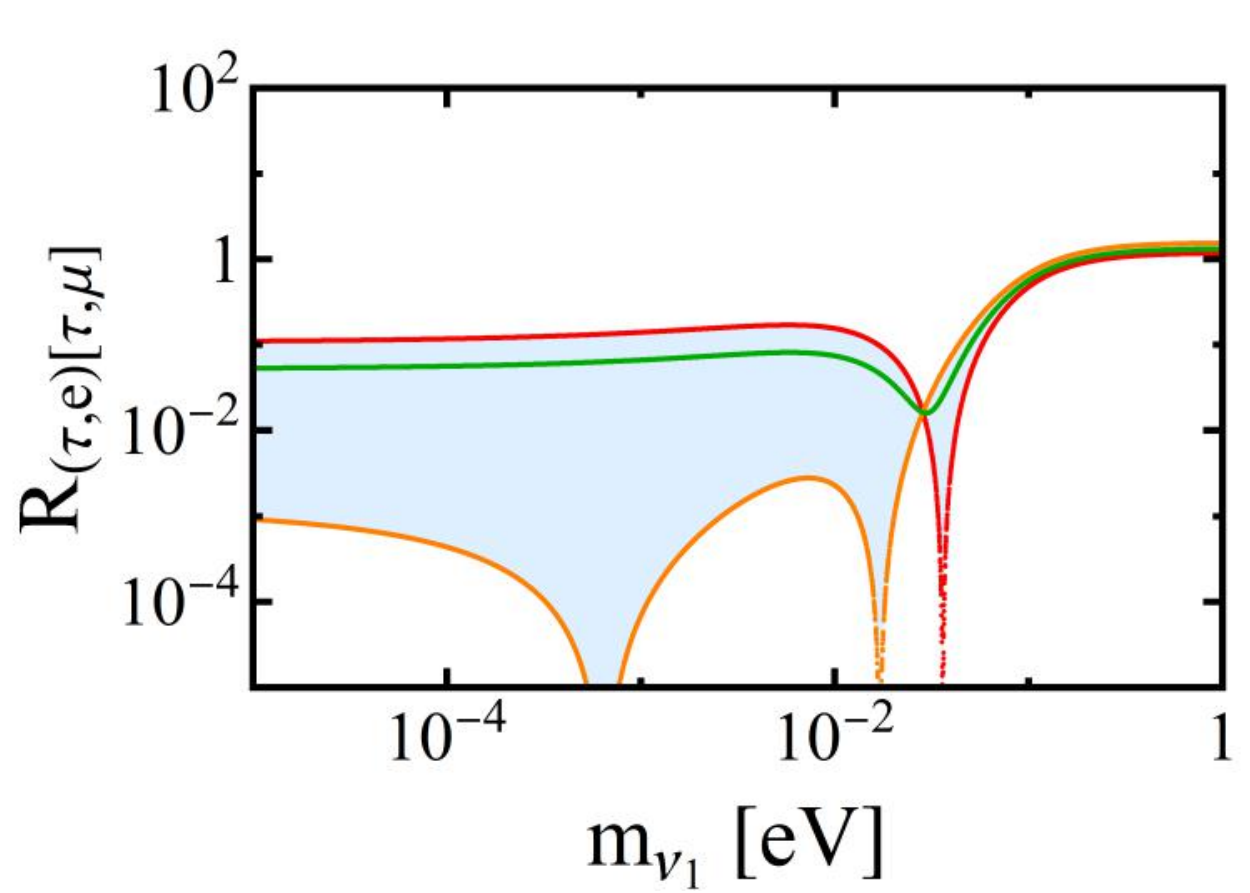}
}
\caption{Plot of the ratios $R_{(i,j)[k,l]}$ for various combinations of flavour initial and final states as a function of the lightest neutrino mass $m_{\nu_1}$. Here the complex parameter $a_2^{\text{c}}$ has been switched off but the low-energy phase $\delta_{CP}$ is varied. Lines in red and orange correspond to the CP-conserving cases $\delta_{CP} = 0 \text{ and } \delta_{CP} = \pi$ respectively. In green is the maximally violating case of $\delta_{CP} = \pi/2 \text{ or } \delta_{CP}=3\pi/2$. The shaded blue region corresponds to $\delta_{CP} = (0,2\pi) \backslash \{\pi/2,\,\pi,\,3\pi/2\}$.}
\label{figure:R-varydel-lin}
\end{figure}

Figure~\ref{figure:R-varydel-lin} plots the three ratios of cLFV observables as in the ISS case for the scenario where $a_2^{\text{c}} = 0$ while $\delta_{CP}$ is varying. Similar predictions on the ratio $R_{(\mu,e)[\tau,\mu]}$ are made compared to the ISS scenario and therefore the branching ratio BR($\tau \rightarrow \mu \gamma$) should be larger by one or two orders of magnitude compared to BR($\mu \rightarrow e \gamma$). Here there is a cancellation in $\tau \rightarrow e \gamma$ for specific values of the lightest neutrino $m_{\nu_1}$ and the phase $\delta_{CP}$. In these regions a strong suppression of this channel is predicted. Outside the regions of strong cancellation the LSS similarly predicts $R_{(\tau,e)[\tau,\mu]}<1$ but always predicts $R_{(\mu,e)[\tau,e]}>1$ unlike the type-I and ISS case where this varied depending on the value of $\delta_{CP}$. As before the introduction of CPV appears to bring the ratios closer together.

\begin{figure}[t]
\centering
{
  \includegraphics[width=0.4\linewidth]{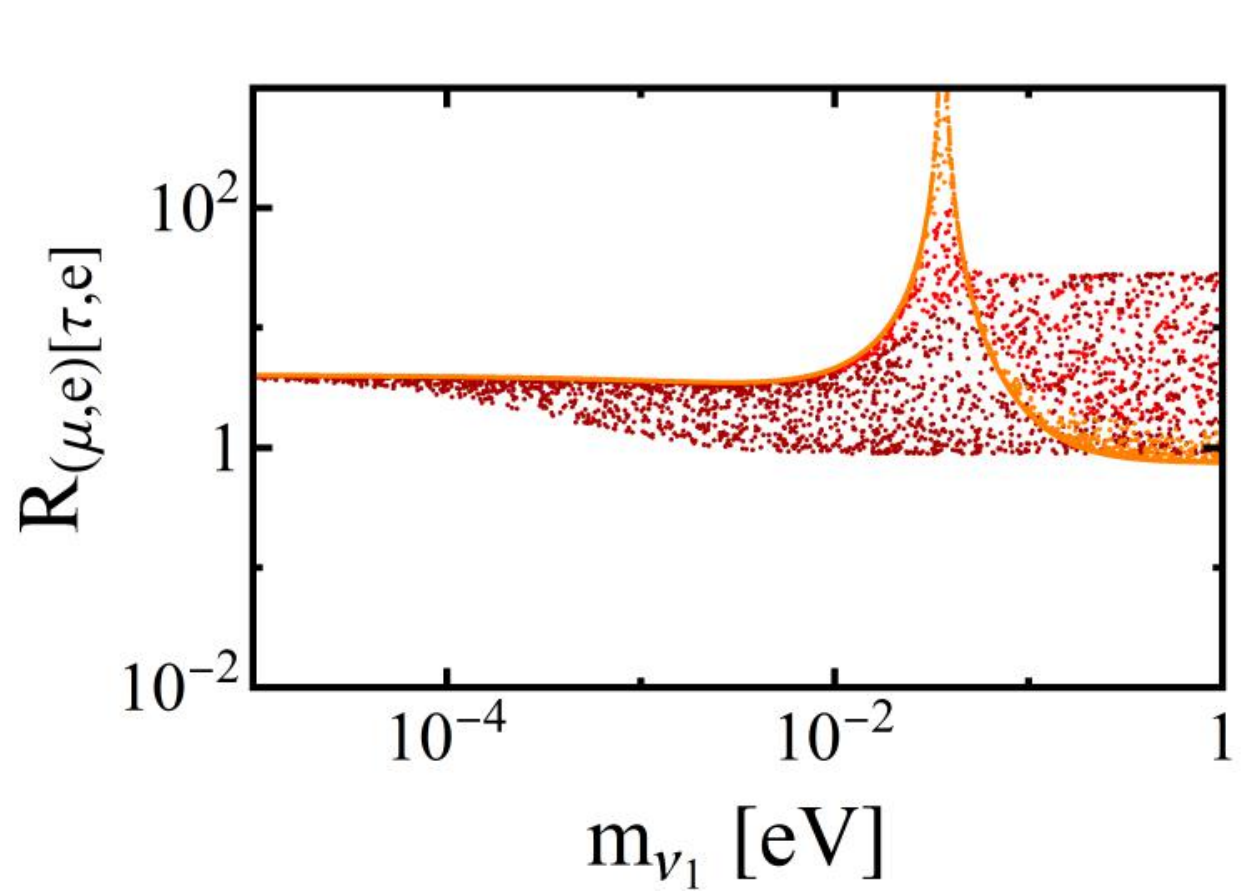}
}
{
  \includegraphics[width=0.4\linewidth]{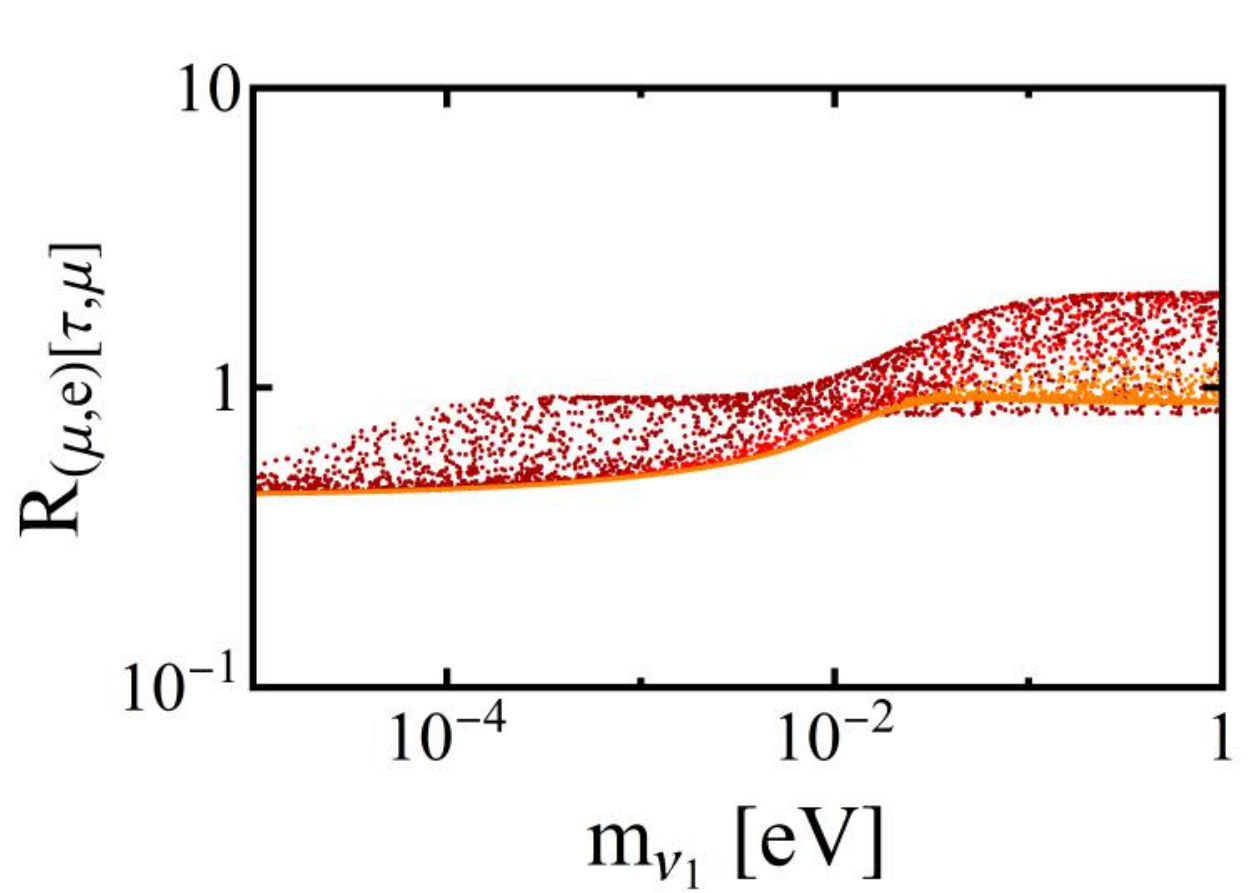}
}
{
  \includegraphics[width=0.4\linewidth]{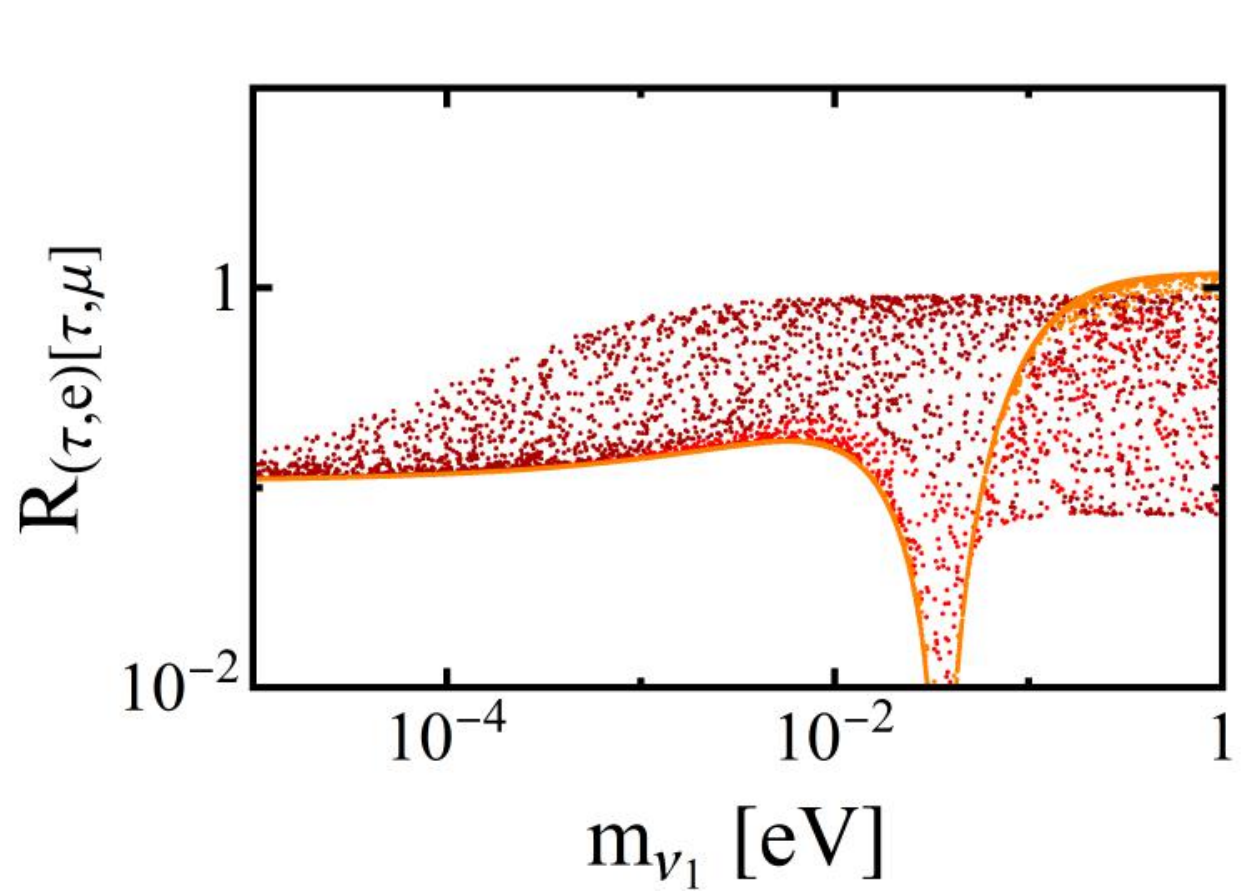}
}
\caption{Plot of the ratios $R_{(i,j)[k,l]}$ for various combinations of flavour initial and final states as a function of the lightest neutrino mass $m_{\nu_1}$. Here the low-energy phase $\delta_{CP}^{\protect\vphantom{T}}$ has been switched off but the complex parameter $a_2^{\text{c}}$ is varied. We find the same generic predictions as in the CP-conserving case of $R_{(\mu,e)[\tau,e]} > 1$ \textbf{(top-left)}  $R_{(\mu,e)[\tau,\mu]} \lesssim 1$ \textbf{(top-right)} and  $R_{(\tau,e)[\tau,\mu]} \lesssim 1$ \textbf{(bottom)}. Similar to the scenario above a cancellation occurs for the process $\tau \rightarrow e \gamma$ for specific values of the lightest neutrino mass $m_{\nu_1}$. In orange $a_2^{\text{c}} < 0.1$, in red $0.1 < a_2^{\text{c}} < 0.3$ and in burgundy $a_2^{\text{c}} > 0.3$.}
\label{figure:R-varyY-lin}
\end{figure}

Figure~\ref{figure:R-varyY-lin} plots the ratios of cLFV observables in the scenario where $\delta_{CP}=0$ and $a_2^{\text{c}}$ is varied. We find that overall the presence of the CPV parameter becomes more impactful for larger values of the lightest neutrino mass, while for a hierarchical spectrum its impact is less significant. Overall similar predictions to the case where $\delta_{CP}$ was varied are obtained and a significant portion of the parameter space is not strongly sensitive to the presence of CPV. This implies the inclusion of CPV necessary for leptogenesis does not significantly modify the predictions given by the CP-conserving case.

\begin{figure}[t]
\centering
{
  \includegraphics[width=0.4\linewidth]{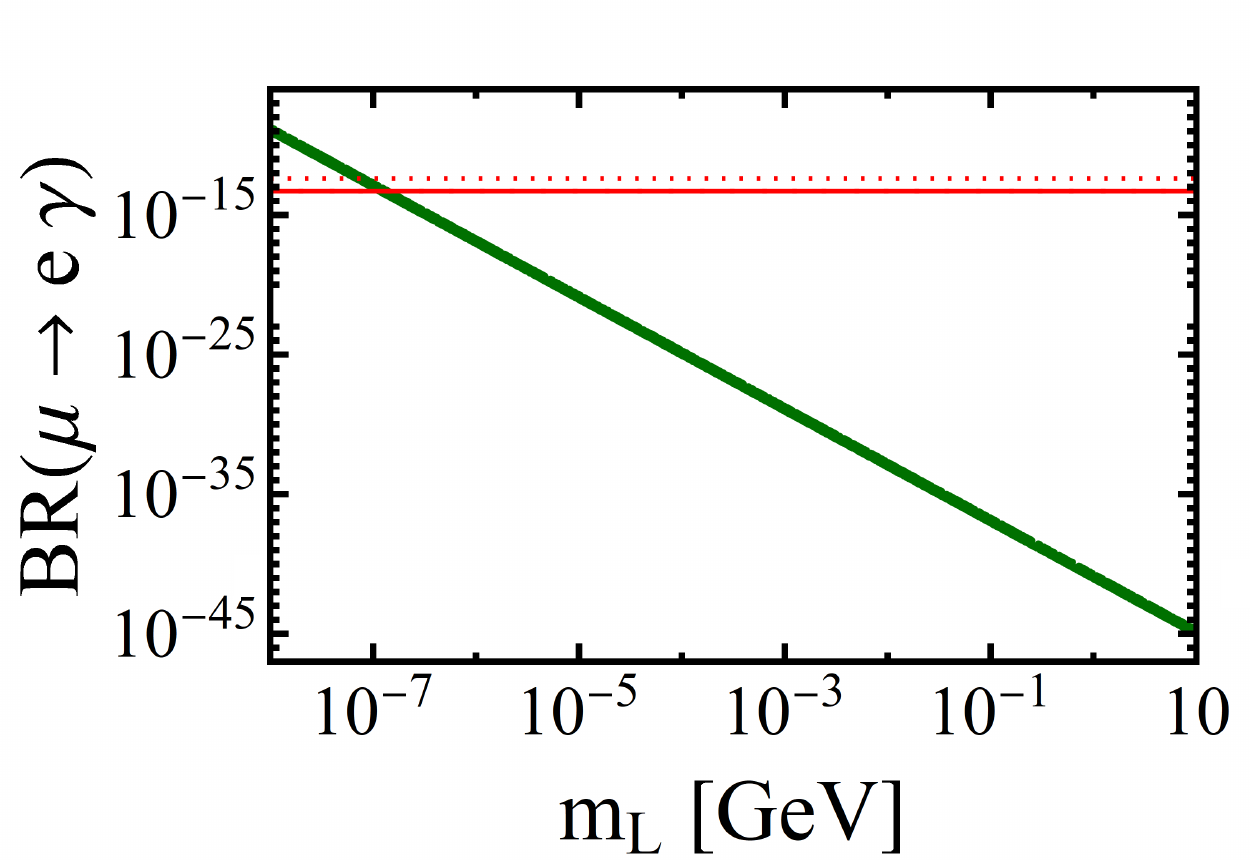}
}
{
  \includegraphics[width=0.4\linewidth]{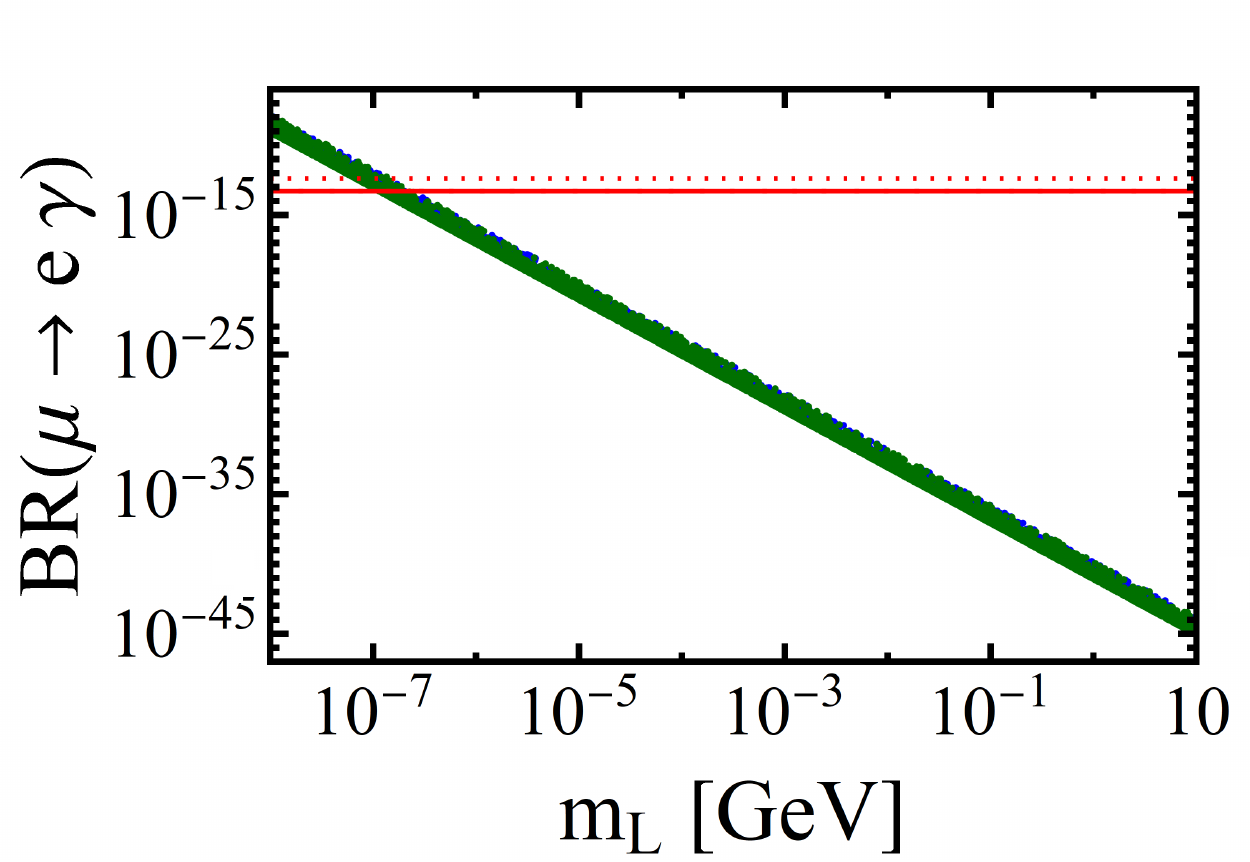}
}
\caption{Plot of the branching ratio of $\mu \rightarrow e \gamma$ for $\theta_2^{\text{c}} = 0$ \textbf{(left)} and $0.1 < a_2^{\text{c}} < 1$ \textbf{(right)} as a function of $m_L^{\protect\vphantom{T}}$. Points in blue correspond to the choice $\delta_{CP}^{\protect\vphantom{T}} = 0$ whereas points in green correspond to $\delta_{CP}^{\protect\vphantom{T}} = 3\pi/2$.  The red dotted(solid) line corresponds to the current(future) sensitivity of MEG and MEG-II~\cite{TheMEG:2016wtm,Cattaneo:2017psr} respectively for this decay mode. Currently very small values of the LNV parameter are probed for MLFV-LSS which correspond to a very large seperation of the LNV and LFV scales. In order to allow for necessary asymmetry generation we require approximately $10^{-5} \lesssim m_L^{\protect\vphantom{T}}/\text{GeV} \lesssim 10^{-3}$ which will not be probed by MEG-II in the near future. The inclusion of CPV of any type does not significantly impact the predictions.}
\label{figure:meg-LSS}
\end{figure}

Finally,~\cref{figure:meg-LSS} plots the predictions for the branching ratio BR($\mu \rightarrow e \gamma$) as a function of the LNV parameter $m_L^{\vphantom{T}}$ for $\Lambda_{\text{LFV}} = 1\text{ TeV}$. As before, we plot cases with no CPV violation and cases in which CPV is present. The dotted (solid) red line corresponds to the current (future) limit placed by MEG(MEG-II). Due to the LSS parameterisation in~\cref{eqn:LSScasas}, the Dirac-mass matrix $m_D^{\vphantom{T}}$ is now inversely related to $m_L^{\vphantom{T}}$ and is more senesitive to parameters within $m_L^{\vphantom{T}}$ changing compared to the Majorana masses in the ISS scenario. While small values of $m_L^{\vphantom{T}}$ are currently constrained, the region $10^{-5} \lesssim \bm{m_L^{\vphantom{T}}/}  \text{GeV} \lesssim 10^{-3}$ required in order to account for baryogenesis will not be probed in near-future experiments. This roughly corresponds with the LNV scale $\Lambda_{\text{LNV}} \simeq (10^{6} - 10^{4})\text{ GeV}$ which will not be probed in the near future. This is in agreement with the estimates found in~\cite{Dinh:2017smk} for the necessary LNV scale for a given LFV scale to be probed. A future measurement of the process $\mu \rightarrow e \gamma$ may rule out LSS as a leptogenesis candidate in the absence of additional physics introduced to lower the overall strength of the washout present for smaller values of $m_L^{\vphantom{T}}$.

\section{Conclusion}
\label{sec:conclusion}

We have studied a well-motivated way in which small mass splittings between heavy SNs from different families may arise, within both the ISS and LSS frameworks, such that leptogenesis is possible despite the strong washout present in the theory. Previously it was found that while a mass splitting naturally exists for the ISS it is not sufficient in order for resonant leptogenesis to occur. For the LSS the degeneracy amongst the SNs at the high scale prevents significant asymmetry generation. While leptogenesis is feasible when all LNV terms are switched on (the ISS+LSS case) we explore the potential for ISS or LSS leptogenesis to occur independently, where additional symmetries may prevent both terms from existing. 

In the context of broken flavour symmetries and the MLFV hypothesis, a degeneracy amongst the heavy SNs is naturally produced, for the purposes of having a predictive theory. The degeneracy is then broken by higher-order spurion VEV contributions, leading to a parameter region consistent with resonant asymmetry generation. In order for the desired splitting the occur in the intended way during cosmological evolution, the critical temperature at which the spurions acquire their non-zero VEVs must be assumed to be above the scale of thermal leptogenesis.

We found that for MLFV-ISS only a small region of very large Majorana masses is able to generate the required asymmetry. Here asymmetry generation requires next-to-leading order corrections to be included, therefore suppressing the overall size of the CP-asymmetry generated per decay of SN. We briefly discussed the impact of CPV on potential low-energy observables, in particular how various ratios of cLFV decays are impacted compared to the CP conserving MLFV scenario. The region compatible with leptogenesis will not, however, be probed by current and near future cLFV experiments.

For MLFV-LSS a large region of parameter space is capable of satisfying the resonance condition simultaneously with the minimised washout required for successful asymmetry generation. Here corrections at lowest order are not flavour aligned and therefore much larger values of the CP-asymmetry are possible compared to the ISS. Similarly, we studied the impact of CPV on cLFV observables and find that small deviations occur due to their inclusion. In this case, relatively small values of the CPV parameters allow for sufficient asymmetry generation allowing for even smaller deviations as compared to the ISS case.

In both cases we estimated the impact of the lightest neutrino mass $m_{\nu_1}$ on the asymmetry generated. We find a clear preference for small values where the light neutrinos are hierarchical and estimate that $m_{\nu_1} \lesssim 10^{-2} \text{ eV}$ is required. Unsurprisingly we find that MLFV leptogenesis favours larger values for the Wilson coefficients related to the Majorana mass corrections in both scenarios.

\begin{acknowledgments}
This work was supported in part by the Australian Research Council. TPD thanks Vincenzo Cirigliano and Luca Merlo for clarifications related to MLFV. TPD is grateful to Uli Felzmann for access to computing hardware on which our scans were conducted.
\end{acknowledgments}

\bibliography{bibliography}

\end{document}